\documentclass[11pt,a4paper]{article}
\usepackage{ccaption}
\usepackage[utf8]{inputenc}
 \usepackage{xr}
\usepackage{rotating}
\usepackage{amsmath,amssymb}
\usepackage{setspace}
\usepackage{afterpage}
\usepackage{geometry}
\usepackage{graphicx}
\usepackage{sectsty}
\sectionfont{\fontsize{12}{15}\selectfont}
\usepackage[english]{babel}
\usepackage[nottoc]{tocbibind}
\usepackage{tocloft}
\cftsetindents{section}{1em}{2em}

\usepackage{fancyhdr}
\usepackage{indentfirst}
\usepackage{multirow}
\usepackage{url}
\usepackage{soul}
\usepackage{cite}
\usepackage{array}
\usepackage{color}
\usepackage{afterpage}
\definecolor{mygray}{gray}{0.6}
\usepackage{ccaption} 
\usepackage[utf8]{inputenc}
\usepackage{mathrsfs}
\usepackage{amsmath,amssymb}
\usepackage{setspace}
\usepackage{afterpage}
\usepackage{geometry}
\usepackage{graphicx}
\usepackage{floatrow}
\usepackage{sectsty}
\sectionfont{\fontsize{12}{15}\selectfont}
\usepackage[english]{babel}
\usepackage{fancyhdr}
\usepackage{indentfirst}
\usepackage{siunitx}
\usepackage{cite}
\usepackage{array}
\newcolumntype{C}[1]{>{\centering\arraybackslash}p{#1}}
\usepackage{flexisym}
\usepackage{longtable}
\usepackage[skip=0pt]{caption}
\usepackage{float}
\floatstyle{plaintop}
\restylefloat{table}
\usepackage{lineno}

\usepackage[skip=0pt]{caption}
\usepackage{multibib}
\newcites{supp}{Supplementary References}

\makeatletter
\renewcommand\part{%
  \secdef\@part\@spart}
\makeatother

\makeatletter
\def\@part[#1]#2{%
  \ifnum \c@secnumdepth >-2\relax \refstepcounter{part}%
    \addcontentsline{toc}{part}{\partname\ \thepart.%
        \protect\partauthor\protect\enspace\protect\noindent#1}%
  \else
    \addcontentsline{toc}{part}{#1}\fi}
\makeatother

\makeatletter
\def\@spart#1{\addcontentsline{toc}{part}%
  {\protect\partauthor\protect\noindent#1}%
    }
\makeatother

\usepackage{minitoc}

\newgeometry{left=1in,right=1in,top=1in,bottom=1in}

\newcommand{\vecm}{\mathbf{m}}
\newcommand{\graph}{G}

\title{\Large AI-Bind: Improving Binding Predictions for Novel Protein Targets and Ligands}

\makeatletter
\renewcommand*{\@fnsymbol}[1]{\ensuremath{\ifcase#1\or \dagger\or *\or  \ddagger\or
    \mathsection\or \mathparagraph\or \|\or **\or \dagger\dagger
    \or \ddagger\ddagger \else\@ctrerr\fi}}
\makeatother    

\author{Ayan Chatterjee$^{1,2}$,\and Robin Walters$^{3}$,\and Zohair Shafi$^{3}$,\and Omair Shafi Ahmed $^{3}$,  \and Michael Sebek$^{1}$, \and Deisy Gysi$^{1,4}$, \and Rose Yu$^{5}$, \and Tina Eliassi-Rad$^{2,3}$ ,\and Albert-L\'aszl\'o Barab\'asi$^{1,4,6}$, \and Giulia Menichetti$^{1,4}$\footnote{Corresponding author. e-mail: menicgiulia@gmail.com}}

\date{%
    $^1${\small Center for Complex Network Research, Northeastern University, Boston, USA}\\
    \vspace{-7pt}
    $^2${\small Network Science Institute, Northeastern University, Boston, USA}\\
    \vspace{-7pt}
    $^3${\small Khoury College of Computer Sciences, Northeastern University, Boston, USA}\\
    \vspace{-7pt}
    $^4${\small Department of Medicine, Brigham and Women's Hospital, Harvard Medical School, Boston, USA}\\
    \vspace{-7pt}
    $^5${\small Department of Computer Science and Engineering, University  of California, San Diego, USA}\\
    \vspace{-7pt}
    $^6${\small Center for Network Science, Central European University, Budapest, Hungary}\\
    \vspace{-7pt} 
}

\begin{document}

\doparttoc 
\faketableofcontents 

\part{} 

\maketitle



\section{ABSTRACT}

Identifying novel drug-target interactions (DTI) is a critical and rate limiting step in drug discovery. While deep learning models have been proposed to accelerate the identification process, we show that state-of-the-art models fail to generalize to novel (i.e., never-before-seen) structures.
We first unveil the mechanisms responsible for this shortcoming, demonstrating how models rely on shortcuts that leverage the topology of the protein-ligand bipartite network, rather than learning the node features. Then, we introduce AI-Bind, a pipeline that combines network-based sampling strategies with unsupervised pre-training, allowing us to limit the annotation imbalance and improve binding predictions for novel proteins and ligands.  We illustrate the value of AI-Bind
by predicting drugs and natural compounds with binding affinity to SARS-CoV-2 viral proteins and the associated human proteins. We also validate these predictions via docking simulations and comparison with recent experimental evidence, and step up the process of interpreting machine learning prediction of protein-ligand binding by identifying potential active binding sites on the amino acid sequence. Overall, AI-Bind offers a powerful high-throughput approach to identify drug-target combinations, with the potential of becoming a powerful tool in drug discovery. 


\section{INTRODUCTION}

The accurate prediction of binding interactions between chemicals and proteins is a critical step in drug discovery, necessary to identify new drugs and novel therapeutic targets, and to reduce the failure rate in clinical trials, and to predict the safety of drugs \cite{Hughes2011, Thafar2019, U.S.Food&DrugAdministration}. While molecular dynamics and docking simulations \cite{Hollingsworth2018, Vivo2016, Meng2011} are frequently employed to identify potential protein-ligand binding, the computational complexity (namely, run-times) of the simulations and the lack of 3D protein structures significantly limit the coverage and the feasibility of large-scale testing. Therefore, machine learning (ML) and artificial intelligence (AI) based models have been proposed to circumvent the computational limitations of the existing approaches \cite{Chen2018,alphafold2021}, leading to the development of models that rely either on deep learning architectures or chemical feature representations \cite{Huang2020,Zhang2020DeepBindPoc,huang2018densely,Verma2019,Cui2019,Zhao2020_exploration,Xia2020}. 

Deep learning frameworks formulate the binding prediction problem as either a binary classification task or a regression task. The successful training of a binary classifier requires positive samples, pairs of proteins and ligands that are known to bind to each other, typically extracted from protein-ligand binding databases like DrugBank \cite{10.1093/nar/gkm958}, BindingDB \cite{10.1093/nar/gkv1072}, Tox21 \cite{doi:10.1021/acs.chemrestox.0c00264}, ChEMBL \cite{Davies2015}, Davis \cite{Davis2011}, or Drug Target Commons \cite{Tang2018}. Training also requires negative samples, i.e., pairs that do not interact or only weakly interact. However, the positive and negative annotations associated with different proteins and ligands are not evenly distributed, but some proteins and ligands have disproportionately more positive annotations than negative ones, and vice-versa, an annotation imbalance learned by the ML models, which then predict that some proteins and ligands bind disproportionately more often than the others. In other words, the ML models learn the binding patterns from the degree of the nodes in the protein-ligand interaction network, neglecting relevant node metadata, like the chemical structures of the ligands or the amino-acid sequences of the proteins \cite{Huang2020,Hu2021,vanLaarhoven2011,ztrk2016}. This annotation imbalance leads to good performance as quantified by the Area Under the Receiver Operating Characteristics (AUROC) and  the Area Under the Precision Recall Curve (AUPRC) for the unknown annotations associated with missing links in the protein-ligand interaction network used for training. A key signal of such shortcut learning is the degradation of the performance of an ML model when asked to predict binding between novel (i.e., never-before-seen) protein targets and ligands. This modeling limitation is in-line with the findings of Geirhos et al. \cite{Geirhos2020}, who showed that deep learning methods tend to exploit shortcuts in training data to achieve good performance. Laarhoven et al. discuss similar bias in drug–target interaction data and its effect on cross-validation performance \cite{vanLaarhoven2014}. Lee et al. \cite{Lee2016} and Wang et al. \cite{Wang2016} proposed approaches that partly address shortcut learning, but fail to generalize to unexplored proteins, i.e., proteins that lack sufficient binding annotations, or originate from organisms with no close relatives in current protein databases. More recently, models such as MolTrans \cite{Huang2020_Moltrans}, MONN \cite{Li2020_Monn}, KGE\_NFM \cite{Ye2021_KGE_NFM}, TransDTI \cite{Kalakoti2022_TransDTI}, HoTS \cite{Lee2022_HoTS}, and DTIHNC \cite{Jiang2022_DTIHNC}, explore innovative structural representations of protein and ligand molecules. Though these models better leverage the molecular structures to predict binding, end-to-end training limits their ability to generalize beyond the molecular scaffolds present in the training data.

Here, we introduce AI-Bind, a pipeline for predicting protein-ligand binding which can successfully generalize to unseen proteins and ligands.
AI-Bind combines network science methods with unsupervised pre-training to control for the over-fitting and the annotation imbalance of existing libraries. We leverage the notion of shortest path distance on a network to identify distant protein-ligand pairs as negative samples. Combining these network-derived negatives with experimentally validated non-binding protein-ligand pairs, we ensure sufficient positive and negative samples for each node in the training data. Additionally, AI-Bind learns, in an unsupervised fashion, the representation of the node features, i.e., the chemical structures of ligand molecules or the amino-acid sequences of protein targets, helping circumvent the model's dependency on limited binding data. Instead of training the deep neural networks in an end-to-end fashion using  binding data, we pre-train the embeddings for proteins and ligands using larger chemical libraries, allowing us to generalize the prediction task to chemical structures, beyond those present in the training data.


\section{RESULTS}

\subsection*{Limitations of existing ML models}

\begin{figure}[t]
    \centering
    \includegraphics[clip,angle=0,width=\textwidth,height=0.8\textheight,keepaspectratio]{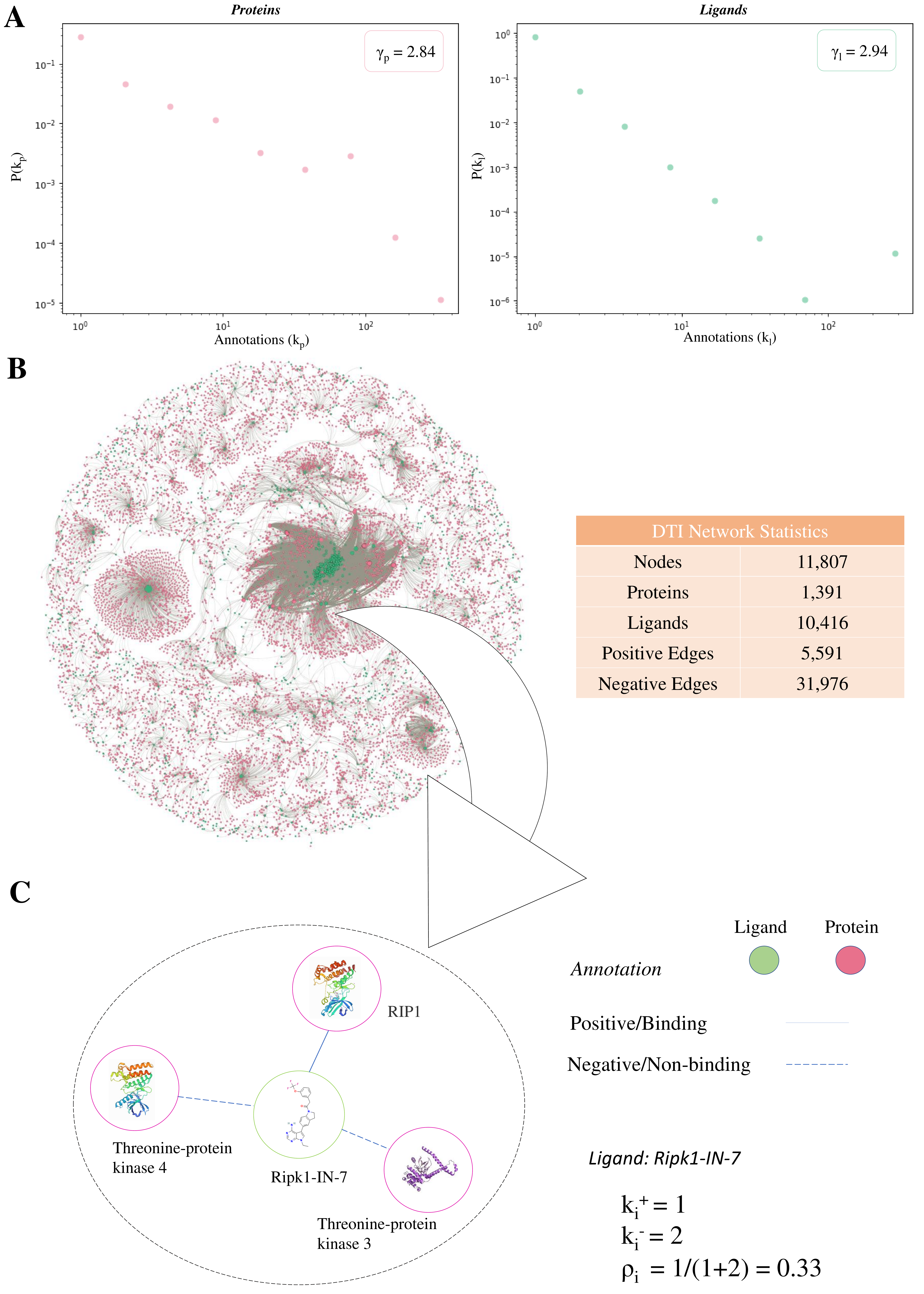}
    \caption{\textbf{Drug-target Interaction Network.} \textbf{(A)} Distributions of the number of annotations in the benchmark BindingDB data are shown in double logarithmic axes (log-log plot), indicate that $P(k_p)$ and $P(k_l)$ are well approximated by power law for both proteins and ligands, with approximate degree exponents $\gamma_p = 2.84$ and $\gamma_l = 2.94$, respectively. \textbf{(B)} The drug-target interaction network used to train the DeepPurpose models, consisting of 10,416 ligands and 1,391 protein targets. Ligands and proteins are represented by green and pink nodes, respectively. \textbf{(C)} Network neighborhood of the ligand Ripk1-IN-7. Solid links represent positive or binding annotations, while dashed links refer to negative or non-binding annotations. Ripk1-IN-7 has one positive and two negative annotations in the training data, implying a degree ratio $\rho$ of 0.33.}
    \label{fig:fig0}
\end{figure}

ML models characterize the likelihood of each node (proteins and ligands) to bind to other nodes according to the features and annotations in the training data. While annotations capture known protein-ligand interactions, features refer to the chemical structures of proteins and ligands which determine their physical and chemical properties, and are expressed as amino acid sequences or 3D structures for proteins, and chemical SMILES \cite{Weininger1988} for ligands. In an ideal scenario, the ML model learns the patterns characterizing the features which drive the protein-ligand interactions, capturing the physical and chemical properties of a protein and of a ligand that determine the mutual binding affinity. Yet, as we show next, multiple state-of-the-art deep learning models, such as DeepPurpose \cite{Huang2020}, ignore the features and rely largely on annotations, i.e., the degree information for each protein and ligand in the drug-target interaction (DTI) network, as a shortcut to make new binding predictions. A bipartite network represents the binding information as a graph with two different types of \textit{nodes}: one corresponding to proteins (also called targets, representing for example a human or a viral protein) and the other corresponding to ligands (representing potential drugs or natural compounds), respectively. A protein-ligand annotation, i.e., evidence that a ligand binds to a protein, is represented as a \textit{link} between the protein and the ligand in the bipartite network \cite{network_science_barabasi}. Experimentally validated annotations define the known DTI network. While binding depends only on the detailed chemical characteristics of the nodes (proteins and ligands), as we show here, many current ML models predictions are primarily driven by the topology of the DTI network. We begin by noticing that the number of annotations linked to a protein or a ligand follows a fat-tailed distribution \cite{barabasi_albert_1999}, indicating that the vast majority of proteins and ligands have only a small number of annotations, which then coexist with a few \textit{hubs}, nodes with an exceptionally large number of binding records \cite{network_science_barabasi}. For example, the number of annotations for proteins follows a power law distribution with degree exponent $\gamma_p = 2.84$ in the BindingDB data used for training and testing DeepPurpose, while the ligands have a degree exponent $\gamma_l=2.94$ (Figure \ref{fig:fig0}A) \cite{alstott_2014}. For these degree exponents, the second moment of the distribution diverges for large sample sizes, implying that the expected uncertainty in the binding information is highly significant, limiting our ability to predict the binding between a single protein and ligand \cite{network_science_barabasi,Yang2020}. Furthermore, positive and negative annotations are determined by applying a threshold on kinetic constants like the constant of disassociation $K_d$. If the kinetic constant associated with a protein-ligand pair is less than a set threshold, we consider that pair as a positive or binding sample; otherwise, the pair is tagged as negative or non-binding. However, $K_d$ is not randomly distributed across the records, but the number of annotations $k$ and the average $K_d$ per $k$ (i.e., $\langle K_d\rangle$) are anti-correlated, indicating stronger binding propensity for proteins and ligands with more annotations ($r_{Spearman}(k_p, \langle K_d\rangle)=-0.47$ for proteins, $r_{Spearman}(k_l, \langle K_d \rangle)=-0.29$ for ligands in the BindingDB data used by DeepPurpose). Alongside this negative correlation, we observe higher variance for $\langle K_d \rangle$ values given a fixed $k$ for the low degree nodes compared to the hubs. As the annotations follow fat-tailed distributions, the observed anti-correlation drives the hub proteins and ligands to have disproportionately more binding records on average, whereas proteins and ligands with fewer annotations have both binding and non-binding examples. This \textit{annotation imbalance} prompts the ML models to leverage degree information (positive and negative annotations) in making binding prediction instead of learning binding patterns from the molecular structures. We term this phenomenon as the \textit{emergence of topological shortcuts} (see Section \ref{SI:topologicalshortcuts}).

To investigate the emergence of topological shortcuts, for each node $i$  with number of annotations $k_i$, we quantify the balance of the available training information via the \textit{degree ratio}, \begin{equation}
\rho_i = \frac{k_{i}^+}{k_{i}^{+} + k_{i}^{-}}=\frac{k_{i}^+}{k_i},
\label{degreeratio}
\end{equation}
where, $k_{i}^{+}$ is the positive degree, corresponding to the number of known binding annotations in the training data, and $k_{i}^{-}$ is the negative degree, or the number of known non-binding annotations in the training data (Figure \ref{fig:fig0}C). As most proteins and ligands lack either binding or non-binding annotations for  (Table \ref{table:1}), the resulting $\{\rho_i \}$ are close to 1 or 0 (Figs. \ref{fig:fig1a}A, B), with extreme $\rho$ values representing the annotation imbalance in the prediction problem. As many state-of-the-art deep learning models, such as DeepPurpose \cite{Huang2020}, uniformly sample the available positive and negative annotations, they assign higher binding probability to proteins and ligands with higher $\rho$ (Figure \ref{fig:fig1a}C, D). Consequently, their binding predictions are driven by topological shortcuts in the protein-ligand network, which are associated with the the positive and negative annotations present in the training data rather than the structural features characterizing proteins and ligands.

The higher binding predictions in DeepPurpose for proteins with large degree ratios (Figure \ref{fig:fig1a}C) prompted us to compare the performance of DeepPurpose with network configuration models,  algorithms that ignore the features of proteins and ligands and instead predict the likelihood of binding  by leveraging only topological constraints derived from the network degree sequence \cite{network_science_barabasi, networks_newman, Menichetti2014, pmid25936214}. In the configuration model (Figure \ref{fig:fig1b}A, Methods), the probability of observing a link is determined only by the the degrees of its end nodes. In a 5-fold cross-validation on the benchmark BindingDB dataset (Table \ref{table:1}), we find that the top-performing DeepPurpose architecture, Transformer-CNN \cite{Huang2020}, achieves AUROC of $0.85$ ($\pm \ 0.005$) and AUPRC of $0.65$ ($\pm \ 0.008$). At the same time, the network configuration model on the same data achieves an AUROC of $0.86$ ($\pm \ 0.005$) and AUPRC of $0.61$ ($\pm \ 0.008$) (Figure \ref{fig:fig1b}B). A consistent behavior is observed for DeepPurpose under node attribute reshuffling, suggesting that both deep neural networks and the configuration model leverage mainly topological information (see Table 3). In other words, the network configuration model, relying only on annotations, performs just as well as the deep learning model, confirming that the topology of the protein-ligand interaction network drives the prediction task. The major driving factor of the topological shortcuts is the monotone relation between $k$ and $\langle K_d \rangle$, which associates a link type with the degree of its end nodes as the $\langle K_d \rangle$ values are directly associated with the link types after thresholding. Moreover, in BindingDB we observe that hubs encounter less variance for $\langle K_d \rangle$ compared to the low degree nodes. Thus, the configuration model is able to achieve good test performance in predicting the link types associated with the hubs. Since hub nodes are associated with the majority of the links in the protein-ligand bipartite network, the configuration model achieves excellent test performance by making correct predictions that mainly leverage the degree information of the hubs. To further investigate this hypothesis, we tested three distinct scenarios: (i) unseen edges (Transductive test), when both proteins and ligands from the test dataset are present in the training data; (ii) unseen targets (Semi-inductive test), when only the ligands from the test dataset are present in the training data; (iii) unseen nodes (Inductive test), when both proteins and ligands from the test dataset are absent in the training data.

We find that both DeepPurpose and the configuration model perform well in scenarios (i) and (ii) (Figures \ref{fig:fig1b}C, D). However, for the inductive test scenario (iii), when confronted with new proteins and ligands, both performances drop significantly (Table \ref{table:2}). DeepPurpose has an AUROC of $0.60$ ($\pm \ 0.066$) and AUPRC of $0.42$ ($\pm \ 0.063$), comparable to the configuration model, for which we have AUROC of $0.50$ and AUPRC of $0.30$ ($\pm \ 0.034$). To offer a final piece of evidence that DeepPurpose disregards node features, we randomly shuffled the chemical SMILES \cite{Weininger1988} and amino acid sequences in the training set, while keeping the same positive and negative annotations per node, an operation that did not change the test performance (Table \ref{table:3}). These tests confirm that DeepPurpose leverages network topology as a learning shortcut and fails to generalize predictions to proteins and ligands beyond the training data, indicating that we must use inductive testing to evaluate the true performance of ML models.

Beyond DeepPurpose, models such as MolTrans \cite{Huang2020_Moltrans} explore different structural representations of protein and ligand molecules. We investigated transductive, semi-inductive, and inductive performances for MolTrans, a state-of-the-art protein-ligand binding prediction model which uses a combination of sub-structural pattern mining algorithm, interaction modeling module, and an augmented transformer encoder to better learn the molecular structures (see Section S8). While the innovative representation of the molecules improves upon DeepPurpose in transductive tests (AUROC of 0.952 ($\pm$ 0.051), AUPRC of 0.872 ($\pm$ 0.131)), the same representation still relies only on the training DTI and fails to generalize to novel molecular structures, as captured by the poor performance in inductive tests (AUROC of 0.575 ($\pm$ 0.059), AUPRC of 0.430 ($\pm$ 0.098)).


\subsection*{AI-Bind and statistics across models}

AI-Bind is a deep learning pipeline that combines network-derived learning strategies with unsupervised pre-trained node features, to optimize the exploration of the binding properties of novel proteins and ligands. Our pipeline is compatible with various neural architectures, three of which we propose here: VecNet, Siamese model, and VAENet. AI-Bind uses two inputs (Figure \ref{fig:fig3}A): For ligands, it takes as input isomeric SMILES, which capture the structures of ligand molecules. AI-Bind considers a search-space consisting of all the drug molecules available in DrugBank and the naturally occurring compounds in the Natural Compounds in Food Database (NCFD) (see Section \ref{SI:SI_sec_NCFD}), and can be extended by leveraging larger chemical libraries like PubChem \cite{pubchem_2020}. For proteins, AI-Bind uses as input the amino acid sequences retrieved from the protein databases Protein Data Bank (PDB) \cite{PDB}, the Universal Protein knowledgebase (UniProt) \cite{uniprot_2020}, and GeneCards \cite{genecards_2016}.

AI-Bind benefits from several novel features compared to the state-of-the-art: (a) It relies on network-derived negatives to balance the number of positive and negative samples for each protein and ligand. To be specific, it uses protein-ligand pairs with shortest path distance $\geq 7$ as negative samples, ensuring that the neural networks observe both binding and non-binding examples for each protein and ligand (see Figure \ref{fig:fig2}, Methods, Section \ref{SI:eigenspokes}). (b) During unsupervised pre-training, AI-Bind uses the node embeddings trained on larger collections of chemical and protein structures, compared to the set with known binding annotations, allowing AI-Bind to learn a wider variety of structural patterns. Indeed, while models like DeepPurpose were trained on $862{,}337$ ligands and $7{,}504$ proteins provided in BindingDB, or $7{,}307$ ligands and $4{,}762$ proteins provided in DrugBank, the unsupervised representation in AI-Bind's VecNet is trained on  $19.9$ million compounds from ZINC \cite{Irwin2012} and ChEMBL \cite{Gaulton2016} databases, and on $546{,}790$ proteins from Swiss-Prot \cite{Bairoch1996}.

We begin the model's validation by systematically comparing the performance of AI-Bind to DeepPurpose and the configuration model on a 5-fold cross-validation using the network-derived dataset for transductive, semi-inductive, and inductive tests. AI-Bind's VecNet model uses pre-trained \texttt{mol2vec} \cite{Jaeger2018} and \texttt{protvec} \cite{Asgari2015} embeddings combined with a simple multi-layer perceptron \cite{haykin1994neural} to learn protein-ligand binding (Figure \ref{fig:fig3}B, see Methods). We find that the configuration model performs poorly in inductive testing (AUROC $0.5$, AUPRC $ 0.469 \pm 0.014$). Due to the network-derived negatives that remove the annotation imbalance, DeepPurpose shows improved performance for novel proteins and ligands (AUROC $0.642 \pm 0.025$, AUPRC $ 0.583 \pm 0.016$). The best performance on unseen nodes is observed for AI-Bind's VecNet, with AUROC of $0.745 \pm 0.032$ and AUPRC of $ 0.729 \pm 0.038$ (see Figure \ref{fig:fig3}C and see Table S3 for a summary of the performances). The unsupervised pre-training for ligand embeddings allows us to generalize AI-Bind to naturally occurring compounds, characterized by complex chemical structures and fewer training annotations compared to drugs (see Section \ref{SI:naturalligands}), obtaining performances comparable to those obtained for drugs (Figure \ref{fig:fig3}D).

Beyond DeepPurpose, AI-Bind's VecNet consistently achieves better inductive performance (AUROC $0.745 \pm 0.032$, and AUPRC $ 0.729 \pm 0.038$) compared to MolTrans (AUROC $0.619 \pm 0.021$, and AUPRC $ 0.480 \pm 0.028$). The comparison between AI-Bind and state-of-the-art models like DeepPrupose and MolTrans validates how unsupervised pre-training of the molecular embeddings improves the generalizability of binding prediction models (see Section S8).

\subsection*{Validation of AI-Bind predictions on COVID-19 proteins}

For a better understanding of the reliability of the AI-Bind predictions, we move beyond standard ML cross-validation and compare our predictions with molecular docking simulations, and in vitro and clinical results on protein-ligand binding. Docking simulations offer a reliable but computationally complex method to predict (or validate) binding between proteins and ligands \cite{Trott2009}. Motivated by the need to model rapid response to sudden health crises, we chose as our validation set the 26 SARS-CoV-2 viral proteins and the 332 human proteins targeted by the SARS-CoV-2 viral proteins \cite{deisy_2021,Gordon2020}. These proteins are missing from the training data of AI-Bind, hence represent novel targets and allow us to rely on recent efforts to understand the biology of COVID-19 to validate the AI-Bind predictions. We could retrieve the amino acid sequences in FASTA format for 16 SARS-CoV-2 viral proteins and 330 human proteins from UniProt \cite{uniprot_2020}, and use them as input to AI-Bind's VecNet. Binding between viral and human proteins is necessary for the virus to synthesize its own viral proteins and to facilitate its replication. Our goal is to predict drugs in DrugBank or naturally occurring compounds that can bind to any of the 16 SARS-CoV-2 or 330 human proteins associated with COVID-19, potentially disrupting the viral infection. After sorting all protein-ligand pairs based on their binding probability predicted by AI-Bind's VecNet ($p_{ij}^{VecNet}$), we tested the predicted top 100 and bottom 100 binding interactions with blind docking simulations using AutoDock Vina \cite{Trott2009},  which estimates binding affinity by considering all possible binding locations on the 3D protein structures (see Methods). Of the 54 proteins present in the top 100 and bottom 100 predicted pairs, 23 had 3D structures available in PDB \cite{PDB} and UniProt \cite{uniprot_2020}, and 51 of the 59  involved ligand structures available on PubChem \cite{pubchem_2020}, allowing us to perform 128 docking simulations (84 involving the top and 44 involving the bottom predictions). We find that 74 out of 84 top predictions from AI-Bind are indeed validated binding pairs. Furthermore, we find that the median binding affinity for the top VecNet predictions is $-7.65$ kcal/mole, while for the bottom ones is $-3.0$ kcal/mole (Figure 6A), confirming that for AI-Bind, the top predictions show significantly higher binding propensity than the bottom ones (Kruskal-Wallis H-test p-value of 2.5*$10^{-5}$) \cite{Kruskal1952,Kruskal1952_python}. As a second test, we obtained the binary labels (binding or non-binding) from docking and AI-Bind predictions using the threshold of $-1.75$ kcal/mole for binding affinities \cite{Smith2012} and the optimal threshold on $p_{ij}^{VecNet}$ corresponding to the highest F1-Score on the inductive test set (see Section \ref{SI:additional_dl_results}, Figure S12). In the derived confusion matrix we observe sensitivity $=0.76$, representing the fraction of binding predictions made by AI-Bind that are true binders, i.e., the ratio $True \ Positives / (True \ Positives + False \ Negatives)$, and F1-Score $=0.82$. These two numbers confirm that the rank list provided by AI-Bind predictions shows a significant similarity to the rank list obtained by binding affinities compared to a random selection (Figure 6B). We further check the stability of these performance metrics by randomly choosing $20$  protein-ligand pairs in a 5-fold bootstrapping set-up and observe F1-Score $= 0.90 \pm 0.02$. Additionally, we find that the predictions made by AI-Bind's VecNet ($p_{ij}^{VecNet}$) and the free energy of protein-ligand binding obtained from docking ($\Delta G$) are anti-correlated with $r_{Spearman}(p_{ij}^{VecNet},\Delta G) = -0.51$. The top 20 VecNet predictions show $r_{Spearman}(p_{ij}^{VecNet},\Delta G) = -0.17$. As lower binding affinity values correspond to stronger binding, these results document the agreement between AI-Bind predictions and docking simulations.

Among the 50 ligands with the highest average binding probability we find two FDA-approved drugs Anidulafungin (NDA\#021948) and Cyclosporine (ANDA\#065017). Experimental evidence \cite{Jeon2020} shows that these drugs have anti-viral activity at very low concentrations in the dose-response curves, and have $IC_{50}$ values of 4.64 $\mu M$ and 5.82 $\mu M$ (see Figure S\ref{fig:SI27}), respectively, measured by immunofluorescence analysis with an antibody specific for the viral N protein of SARS-CoV-2. These low $IC_{50}$ values support anti-viral activity, confirming that Anidulafungin and Cyclosporine bind to COVID-19 related proteins \cite{Cour2020_Cyclosporine}, and the activity at low concentrations indicate that they are safe to use for treating COVID-19 patients \cite{Hughes2011_early_drug_discovery}. Anidulafungin binds to the SARS-CoV-2 viral protein nsp12, a key therapeutic target for coronaviruses \cite{Dey2021}.

AI-Bind also offers several novel predictions with potential therapeutic relevance. For example, it predicts that the naturally occurring compounds Spironolactone, Oleanolic acid, and Echinocystic acid are potential ligands for COVID-19 proteins, all three ligands binding to Tripartite motif-containing protein 59 (TRIM59), a human protein to which the SARS-CoV-2  viral proteins ORF3a and NSP9 bind \cite{Kondo2012,Li2011}. AutoDock Vina supports these predictions, offering binding affinities -7.1 kcal/mole, -8.0 kcal/mole, and -7.6 kcal/mole, respectively. 

Spironolactone, found in rainbow trout \cite{duke}, has been suggested to reduce COVID susceptibility \cite{Jeon2021,Cadegiani2020_1,Cadegiani2020_2,Liaudet2020}. Oleanolic acid is present in apple, tomato, strawberry, and peach, and has been proposed as a potential anti-viral agent for COVID-19 \cite{Paweczyk2020,Carino2020}. Oleanolic acid, which passed the drug efficacy benchmark ADME (Absorption, Distribution, Metabolism, and Excretion), plays an important role in controlling viral replication of SARS-CoV-2 \cite{Kumar2020} and is  effective in preventing virus entry at low viral loads \cite{Carino2020}.
Finally, Echinocystic acid, found in sunflower, basil, and gala apples, is known for its anti-inflammatory \cite{Joh2012,Joh2013,Ryu2013} and anti-viral activity \cite{Tong2004,Deng2015}, but its potential anti-viral role in COVID-19 is yet to be validated.

\subsection*{Identifying active binding sites}

Beyond predicting binding probability, AI-Bind can also be used to identify the probable active binding sites on the amino acid sequence, even in absence of a 3D protein structure. Specifically, we can use AI-Bind to identify which  trigrams in the amino acid sequence play the most significant role in binding predictions, indicative of potential protein-ligand binding locations. We perturb each amino acid trigram in the sequence and observe the changes in AI-Bind prediction (see Section \ref{SI:trigram_study}). Valleys in the obtained binding probability profile represent the trigrams most predictive of binding locations on the amino acid sequence. To validate the AI-Bind predicted binding sites we focus on the human protein TRIM59, a protein for which we have results from multiple docking simulations, using PyMOL \cite{pymol} and identified the amino acid residues binding to the ligand molecules (Figure 6C). We find that the amino acid residues responsible for binding directly map to the valleys in the binding probability profile identified by AI-Bind. By visualizing the docking results for Pipecuronium, Buprenorphine and Voclosporin, ligands that bind to three different pockets on TRIM59, we mark the valleys corresponding to the respective binding sites on the binding probability profiles (Figure 6C). For example, pocket 1, where Pipecuronium binds, corresponds to four AI-Bind predicted valleys marked by 1A, 1B, 1C and 1D.

Since not all valleys in the binding probability profile map to binding sites that we could match with ligands, we use the protein secondary structure to prioritize the valleys. We predict the secondary structure from the amino acid sequence using S4PRED \cite{Moffat2021} and identify the regions with $\alpha$-helix, $\beta$-sheet and coil. In particular, $\alpha$-helices prefer non-solvent accessible environments \cite{Kutchukian2009}, contain non-polar amino acid residues \cite{Fujiwara2012}, and consist of weaker inter-molecular interactions \cite{Cheng2013}. Thus, helices show less propensity for protein-ligand binding. In particular, $\alpha$-helices prefer non-solvent accessible environments \cite{Kutchukian2009}, contain non-polar amino acid residues \cite{Fujiwara2012}, and consist of weaker inter-molecular interactions \cite{Cheng2013}. Thus, the presence of alpha-helices reduce the chances of binding between a ligand and a protein. In contrast, $\beta$-sheets and non-regular coil regions (unstructured regions) are preferred by ligands as active binding sites since they provide more binding opportunity to other molecules \cite{Remaut2006}. Indeed, most of the valleys in Figure 6C where ligands bind map to $\beta$-sheets and coils on TRIM59, associated with pockets 1 and 2 (27 out of 34 ligands validated by docking). Valleys which have large overlap with the $\beta$-sheets and coils provide most of the predicted binding. By combining the binding probability profile predicted by AI-Bind and the secondary structure predicted by S4PRED, we can create an optimal search grid for the subsequent docking simulations, drastically reducing runtime.

We pursued further validation of AI-Bind predicted binding sites with a gold standard protein binding dataset \cite{Cheng2009_Gold} and with p2rank, another state-of-the-art binding prediction model \cite{Krivk2018_p2rank}, to extensively assess the reliability of the AI-Bind pipeline (see Section S13).

In summary, ML models often fail in real world settings when making predictions on data that they were not explicitly trained upon despite achieving good test performance based on traditional ML-based metrics \cite{Heaven_2020_tech_review}. It is therefore necessary to validate the applicability of these models before deploying them. The documented validation of the AI-Bind predictions with molecular dynamic simulations and in vitro experiments offers us confidence AI-Bind is an effective prioritization tool in diverse settings. 


\section{DISCUSSION}

The accurate prediction of drug-target interactions is an essential precondition of drug discovery.
Here we showed that by taking topological shortcuts, existing deep learning models significantly limit their predictive power. Indeed, a mechanistic and quantitative understanding of the origins of these shortcuts indicates that uniform sampling in presence of annotation imbalance drives ML models to disregard the features of proteins and ligands, limiting their ability to generalize to novel protein targets and ligand structures.  To address these shortcomings, we introduced a new pipeline, AI-bind, which mitigates the annotation imbalance of the training data by introducing network-derived negative annotations inferred via shortest path distance, and improves the transferability of the ML models to novel protein and ligand structures by unsupervised pre-training. The proposed unsupervised pre-training of node features also influences the quality of false predictions, removing potential structural biases towards specific protein families (see Section \ref{SI:genephylogeny}).
Once we improved the statistical sampling of the training data and generated the node embeddings in an unsupervised fashion, we observed an increase in performance compared to DeepPurpose, resulting in commendable AUROC ($24\%$ improvement) and AUPRC ($74\%$ improvement) and, most importantly, an ability to predict beyond proteins and ligands present in the training dataset. 

A major limitation of using binding  predictions in drug discovery is that binding to disease-related protein targets  does not always imply a therapeutic treatment. As future work, we plan to extend our implementation by introducing an ML-based classifier to sort the list of potential ligands according to their pharmaceutical (therapeutic) effects, combining the current node features with additional metrics derived from traditional network medicine approaches \cite{Guney2016,doValle2021}. 

AI-Bind leverages ligands' Morgan fingerprints and proteins' amino acid sequences, which encode relevant properties of the molecules: from the presence of hydrogen donors, hydrogen acceptors, count of different atoms, chirality, and solubility for ligands, to the existence of R groups, N or C terminus in proteins. All these properties influence the mechanisms driving protein-ligand binding (see Section \ref{SI:optimal_features}) \cite{pmid29308120}. Yet, the binding phenomenon is largely dependent on the 3D structures of the molecules, which determines the binding pocket structures and the rotation of the bonds. We plan to embed the 3D structures of protein and ligand molecules, which will take into account higher order molecular properties driving protein-ligand binding and refine the predictive power of AI-Bind. To maximize generalization across 3D structure, we will use $\mathrm{SE}(3)$ equivariant  networks to learn embeddings.  Equivariance has proven to be a powerful tool for improving generalization over molecular structure \cite{satorras2021n, fuchs2020se, jumper2021highly, https://doi.org/10.48550/arxiv.2202.05146}. We also plan to explore the performance of AI-Bind over the entire druggable genome \cite{Finan2017}, allowing us to predict for each protein, which domains are responsible for the binding predictions, revealing binding locations of the ligands and the proteins. Finally, we envision enabling AI-Bind to predict the kinetic constants $K_d$, $K_i$, $IC_{50}$, and $EC_{50}$ by formulating a regression task over these variables.

The existing docking infrastructures allow screening for a specific protein structure against wide chemical libraries. Indeed, VirtualFlow \cite{Gorgulla2020}, an open-source drug discovery platform offers virtual screening over more than 1.4 billion commercially available ligands. However, running docking simulations over these vast libraries incurs high costs for data preparation and computation time and are often limited to only proteins with 3D structures \cite{PDB}. For example, in our validation step, only half (23 out of 54) of the 3D structures of the proteins associated with COVID-19 were available. Since AI-Bind only requires the chemical SMILES for ligands\cite{Weininger1988} and amino acid sequences for proteins, it can offer fast screening for large libraries of targets and molecules without requiring 3D structures, guiding the computationally expensive docking simulations on selected protein-ligand pairs. 


\section{METHODS}

\subsection*{Data Preparation}

We use InChIKeys \cite{Heller2015} and amino acid sequences as the unique identifiers for ligands and targets, respectively. Positive and negative samples are selected from DrugBank, BindingDB and DTC (see Section \ref{SI:databases}). We consider samples from BindingDB and DTC to be binding or non-binding based on the kinetic constants $K_i$, $K_d$, $IC_{50}$, and $EC_{50}$. We use thresholds of $\leq 10^3nM$ and $\geq 10^6 nM$ to obtain positive and (absolute) negative annotations, respectively \cite{Smith2012}. We then filter out all samples outside the temperature range 20$^{\circ}$C-45$^{\circ}$C to remove ambiguous pairs. All amino acid sequences were obtained from UniProt \cite{uniprot_2020}.

\subsubsection*{Positive Samples}

We consider the binding information from DrugBank as positive samples. From these annotations, we removed 53 pairs which are available in BindingDB and have kinetic constants $\geq 10^6 nM$. To obtain additional positive samples for drugs, we searched in BindingDB using their InChIKeys. We obtained $4{,}330$ binding annotations from BindingDB related to the drugs in DrugBank. Overall, we gathered a total of $28{,}188$ positive samples for drugs. We identified naturally occurring/food-borne compounds by leveraging the Natural Compounds in Food Database (NCFD) database (see Section \ref{SI:SI_sec_NCFD}). We queried BindingDB and DTC with the associated InChIKeys, obtaining a total of $1{,}555$ positive samples. 

\subsubsection*{Network-Derived Negative Samples} \label{network-derived-negatives}

To generate annotation-balanced training data for AI-Bind, we merged the positive annotations derived from DrugBank, BindingDB, and DTC, for a total of $5{,}104$ targets and $8{,}111$ ligands, of which 485 are naturally occurring, and calculated the shortest path distribution. All odd-path lengths in the bipartite network correspond to protein-ligand pairs (Figure \ref{fig:fig2}C). Overall, the longer the shortest path distance separating a protein and a ligand, the higher the kinetic constant observed in BindingDB (Figure \ref{fig:fig2}D).
In particular, pairs more than $7$ hops apart have, on average, kinetic constants $K_i\ge 10^6nM$, which is generally considered above the protein-ligand binding threshold \cite{Smith2012} (see Section \ref{SI:eigenspokes}). We randomly selected a subset of protein-ligand pairs which are 7 hops apart as negative samples, to create an overall class balance between positive and negative samples in the training data.  Finally, we removed all nodes with only positive or only negative samples and obtained the \emph{network-derived negative samples}.\\
We performed testing and validation on $\geq 11$-hop distant pairs. 
Additionally, we included in testing and validation the absolute non-binding pairs derived from BindingDB by thresholding the kinetic constants ($K_i$, $K_d$, $IC_{50}$, and $EC_{50}$).


\subsection*{Network Configuration Model} 

\subsubsection*{Overview}
Protein-ligand annotations are naturally embedded in a bipartite duplex network, consisting of a set of  nodes, comprising all proteins and ligands, interacting in  two layers, each reflecting a distinct type of interaction linking the same pair of nodes \cite{Menichetti2014}. More specifically, one layer (Layer 1) captures the positive or binding annotations, while the second layer (Layer 2) collects the negative or non-binding annotations (Figure \ref{fig:fig1b}A). A multilink $\vecm$  between two nodes encodes the pattern of links connecting these nodes in different layers. In particular, $\vecm=(1,0)$ indicates positive interactions, $\vecm=(0,1)$ refers to negative interactions, $\vecm=(0,0)$ represents the absence of any type of annotations, and $\vecm=(1,1)$ is mathematically forbidden, as binding and non-binding cannot coexist for the same pair of protein and ligand. 

We developed a canonical bipartite duplex null model that conserves on average the number of positive and negative annotations of each node, while correctly rewiring positive and negative links and avoiding forbidden configurations. By means of entropy maximization with constraints, we derive the analytical formulation of each multilink probability and the conditional probability of observing positive binding once an annotation is reported.

\subsubsection*{Mathematical Formulation}
Let $A_{ij}^{\vecm}$ be the multi-adjacency matrix representing the bipartite duplex of ligands ($\{i\}$) and proteins ($\{j\}$), with elements  equal to 1 if there is a multilink $\vecm$ between $i$  and $j$  and zero otherwise. We define the multidegree of ligand $i$ and target $j$ as
\begin{equation}
k_i^{\vecm} =\sum_{j=1}^{N_T} A_{ij}^{\vecm}, \qquad
t_j^{\vecm} = \sum_{i=1}^{N_L} A_{ij}^{\vecm},
\label{multidegree}
\end{equation}
where $N_T$ is the number of targets and $N_L$ is the number of ligands.

A bipartite duplex network ensemble can be defined as the set of all duplexes satisfying a given set of constraints, such as the expected multidegree sequences defined in Eq. \ref{multidegree}. We determine the probability of observing a bipartite duplex network $P(\vec\graph)$ by entropy maximization with multidegree constraints $\{k_i^{(1,0)}\}$, $\{k_i^{(0,1)}\}$, $\{t_j^{(1,0)}\}$, and $\{t_j^{(0,1)}\}$, and corresponding Lagrangian multipliers $\{\lambda_i^{(1,0)}\}$, $\{\lambda_i^{(0,1)}\}$, $\{\mu_j^{(1,0)}\}$, and $\{\mu_j^{(0,1)}\}$ \cite{Menichetti2014, pmid25936214, C5MB00143A}. The probability $P(\vec\graph)$ factorizes as

\begin{equation}
P(\vec\graph) = \frac{1}{Z} \prod_{ij} \exp{\left[- \sum_{\vecm \neq (0,0),(1,1)} (\lambda_i^{\vecm} + \mu_j^{\vecm}) A_{ij}^{\vecm}\right]},
\end{equation}
with
\begin{equation}
Z= \prod_{ij} \left[1+ \sum_{\vecm \neq (0,0),(1,1)} e^{-(\lambda_i^{\vecm} + \mu_j^{\vecm})}\right].
\end{equation}

Multilink probabilities $p_{ij}^{\vecm}$ are determined by the derivatives of $\log (Z)$ according to $(\lambda_i^{\vecm} + \mu_j^{\vecm})$. For instance, the probability of observing a positive annotation is
\begin{equation}
 p_{ij}^{(1,0)} = \frac{e^{-(\lambda_i^{(1,0)} + \mu_j^{(1,0)})}}{1+e^{-(\lambda_i^{(1,0)} + \mu_j^{(1,0)})}+e^{-(\lambda_i^{(0,1)} + \mu_j^{(0,1)})}},  
 \label{positive}
\end{equation}
while the probability of observing a negative annotation follows
\begin{equation}
p_{ij}^{(0,1)} = \frac{e^{-(\lambda_i^{(0,1)} + \mu_j^{(0,1)})}}{1+e^{-(\lambda_i^{(1,0)} + \mu_j^{(1,0)})}+e^{-(\lambda_i^{(0,1)} + \mu_j^{(0,1)})}},
 \label{negative}
\end{equation}
with $p_{ij}^{(1,0)}+p_{ij}^{(0,1)}+p_{ij}^{(0,0)}=1$.

In this theoretical framework, binding prediction is inherently conditional, as for each ligand $i$ and protein $j$, we test only the presence of positive and negative annotations. Consequently, $p_{ij}^{(1,0)}$ and $p_{ij}^{(0,1)}$ are normalized by the probability of observing a generic annotation $p_{ij}^{(1,0)}+p_{ij}^{(0,1)}$.\\
In case of unseen edges, binding prediction is determined by
\begin{equation}
p_{ij}^{\mathrm{conditional}}=
\frac{p_{ij}^{(1,0)}}{p_{ij}^{(1,0)}+p_{ij}^{(0,1)}},
\label{pcondititonal_transductive}
\end{equation} 
while in case of unseen target $j^*$, the binding probability towards a known compound $i$ follows
\begin{equation}
p_{ij^*}^{\mathrm{conditional}}=
\frac{\langle p_{ij}^{(1,0)} \rangle_j}{\langle p_{ij}^{(1,0)}\rangle_j+\langle p_{ij}^{(0,1)}\rangle_j}=\rho_i,
\label{pcondititonal_semi_inductive}
\end{equation}
where $\langle \cdot \rangle_j$ denotes the average over all known targets, and $\rho_i$ follows from Eq. \ref{degreeratio}.\\
In case of unseen ligand $i^*$ and  target $j^*$, the binding probability is determined by the overall number of positive ($L^{(1,0)}$) and negative ($L^{(0,1)}$) annotations, i.e.,
\begin{equation}
p_{i^*j^*}^{\mathrm{conditional}}=
\frac{\langle p_{ij}^{(1,0)} \rangle_{ij}}{\langle p_{ij}^{(1,0)}\rangle_{ij}+\langle p_{ij}^{(0,1)}\rangle_{ij}}=\frac{L^{(1,0)}}{L^{(1,0)}+L^{(0,1)}}
\label{pcondititonal_inductive},
\end{equation}
where $\langle \cdot \rangle_{ij}$ indicates the average over all known pairs of ligands and targets.


\subsection*{Novel Deep Learning Architectures}

\subsubsection*{VecNet}

VecNet uses the pre-trained \texttt{mol2vec} \cite{Jaeger2018} and \texttt{protvec} \cite{Asgari2015} models (Figure \ref{fig:fig3}B). These models create 300- and 100-dimensional embeddings for ligands and proteins, respectively. Based on \texttt{word2vec} \cite{mikolov2013distributed},  they treat the Morgan fingerprint \cite{Rogers2010} and the amino acid sequences as sentences, where words are fingerprint fragments or amino acid trigrams. The training is unsupervised and independent from the following binding prediction task.

\subsubsection*{VAENet}

VAENet uses a Variational Auto-Encoder \cite{doersch2016tutorial}, an unsupervised learning technique, to embed ligands onto a latent space. The Morgan fingerprint is directly fed to convolutional layers. The auto-encoder creates latent space embeddings by minimizing the loss of information while reconstructing the molecule from the latent representation. We train the Variational Auto-Encoder on $9.5$ million chemicals from ZINC database \cite{Irwin2012}, and all drugs and natural compounds in our binding dataset. Similar to VecNet, we use ProtVec for target embeddings. 

\subsubsection*{Siamese Model}

The Siamese model embeds ligands and proteins into the same space using a one-shot learning approach \cite{koch2015siamese}. We construct triplets of the form $\langle$protein target, non-binding ligand, binding ligand$\rangle$ and train the model to find an embedding space that maximizes the Euclidean distances between non-binding pairs, while minimizing it for the binding ones. 

\section*{File Preparation for Docking Simulations}

The steps to implement docking simulations in AutoDock Vina \cite{Trott2009} include:
\begin{enumerate}
    \item Obtain the 3D ligand structures in SDF format from PubChem and save it in .pdb format with PyMOL for use in AutoDockTools.
    \item Download the 3D protein structures in .pdb format and load them into AutoDockTools to remove water molecules from the protein structure, add all hydrogen atoms, and the Kollman charge to the protein.
    \item Save both the protein and the ligand structures in .pdbqt format using AutoDockTools. 
    \item Create the grid for docking that encompasses the whole protein structure. This grid selection ensures a blind docking set-up, so that all locations on the protein are considered for determining the binding affinities. The selected grid sizes are available in gridsizes.txt (see Data and Code availability).
    \item Create the configuration files with the grid details for each protein and launch the docking simulation. We consider the protein molecules to be rigid, whereas the ligand molecules are flexible, i.e., we allow rotatable bonds on the ligands. 
\end{enumerate}


\section*{Acknowledgement}

We thank Noah DeMoes from Lincoln Laboratory for helping out with setting up the DeepPurpose models. Christian De Frondeville from the Bioinformatics department at Northeastern University has helped with the gold standard validation of binding probability profile.

\section*{Author Contributions}

A.C. contributed to writing the manuscript, data curation and preparation, generating the predictions for the network configuration model, performing experiments to identify the emergence of topological shortcuts, implementing negative sample generation, developing and testing of VecNet and VAENet, running docking simulations and developing the method to predict the active binding sites.

R.W. contributed to writing the manuscript, generating the predictions for the network configuration model, designing and training VecNet and VAENet. 

Z.S. contributed to training and testing of all the deep learning models, and designing the Siamese model.

O.S.A. contributed to the deep learning literature review, running the DeepPurpose models, implementing negative sample generation, and training VAENet.

M.S. contributed to exploring the optimal representation of molecules and developing the method to predict the active binding sites.

D.G. contributed to the data curation and preparation, and performed the gene phylogeny study. 

R.Y., T.E.R., and A.L.B. have provided guidance on designing the experiments and writing the manuscript.

G.M. conceived the project, developed the duplex configuration model, designed experiments to identify the emergence of topological shortcuts, contributed to data preparation, data analysis, and writing the manuscript.

\section*{Competing Interests}

A.L.B. is the founder of Scipher Medicine and Naring Health, companies that explore the use of network-based tools in health, and Datapolis, that focuses on urban data.

\section*{Materials \& Correspondence}
Correspondence and requests for materials should be addressed to G.M.

\section*{Data and Code availability}

The codes that support the findings of this study are openly available at our GitHub page at \url{https://github.com/Barabasi-Lab/AI-Bind}.
The data files are shared via Zenodo at \url{https://zenodo.org/record/7226641}.
Top binding predictions from AI-Bind's VecNet on the COVID-19 related proteins, arranged in the descending order of predicted probabilities and validated by docking, are available at 
\url{https://github.com/Barabasi-Lab/AI-Bind/blob/main/Validation/Predictions.csv}.

\pagebreak
\newpage




\newpage

\newcolumntype{F}[1]{%
    >{\raggedright\arraybackslash\hspace{0pt}}p{#1}}%
\newcolumntype{T}[1]{%
    >{\centering\arraybackslash\hspace{0pt}}p{#1}}%

\begin{table}[h!]
    \centering
    \caption{\textbf{BindingDB Training Data for DeepPurpose.}  Most ligands and proteins in DeepPurpose training data have either binding or non-binding annotations, which creates imbalance in the degree ratio (see Eq. \ref{degreeratio}).}
    \renewcommand{\arraystretch}{1.0}
    \small
    \begin{tabular}{|T{0.1\textwidth}|T{0.2\textwidth}|T{0.2\textwidth}|T{0.2\textwidth}|T{0.15\textwidth}|}
    \hline
    Node Type & Has Only Positive Annotations & Has Only Negative Annotations & Has both annotations & Total Node Count \\ [0.5ex]
    \hline\hline
    Ligand & $3,084$ & $6,539$ & $793$ & $10,416$ \\ 
    \hline
    Protein & $168$ & $556$ & $667$ & $1,391$\\
    \hline
    \end{tabular}
    \\ [1ex] 
    \label{table:1}
\end{table}

\begin{table}[h!]
  \centering
  \caption{\textbf{DeepPurpose and Duplex Configuration Model Performances on BindingDB dataset.} DeepPurpose and the duplex configuration model perform well in both transductive and inductive tests on the benchmark BindingDB data. Both models fail to achieve good performance in the inductive test, i.e., while predicting over both unseen proteins and ligands. }
  \renewcommand{\arraystretch}{1.0}
  \small
  \begin{tabular}{|T{0.14\textwidth}|T{0.11\textwidth}|T{0.11\textwidth}|T{0.11\textwidth}|T{0.11\textwidth}|T{0.11\textwidth}|T{0.11\textwidth}|T{0.11\textwidth}|T{0.11\textwidth}|}
    \hline
    \multirow{2}{*}{Model} &
      \multicolumn{2}{c|}{Transductive} &
      \multicolumn{2}{c|}{Semi-inductive} &
      \multicolumn{2}{c|}{Inductive} \\
      & {AUROC} & {AUPRC} & {AUROC} & {AUPRC} & {AUROC} & {AUPRC} \\
      \hline
    DeepPurpose & $0.82 \pm 0.003$ & $0.48 \pm 0.004$ & $0.76 \pm 0.036$ & $0.69 \pm 0.064$ & $0.60 \pm 0.066$ & $0.42 \pm 0.063$ \\
    \hline
    Config. Model & $0.83 \pm 0.009$ & $0.5 \pm 0.011$ & $0.77 \pm 0.048$ & $0.71 \pm 0.065$ & $0.50 \pm 0.00$ & $0.30 \pm 0.034$  \\
    \hline
  \end{tabular}
  \label{table:2}
\end{table}

\begin{table}[h!]
    \centering
    \caption{\textbf{Assigning SMILES and Amino Acid Sequences Randomly.} A random reshuffle of SMILES and amino acid sequences does not affect the performance of DeepPurpose. This outcome suggests the limitation of DeepPurpose in learning chemical structures.}
    \renewcommand{\arraystretch}{1.0}
    \small
    \begin{tabular}{|T{0.2\textwidth}|T{0.2\textwidth}|T{0.2\textwidth}|}
    \hline
    Version & AUROC & AUPRC \\ [0.5ex]
    \hline\hline
    Original & $0.85 \pm 0.005$ & $0.64 \pm 0.008$ \\ 
    \hline
    Randomized & $0.85 \pm 0.005$ & $0.63 \pm 0.008$ \\
    \hline
    \end{tabular}
    \\ [1ex] 
    \label{table:3}
\end{table}


\newpage

\begin{figure*}[ht!]
    \centering
    \includegraphics[trim=0 10 0 25,clip,angle=0,width=0.9\textwidth,height=0.7\textheight,keepaspectratio]{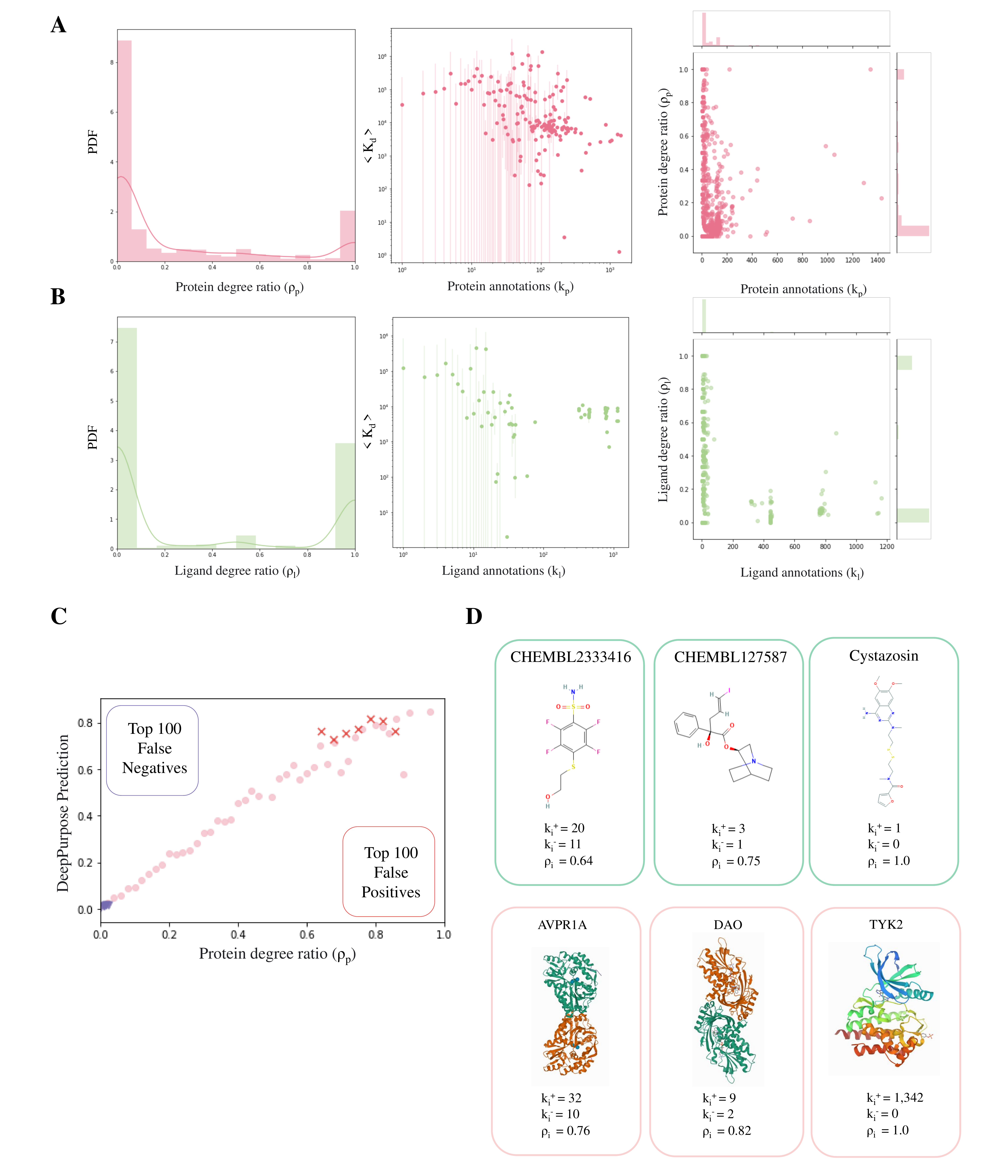}
    \caption{\textbf{Annotation bias in BindingDB training data and DeepPurpose predictions.} \textbf{(A)-(B)} The distribution of degree ratios $\{\rho_p\}$ for the proteins in the original DeepPurpose training set (in a fold from the 5 fold cross-validation). Degree ratio defined in Eq. \ref{degreeratio} refers to the ratio of positive annotations to the total annotations for a given node in the protein-ligand interaction network. The average $K_d$ for different degree values $\{k_p\}$ are negatively correlated with $r_{Spearman}(k_p,\langle K_d \rangle) \approx -0.47$. We observe larger variance in $\langle K_d \rangle$ for the low degree nodes. After thresholding $K_d$ values associated with each link to create the binary labels, the hubs get many positive or binding annotations, whereas the low degree nodes get both binding and non-binding annotations. As the hubs are associated with many links in the network, learning the type of binding from degree information helps ML models to achieve good performance leveraging shortcut learning. We observe similar association patterns for the ligands with $r_{Spearman}(k_p,\langle K_d \rangle) \approx -0.29$. \textbf{(C)} Protein degree ratios $\{\rho_p\}$ and DeepPurpose predictions are highly correlated with $r_{Spearman}=0.94$. We observe that the top 100 false positive predictions include the proteins with large $\{\rho_p\}$ represented by the red crosses, whereas the false negatives are contributed by the proteins with small \{$\rho_p$\} which are represented by the blue dots. \textbf{(D)} Examples of proteins and ligands with large degree ratios and contributing to false positive predictions.}
    \label{fig:fig1a}
\end{figure*}

\begin{figure}[ht!]
    \centering
    \includegraphics[trim=0 10 0 0,clip,angle=0,width=\textwidth,height=\textheight,keepaspectratio]{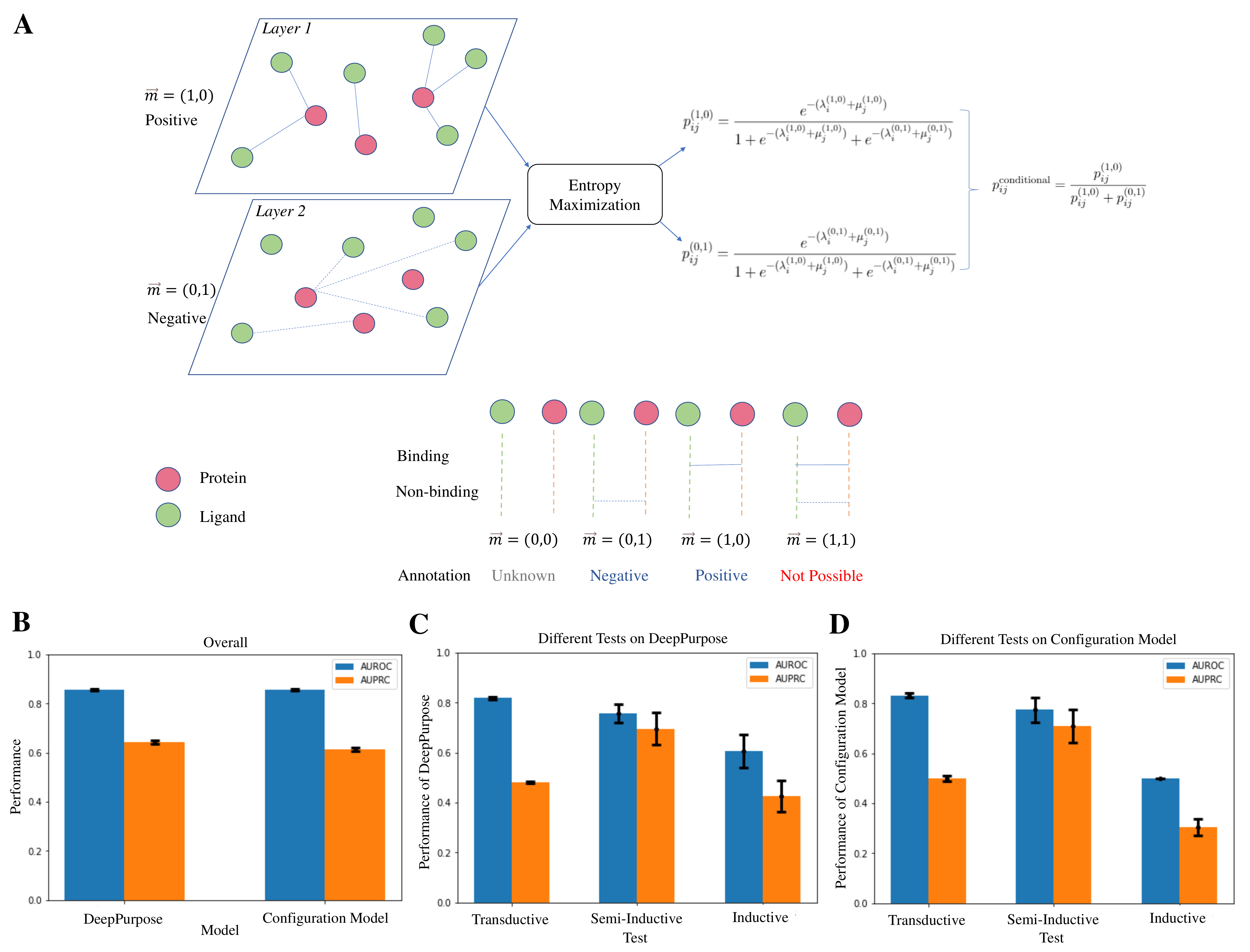}
    \caption{\textbf{Comparing DeepPurpose and the Duplex Configuration Model.} \textbf{(A)} The duplex configuration model includes two layers corresponding to binding and non-binding annotations. Positive and negative link probabilities are determined by entropy maximization (see Methods), and used to estimate the conditional probability in transductive (Eq. \ref{pcondititonal_transductive}), semi-inductive (Eq. \ref{pcondititonal_semi_inductive}), and inductive (Eq. \ref{pcondititonal_inductive}) scenario.
     \textbf{(B)-(D)} The configuration model achieves similar test performance as DeepPurpose on the the benchmark BindingDB data in a 5-fold cross-validation. Breakdown of performances shows good predictive performance on unseen edges and unseen targets. But the same models have poor predictive performance on unseen nodes.}
    \label{fig:fig1b}
\end{figure}

\begin{figure}[ht!]
    \centering
    \includegraphics[trim=0 10 0 0,clip,angle=0,width=\textwidth,height=\textheight,keepaspectratio]{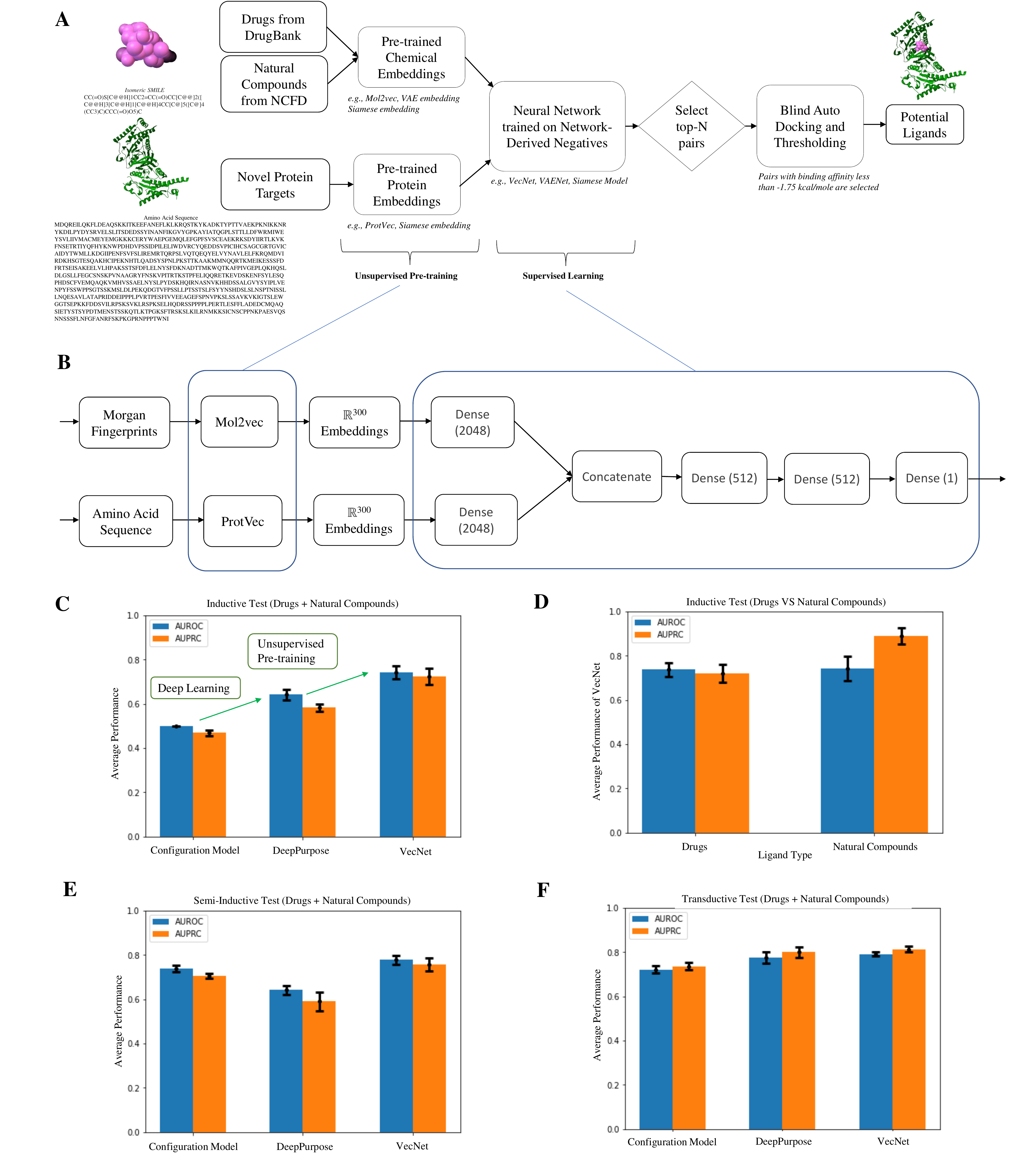}
    \caption{\textbf{AI-Bind pipeline: VecNet Performance and Validation.} \textbf{(A)} AI-Bind pipeline generates embeddings for ligands (drugs and natural compounds) and proteins using unsupervised pre-training. These embeddings are used to train the deep models. Top predictions are validated using docking simulations and are used as potential binders to test experimentally. \textbf{(B)} AI-Bind's VecNet architecture uses Mol2vec and ProtVec for generating the node embeddings. VecNet is trained in a 5-fold cross-validation set-up. Averaged prediction over the 5 folds is used as the final output of VecNet. \textbf{(C)-(F)} 5-fold cross-validation performance of VecNet, DeepPurpose, and Configuration Model. All the models perform similarly in case of predicting binding for unseen edges and unseen targets. The advantage of using deep learning and unsupervised pre-training is observed in the case of unseen nodes (inductive test). AI-Bind's VecNet is the best performing model across all the scenarios. Additionally, we observe similar performance of VecNet for both drugs and natural compounds.}
    \label{fig:fig3}
\end{figure}

\begin{figure}[ht!]
    \centering
    \includegraphics[trim=0 10 0 0,clip,angle=0,width=\textwidth,height=\textheight,keepaspectratio]{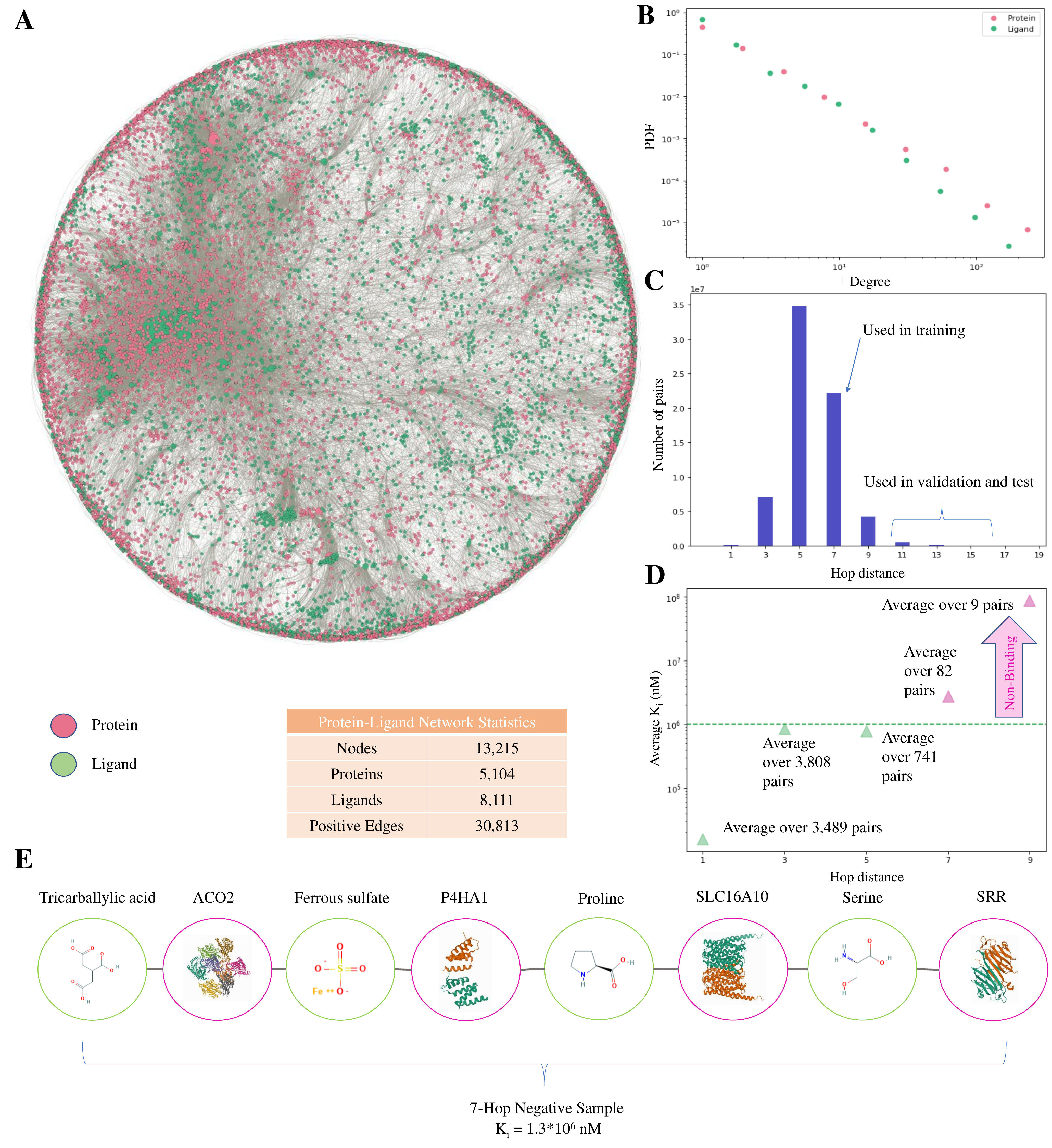}
    \caption{\textbf{Network-Derived Negatives.} \textbf{(A)} Protein-ligands bipartite network consisting of only binding (positive) annotations for drugs and natural compounds. \textbf{(B)} Degree distributions of ligands and proteins are fat-tailed in nature. \textbf{(C)} Shortest-path length distribution capturing all possible protein-ligand pairs. We use protein-ligand pairs with shortest path distance of 7 for training, while absolute negatives obtained from BindingDB and pairs with shortest path distances $\ge$ 11 are used for validation and test.  \textbf{(D)} Average experimental kinetic constant as a function of the shortest path distance. Higher path distance corresponds to higher $K_i$ in BindingDB. Beyond $7$ hops, the expected constant exceeds the binding threshold of $10^6nM$. \textbf{(E)} An example of a protein-ligand pair which is $7$ hops apart and is used as a negative sample in the AI-Bind training set.}
    \label{fig:fig2}
\end{figure}

\begin{figure*}[ht!]
    \centering
    \includegraphics[trim=0 10 0 0,clip,angle=0,width=\textwidth,height=\textheight,keepaspectratio]{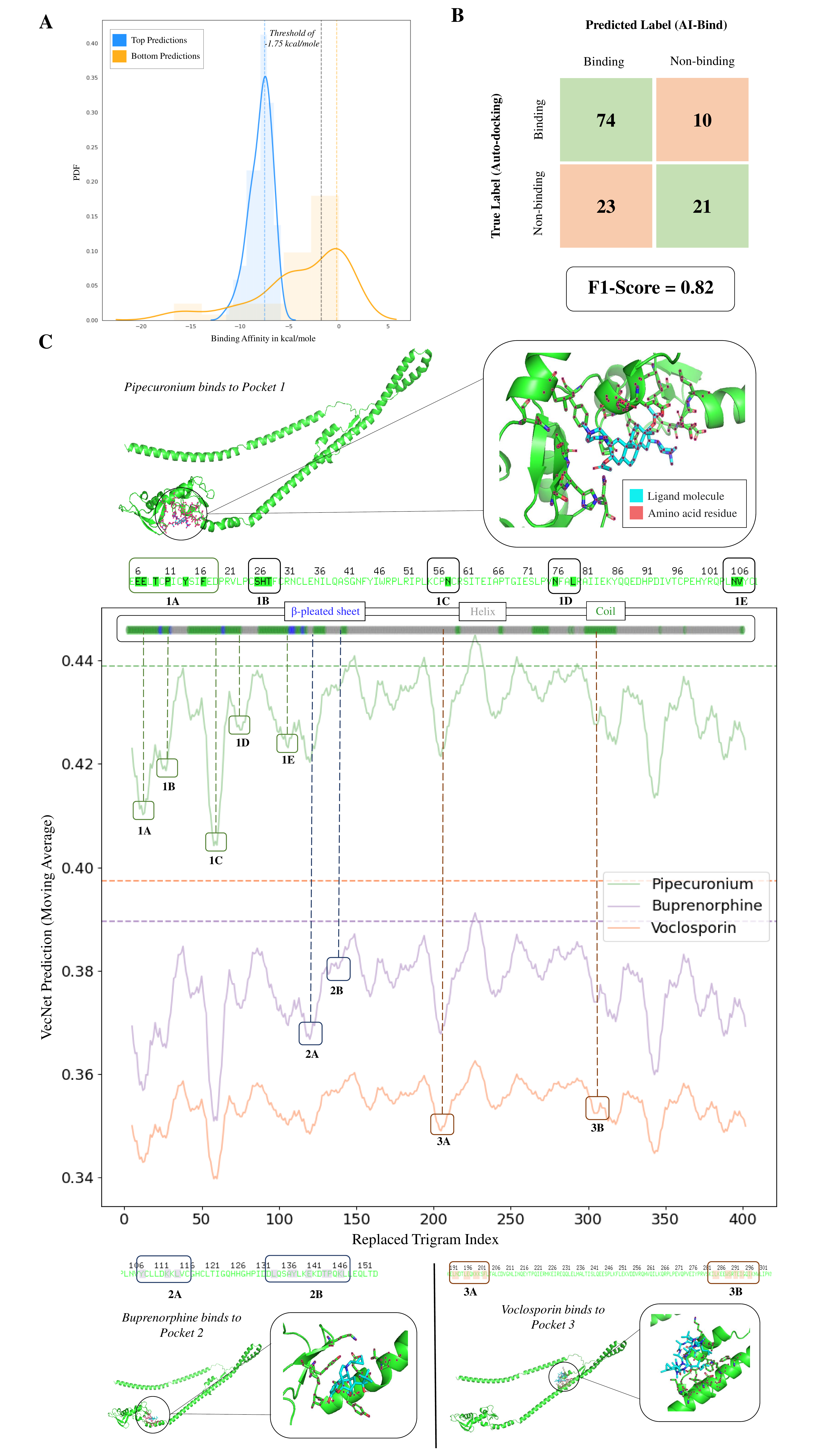}
\end{figure*}

\begin{figure*}[ht!]
    \centering
    \caption{\textbf{Validating and interpreting AI-Bind predictions.} \textbf{(A)} Distribution of binding affinities for top and bottom 100 predictions made by AI-Bind's VecNet over viral and human proteins associated with COVID-19. We ran docking on top 84 predictions and bottom 44 predictions. We observe that the top binding predictions of AI-Bind show lower binding energies (better binding) compared to the bottom predictions. Considering the binding threshold of $-1.75$ kcal/mole, $88\%$ of the top predicted pairs by AI-Bind are inline with the docking simulations.
    \textbf{(B)} We construct the confusion matrix for the top and bottom predictions from AI-Bind. We obtain the true labels using the threshold of $-1.75$ kcal/mole on the binding affinities from docking. We observe that AI-Bind predictions produce excellent F1-Score, offering predictions significantly better than random selection.
    \textbf{(C)} Binding probability profile for human protein TRIM59. Multiple valleys in the profile directly map to the amino acid residues to which the ligands bind and are indicative of the active binding sites on the amino acid sequence. We identify the valleys on the binding probability profiles for three ligands Pipecuronium, Buprenorphine and Voclosporin, which bind at different pockets on TRIM59. Valleys for these pockets have been mapped back to the amino acid sequence (valleys 1A, 1B, 1C and 1D for pocket 1, valleys 2A and 2B for pocket 2, and valleys 3A and 3B for pocket 3). Furthermore, we highlight the secondary structure of TRIM59 from the amino acid sequence. Valleys containing the $\beta$-pleated sheets and the coils are more prone to binding compared to the one with the $\alpha$-helices \cite{Kutchukian2009,Fujiwara2012,Cheng2013,Remaut2006}. Combining the binding probability profile and the secondary protein structure allows us to identify active binding sites, guiding the design of an optimal search grid for docking simulations.}
    \label{fig:fig4}
\end{figure*}


\newpage

\newpage
\pagebreak






\captionsetup[table]{name=Table SI.}
\captionsetup[figure]{name=Figure SI.}

\newgeometry{left=1in,right=1in,top=1in,bottom=1in}

\newpage
\appendix
\addcontentsline{toc}{section}{Appendix} 
\renewcommand{\thesection}{\Alph{section}}
\part{Appendix} 

{\Large {\begin{center} SUPPLEMENTARY INFORMATION \end{center}} } 

\parttoc 

\newpage

\setcounter{section}{0} 

\renewcommand\thesection{S\arabic{section}}


\section{Emergence of topological shortcuts}
\label{SI:topologicalshortcuts}

Decision rules learned by many machine learning (ML) models tend to perform well on benchmark datasets, but fail to generalize well when given never-before-seen data. Instead of learning generalizable patterns from features observed during training, these models leverage shortcuts in the data to maximize transductive performance, i.e., the performance on seen data \cite{Geirhos2020}. In this section, we investigate how the properties of the network data used in training can drive ML models to
learn topological shortcuts, rather than taking into account node features that would allow better generalizability to unseen data. We assess the emergence of topological shortcuts by null models (configuration model) achieving good transductive test performance.

\subsection*{BindingDB data observations}

First, we investigated the statistical properties of the training database used by DeepPurpose \cite{Huang2020}, a modeling pipeline offering state-of-the-art neural architectures to predict protein-ligand binding. The training data is based on all the records in BindingDB \cite{Liu2007} characterized by a kinetic disassociation constant $K_d$. The distribution of the number of annotations per protein  $P(k_p)$ is well fitted by a power law distribution using \cite{alstott_2014}
\begin{equation}
    P(k_p) \sim k^{-\gamma_p}, 
\end{equation}
with $\gamma_p = 2.84$, $k_p^{min} = 1$, and $k_p^{max} = 1{,}426$ (Figure 1A in the main text). We observe similar results for the ligands, with $\gamma_l = 2.94$, $k_l^{min} = 1$, and $k_l^{max} = 1{,}161$.

From the original annotations in BindingDB, a binary classification dataset is derived using a threshold of $30nM$ for $K_d$ \cite{Huang2020}. Protein-ligand pairs with $K_d < 30nM$ are binding or positive pairs, and everything else is labeled as non-binding or negative. Overall, we observe that $16\%$ of the records are labeled as positive, a characteristic of the database that we summarize as $p_{bind}=0.16$, the probability to observe a binding annotation independently from the identity of the protein and the ligand.

Additionally, we find that the number of annotations $k$ and the average disassociation constant $\langle K_d \rangle$ per degree $k$ are not independent but show a negative rank correlation. In particular, for proteins we find $r_{Spearman}(k_p,\langle K_d \rangle) = -0.47$, and for ligands $r_{Spearman}(k_l,\langle K_d \rangle) = -0.29$ (see Figs. \ref{fig:fig1a}A, B). Alongside this negative correlation, we observe higher variance for $\langle K_d \rangle$ values for the low degree nodes compared to higher degree nodes (Figs. \ref{fig:fig1a}A, B). We find that it is easier to capture the properties of the $K_d$ distribution in BindingDB by modeling it as a log-normal (Figure S\ref{fig:SI8}A, B). Thus, we select the $K_d$ values in the log-space for exploring the emergence of topological shortcuts. Consistently, we observe that in the $\log K_d$ space the variance is larger for the lower degree nodes compared to the hubs with $r_{Spearman}(k,  \sigma_{\langle \log K_d \rangle})=-0.71$, where $\sigma_{\langle \log K_d \rangle}$ is the standard deviation of $\langle \log K_d \rangle$ (Figure S\ref{fig:SI8}C).
This means that the variance is progressively decreasing for higher degree nodes and hubs. Indeed, the hubs are less in number and show similar kinetic features leading to less variance in both $\langle K_d \rangle$ and $\langle \log K_d \rangle$.
The relation between degree and kinetic constant makes the link prediction task easier for the hubs compared to the low degree nodes, using only degree information. Since most links in the network are associated with the hubs, the configuration model is able to achieve excellent transductive test performance.

\subsection*{Toy model set-up}

In order to test our hypotheses regarding the creation of topological shortcuts, we simulate synthetic network data that we call \textit{toy models} (Figure S3). We create a duplex of unipartite networks with features inspired by the protein sample captured in BindingDB, as similar considerations extend to bipartite networks.
Specifically, we vary the degree distribution $P(k)$ and  $r_{Spearman}(k,\langle K_d \rangle)$ to explore when the output of the duplex configuration model $\{p_{ij}^{conditional}\}$ (Eq. 7) becomes highly variable and thus informative, creating the potential for topological shortcuts (see Figure 3A and Methods). In other words, the closer $\{p_{ij}^{conditional}\}$ gets to a Naive Bayes classifier, the less attractive it will be for machine learning models learning a classification task as the predictions would provide information with no discrimination power.

We generate random duplexes of unipartite networks with Poisson or power law degree distributions and different correlations $r_{Spearman}(k,\langle K_d \rangle)$ according to four different specifications:
\begin{itemize}
    \item Poisson degree distribution 
    and $r_{Spearman}(k,\langle K_d \rangle)\approx -0.47$;   
    \item Poisson degree distribution 
    and $r_{Spearman}(k,\langle K_d \rangle)\approx 0$;  
    \item Power law degree distribution 
    and $r_{Spearman}(k,\langle K_d \rangle)\approx -0.47$;      
    \item Power law degree distribution 
    and $r_{Spearman}(k,\langle K_d \rangle)\approx 0$.
\end{itemize}

We generate random unipartite toy networks inspired by the topological and kinetic features of the protein training data used in DeepPurpose. We fix the size of the network to $N=1,507$ and we create randomized networks using the same degree sequence as in BindingDB. For the Poisson case, the link density is constrained by the average number of annotations in the original network. The weight $K_d^{(i,j)}$ assigned to edge $(i,j)$ represents a kinetic constant, and it is derived as the geometric mean of the contribution $K_d^i$ from node $i$ and the contribution $K_d^j$ from node $j$, namely,
\begin{equation}
    K_{d}^{(i,j)} = \sqrt{K_{d}^i K_{d}^j}. 
    \label{equation:edge_kd}
\end{equation}
We explore multiple scenarios to validate our hypothesis on the emergence of topological shortcuts: in presence of a complex correlated relation between $k$ and $\langle K_d \rangle$ as observed in real-world scenarios, affecting both average values and fluctuations, only power law networks will lead to topological shortcuts.
To simplify the modeling of different correlation structures, we use the log-space of kinetic constants, and explore three different sampling strategies: 
(a) sampling without any variance in the $\log K_d^i$ values contributed by node $i$,
(b) sampling with variance in the $\log K_d^i$ values, independent from the degree $k$ of node $i$ and equal to logarithmic variance of BindingDB annotations, (c) sampling with variance in the $\log K_d^i$ values, decreasing as a function of $k$,  as observed in the BindingDB data. 

According to the sampling strategy, each node contributes to Eq. \ref{equation:edge_kd} for all the associated edges with a different extent of variability. In particular, in (a) each node is assigned to a single $\log K_d^i$ for all its edges, sampled according its degree $k$. For a fixed degree $k$, $\log K_d^i$ is sampled from a normal distribution with mean $m= \mu + r_{Spearman}*\sigma*ECDF(k)$ and standard deviation $s=\sqrt((1-r_{Spearman}^2)*\sigma^2)$, where $\mu$ and $\sigma$ are the mean and standard deviation of the $\log K_d$ values in BindingDB. In (b) we follow a similar approach to (a), but instead of sampling a single value, we assign to each node a sample of $5{,}000$ i.i.d. $\log K_d$ instances. Thereafter, when assigning $\log K_d^i$ to each link, we sample uniformly from the generated list of values. In (c), the scenario observed in real data, we first calculate the mean $\langle \log K_d \rangle$ and the standard deviation $\sigma_{\langle \log K_d \rangle}$ for all unique $k$ values. Then, for each link associated with node $i$ we sample a $\log K_d^i$ value from a normal distribution with mean and standard deviation equal to the parameters corresponding to the degree of node $i$. This methodology enforces the same type of complex correlated scenario observed in the original data.

We select as threshold for $K_d^{(i,j)}$ the value for which 16$\%$ of the annotations become positive (binding), enforcing the constraint on the observed $p_{bind}=0.16$. Based on this threshold, we generate the duplex layers with positive and negative edges and calculate the multilink degree sequences, input to the configuration model (see Methods). For the uncorrelated version, we fix the topology while shuffling the $K_d^{(i,j)}$ values at random, which removes any correlation between $k$ and $\langle K_d \rangle$.

\subsection*{Mathematical formalism for the uncorrelated scenario}

When $\langle K_d \rangle$ and $k$ are independent, we can analytically derive the statistical behavior of positive degree $k^{+}$, negative degree $k^{-}$, and degree ratio $\rho$ (Eq. 1 in the main text). 
For each node, the probability of observing $k^{+}$ positive annotations out of $k$ links is binomial
\begin{equation}
   P(k^{+} \lvert k) = {k \choose k^{+}} p_{bind}^{k^{+}} (1-p_{bind})^{(k-k^{+})},
\end{equation}
where $p_{bind}$ encodes the percentage of positive records observed in the database. 

The distribution of positive annotations $k^{+}$ for the whole database is then a compound distribution
\begin{equation}
    P(k^{+}) = \int P(k) P(k^{+} \lvert k) dk,
\end{equation}
where $P(k)$ is the candidate probability distribution for the number of annotations $k$.

From the laws of total expectation and total variance we derive
\begin{eqnarray}
    \langle k^{+} \rangle &=& p_{bind}  \langle k \rangle, \\
    \sigma^2(k^{+})&=&p_{bind} (1- p_{bind}) \langle k \rangle + p_{bind}^2 (\langle k^2 \rangle-\langle k \rangle^2),
\end{eqnarray}
where similar equations hold for $k^{-}$, with $(1- p_{bind})$ replacing $ p_{bind}$. When $P(k)$ is fat-tailed, $(\langle k^2 \rangle-\langle k \rangle^2)$ becomes dominant, and the random variable $k^{+} \approx p_{bind}  \langle k \rangle$.
This formulation suggests that, even in the presence of fat-tailed $P(k)$, the lack of correlations between $\langle K_d \rangle$ and $k$ would determine a distribution of degree ratio $\rho$ well represented by the average
\begin{equation}
 \langle \rho \rangle \approx \frac{\langle k^{+} \rangle}{\langle k \rangle}=p_{bind},
\end{equation}
with noise determined by $p_{bind}$ and $P(k)$. As the duplex configuration model constrains the degree ratio sequence $\{\rho_i\}$, the variability of $\{p_{ij}^{conditional}\}$ drops significantly in absence of  correlation between $\langle K_d \rangle$ and $k$, bringing the model closer to a Naive Bayes classifier (Figure S\ref{fig:SI10_1}).

We can clearly derive the behavior of $p_{ij}^{conditional}$ in the case of \textit{uncorrelated networks}, i.e., networks with no degree correlation and an upper bound for the maximum degree equal to $\sqrt{\langle k \rangle N}$, where $N$ is the size of the unipartite network (Advanced Topics 7.B in \cite{network_science_barabasi}). In this scenario the Lagrangian multipliers satisfy:
\begin{eqnarray}
    e^{-\lambda_i^{(1,0)}} &=& \frac{ k^{+}_i}{\sqrt{\langle k^{+} \rangle N}} \approx \frac{p_{bind} k_i}{\sqrt{p_{bind}\langle k \rangle N}}, \\
e^{-\lambda_i^{(0,1)}} &=& \frac{ k^{-}_i}{\sqrt{\langle k^{-} \rangle N}}\approx \frac{(1-p_{bind}) k_i}{\sqrt{(1-p_{bind})\langle k \rangle N}},\\
p_{ij}^{(1,0)} &=& e^{-(\lambda_i^{(1,0)}+ \lambda_j^{(1,0)} )},\\
p_{ij}^{(0,1)} &=& e^{-(\lambda_i^{(0,1)}+ \lambda_j^{(0,1)} )}.
\end{eqnarray}
It follows that Eq. 7 in the main text for $p_{ij}^{conditional}$ in the transductive test becomes independent from the identity of node $i$ and $j$, as the product $k_i k_j$ simplifies, leading to $p_{ij}^{conditional} \approx p_{bind}$.

\subsection*{Observations}

The major driving factor is the emergence of topological shortcuts in the relation between $k$ and $\langle K_d \rangle$. The monotonicity of the relation between $k$ and $\langle K_d \rangle$ helps the configuration model to predict the link probabilities using the degree sequence as the $K_d$ values are directly associated with the link types after thresholding. When $k$ and $\langle K_d \rangle$ values are anti-correlated and $\langle K_d \rangle$ values have negligible fluctuations for a fixed $k$, degree becomes a strong predictor of $K_d^{(i,j)}$ and subsequently the link types. Hence we observe excellent transductive test performance of the configuration model, for any topology. But when we introduce variance over the $\langle K_d \rangle$ values, the monotonic relation between $k$ and the link types is disrupted. Hence it is difficult for the configuration model to predict the link type only using the degree information. This observation is consistent for networks with both power law and Poisson degree distributions. Yet, we observe that the variance of $\langle K_d \rangle$ is not uniform for different $k$ values in BindingDB. The hubs encounter less variance in $\langle K_d \rangle$ compared to the low degree nodes. Thus, the configuration model is able to predict the link types associated with the hubs. Since these hubs are associated with the majority of the links in the protein-ligand interaction network, making correct predictions using only the degree information of the hubs helps the configuration model achieve commendable transductive test performance. The performance drops for networks with Poisson degree distributions, where hubs are absent, despite enforcing the same type of correlation structure. When we remove the anti-correlation between $k$ and $\langle K_d \rangle$, irrespective of the variance of $\langle K_d \rangle$ values, the configuration model fails to predict the link types using only the degree information. In this scenario, the configuration model performs similar to a Naive Bayes classifier. Related observations are summarized in Table S\ref{table:annotation_distribution_correlation}. Given the observed correlation structure in BindingDB (real world scenario), which affects both expected values and fluctuations in the kinetic constants, topological shortcuts emerge in presence of power law.

\section{Naturally occurring ligands} 
\label{SI:naturalligands}

We extend the drug repurposing task to additional ligands which are not necessarily considered drugs but may nonetheless bind to protein targets. Specifically, we look into the \textit{Natural Compounds in Food Database (NCFD)}  (see Section \ref{SI:SI_sec_NCFD}), which contains food-borne natural compounds, some of which are potential protein binders.  Although these ligands have known chemical structures, they lack adequate binding annotations for training ML models. Binding predictions for these ligands largely depend on comparing their chemical features to other ligands, for which more binding data is available. Figure S\ref{fig:SI1} shows that the naturally occurring compounds in NCFD are larger in size and are more diverse in terms of atomic constituents compared to the drug molecules in DrugBank. This suggests that the binding prediction task on these natural compounds is challenging, which we tackle by maximizing the amount of training data for these natural compounds, and pre-training the chemical embeddings on large chemical libraries. 


\section{DeepPurpose false negative predictions due to annotation imbalance}
\label{SI:falsenegative}

A false negative prediction corresponds to a low binding probability output by the ML model  for a protein-ligand pair which does, in fact, bind. In Figures S\ref{fig:SI5}A and S\ref{fig:SI5}B, we observe that DeepPurpose produces false negative predictions more often for ligands and proteins with low degree ratios. We notice the opposite for the false positives; nodes with high degree ratios contribute more to the false positive predictions in DeepPurpose predictions (see Figure 2C in the main text).



\section{Databases} 
\label{SI:databases}

AI-Bind combines data from four databases: DrugBank, Drug Target Commons (DTC), BindingDB, and Natural Compounds in Food Database (NCFD).  

\subsection{DrugBank}
\label{SI:SI_sec_DrugBank}

DrugBank \cite{10.1093/nar/gkm958} consists of $7{,}307$ drugs and $4{,}762$ protein targets, which form $25{,}373$ drug-target binding pairs. 167 of these drugs are found in NCFD, and we classify them as naturally occurring and food-borne. We consider all reported protein-ligand pairs from DrugBank as positive samples in our dataset, except 53 pairs which have kinetic constants $\geq 10^6 nM$ in BindingDB. The protein sequences included in DrugBank are derived from a wide variety of organisms, including human and different viruses. 

We observe that the annotation distribution of the proteins and the ligands in DrugBank is fat-tailed (see Figure S\ref{fig:SI2}A). This observation is similar to the annotation distributions in BindingDB. The fat-tailed nature of the degree distribution in the binding datasets is a result of the experimentation associated with studying protein-ligand binding. Some proteins and ligands are indeed studied more than others, and hence appear as hubs in such datasets.

\subsection{Drug Target Commons} 
\label{SI:SI_sec_DTC}

We use Drug Target Commons (DTC) \cite{Tang2018} for obtaining binding information related to the natural compounds in NCFD. The intersection of NCFD and DTC contains $1{,}820$ natural ligands and $466$ associated proteins. 

\subsection{BindingDB}
\label{SI:SI_sec_BindingDB}

BindingDB \cite{Liu2007} consists of protein-ligand pairs along with associated kinetic constants and physical conditions related to the reactions such as pH and temperature. We use BindingDB to extend the number of binding pairs in our training data, filter out the non-binding ones from DrugBank, and obtain absolute negative samples. 

\subsection{Natural Compounds in Food Database} 
\label{SI:SI_sec_NCFD}

Multiple existing databases contain information about the compounds present in different food items. As a part of the Foodome project at Center for Complex Network Research (CCNR), we curated external databases like FooDB \cite{FooDB}, Dictionary of Food Compounds (DFC) \cite{Yannai2012}, and KNApSAcK \cite{Afendi2011} to gather information about the compounds in food. Metabolomic experiments were performed to further enrich the database. NCFD contains $20{,}700$ compounds found in different food items, among which $\approx19{,}000$ contain isomeric SMILES \cite{Weininger1988}, a plain-text encoding of the chemical structures of each molecule\footnote[1]{NCFD data was accessed on 7.14.2021. As this database undergoes constant change, we have included a description of the dataset at the time of download in the SI.}.  AI-Bind uses SMILES as input to its ML models for learning useful chemical embeddings.

Figure S\ref{fig:SI2} shows the detailed breakdown of the protein-ligand binding pairs obtained from different databases. 


\section{7-hop threshold for network-derived negatives}
\label{SI:eigenspokes}

We use shortest path distances to generate negative samples.  We consider the node pairs which have shortest path distance $\geq 7$ in the network as non-binding. We derive this 7-hop threshold based on two observations.  First, 7 hops is the minimum shortest path distance at which the average kinetic constant value is above the non-binding threshold of $10^6 nM$ (See Figure 5D in the main text).  
Second, 7 hops is small enough that the negative samples for a given node are not easily distinguishable from positive samples, making the learning task more complex, which helps to defeat shortcut learning in ML models.  
The latter observation is based on EigenSpokes  \cite{Prakash2010} analysis, a network-based dimensionality reduction procedure inspired by Principal Component Analysis (PCA).
Let $A$ be the square adjacency matrix of the protein-ligand network.  Since $A$ is real symmetric, it is orthogonally diagonalizable.  Let $e_1,\ldots,e_n$ be the eigenvectors of $A$ sorted by eigenvalue magnitude $|\lambda_1| \geq |\lambda_2| \geq \ldots \geq |\lambda_n|$.  Given a node $i$, we write the row $a_i$ of $A$ in terms of the eigenbasis $a_i = u_{i1} e_1 + \ldots + u_{in} e_n$.  Truncating after the first 5 eigenvectors (with highest eigenvalue magnitudes) gives a low-dimensional embedding $\bar{u}_i =(u_{i1},u_{i2},u_{i3},u_{i4},u_{i5})$ of each node.   The choice of 5 dimensions gives a useful low-dimensional embedding, while still capturing the most significant degrees of variation.  

Now, consider a fixed protein $i$.  Then ligands $\{j_1, j_2, ...\}$ which bind to $i$ (1-hop) have high magnitude and variance in this 5-dimensional space.  On the other hand, ligands $\{k_1, k_2, ...\}$ that are at a distance of 13 hops from $i$ have $\bar{u}_k$ very close to the origin (Table S\ref{tab:eigenspokes_mean_var}). When 13 hops is chosen as the threshold for negative samples, it would thus be trivial for ML models to distinguish nodes  $\{j_1, j_2, ...\}$ 1 hop away apart from nodes $\{k_1, k_2, ...\}$ 13 hops away, resulting in shortcut learning. Indeed, the same low-degree ligands on the periphery of the network would become negative samples for all the proteins.

We observe a similar behavior for 11-hop and 9-hop thresholds.  However, at 7 hops, we see significantly higher magnitude and variance in $\bar{u}_k$, indicating more diverse negative samples for each protein. In Figure S\ref{fig:eigenspokes}A, we visualize $(u_3,u_4)$ for ligands, colored based on the hop-distances from the example protein BPT4.  We see that at shortest path distances $\geq 7$, most nodes are very close to the origin.  In Figure S\ref{fig:eigenspokes}B, we show the mean of all $\|\bar{u}_j\|$ values averaged over all pairs $(i,j)$ of a given path length. Similarly to what we observed for BPT4, we observe a significant fall-off in magnitude as the shortest path length increases.


\section{Novel deep learning models}

We observe that neural networks exploit the topology of the protein-ligand bipartite network used in training to achieve good performance, and lack node-level generalizability when trained in an end-to-end fashion. AI-Bind circumvents these issues by training its ML models in two phases.
First, AI-Bind learns the node features using unsupervised pre-training, and then it separately trains its classifiers in a supervised manner to predict binding. To show that AI-Bind is not specific to a certain neural network architecture, we experiment with 3 two-phase networks: VecNet, Siamese model, and VAENet.  
AI-Bind first trains a neural network in an unsupervised manner to embed the nodes into a low-dimensional latent space, learning generalizable node representations based on the node features alone (chemical structures of ligands and amino acid sequences of proteins). For example, one of the AI-Bind architectures, VecNet, uses unsupervised node representations from Mol2vec \cite{jaeger2018mol2vec} and ProtVec \cite{Asgari2015}, which are trained separately from each other and from the protein-ligand bipartite network used in training.  Mol2vec and ProtVec are both based on Word2Vec \cite{mikolov2013efficient}, and are designed to create low-dimensional vector representations which retain contextual information for ``words'' in ``sentences'', where the ``sentences'' are formed by molecular sequence descriptions such as Morgan fingerprints \cite{rogers2010extended} or protein sequences. In the second phase, these node representations are passed as input to a binding prediction network, which is trained in a supervised manner.  In AI-Bind's VecNet, the binding prediction network uses fully-connected layers and ReLU non-linearities. 

The Siamese model uses triplet similarity  to find a representation for the node (protein and ligand) features based on their common bindings. The embeddings are then used as inputs to a multilayer perceptron, which learns bindings in a separate supervised training.  The last of AI-Bind's three models, VAENet, uses a Variational Auto-Encoder \cite{kingma2013auto} in order to learn unsupervised ligand representations.

\subsection{VecNet}

We use the pre-trained Mol2vec \cite{Jaeger2018} and ProtVec \cite{Asgari2015} models for node representations. The pre-trained Mol2vec and ProtVec models create 300 and 100 dimensional embeddings for ligands and proteins, respectively.  They are based on Word2Vec \cite{mikolov2013distributed}, and treat the Morgan fingerprints \cite{Rogers2010} and amino acid sequences as sentences in which substructure fingerprints (fragments) and trigrams are the words, respectively.  They are trained in an unsupervised manner to create the representations independent of the binding information.  Namely, they are trained to predict which words appear near each other in sentences.  

Given a fingerprint $x^0$ and an amino acid sequence $x^1$, we encode them using Mol2vec and ProtVec, and then pass them through a simple decoder. We experimented with different neural network architectures with differing number of layers (up to 6 dense layers) and number of neurons per layer (selected from powers of 2 starting at 128 to 2048) and picked one that performed best in inductive tests. This architecture is shown in Figure S\ref{fig:Deep_Architectures}A.

More formally, VecNet computes : 
\begin{align*}
    \bar{x}^0 &= \mathtt{mol2vec}(x^0) \in \mathbb{R}^{300}, \qquad \qquad \bar{x}^1 = \mathtt{protvec}(x^1) \in \mathbb{R}^{100} \\
    \tilde{x}^0 &= \mathtt{ReLU}(W^0\bar{x}^0) \in \mathbb{R}^{2048}, \qquad \qquad \tilde{x}^1 = \mathtt{ReLU}(W^1\bar{x}^1) \in \mathbb{R}^{2048} \qquad \qquad 
\end{align*}
\begin{align*}
     h^0 & =  \mathtt{Concatenate}(\tilde{x}^0, \tilde{x}^1  ) \in \mathbb{R}^{4096} \\
     h^1 & =  \mathtt{ReLU}( W^2 h^0) \in \mathbb{R}^{512} \\
     h^2 & =  \mathtt{ReLU}( W^3 h^1) \in \mathbb{R}^{512} \\
    \hat{y} &= \sigma(W^4 h^2 ) \in [0, 1]
\end{align*}
where $\sigma$ is the sigmoid function and $\sigma (x) = \frac{1}{1 + e^{-x}}$.

\paragraph{Prior use of Mol2Vec and ProtVec in binding prediction.} 

Mol2Vec has previously been used for binding prediction, but only for pre-specified proteins \cite{jaeger2018mol2vec}, where the ML model is trained on one protein at a time.  No information is encoded regarding the protein except for its binding scores with other chemicals in the training data.  In contrast, AI-Bind's VecNet attempts to generalize for different proteins, which we encode using ProtVec. Jaeger et al. \cite{jaeger2018mol2vec} also propose PCM2vec, in which they predict properties of proteins by concatenating Mol2Vec and ProtVec vectors for the same protein read in as a molecule and amino acid sequence, respectively. However, they do not attempt to combine these vectors for different inputs corresponding to a protein-ligand pair.

\subsection{Siamese model}
The Siamese model uses one-shot learning to embed proteins and ligands into the same latent space \cite{koch2015siamese}. For a given node, the Siamese model minimizes the Euclidean distances of that node from the nodes which bind to it, while maximizing the distances to the nodes which do not.  This process is executed in triplets of the forms $\langle$protein, non-binding ligand, binding ligand$\rangle$. For the first kind, AI-Bind trains the network to maximize the Euclidean distance between the protein target and the non-binding ligand, while minimizing the distance of the target from the binding ligand (Figure S\ref{fig:model_flow}A). AI-Bind uses these embeddings, generated by the Siamese architecture, to train a separate model for the downstream classification task of predicting binding. We studied the inductive test performance by changing the number of convolutional layers and the number of embedding dimensions. The final Siamese model consists of 4 convolutional layers and creates 128-dimensional output vectors. The classification network concatenates the embeddings for a protein and a ligand, and then passes it through two fully connected layers, similar to VecNet, to predict the binding probabilities (Figure S\ref{fig:Deep_Architectures}C).

\subsection{VAENet}

VAENet uses a Variational Auto-Encoder \cite{doersch2016tutorial}, an unsupervised learning technique, to embed ligands onto a latent space. Morgan fingerprints are directly fed into a convolutional neural network. The auto-encoder minimizes the loss of structural information while reconstructing the molecule back from the latent representation (Figure S\ref{fig:model_flow}B).
We generate 300-dimensional ligand embeddings using the auto-encoder, which is consistent with the dimensionality of the Mol2vec embeddings used in VecNet. 
The variational nature of this 300-dimensional space allows it to be continuous, allowing for better generalizability. We achieve this generalizability by using the re-parametarization trick from \cite{NIPS2015_5666} to sample from the latent space, instead of directly connecting the latent space to the decoder. Having a generalizable continuous space allows us to map novel ligands into the latent space.

The downstream classification task is achieved by training a fully connected neural network on the concatenated embeddings generated from the Variational Auto-Encoder and ProtVec. The non-end-to-end nature of this architecture ensures that the learned molecular features are independent from the classification task, which has a tendency to exploit shortcuts related to the topology of the protein-ligand interaction network. We observe lower performance for VAENet compared to VecNet (Table S\ref{table:model_results}) mainly for two reasons: (i) VAE has a smaller training dataset of $\sim9$ million chemicals from ZINC, whereas Mol2vec uses 19.9 million chemicals in training. Thus, Mol2vec is better at generalizing to unknown ligand structures. (ii) VAE uses an auto-encoder to embed the ligand molecules, which is a dimensionality reduction approach. Mol2vec uses skipgrams to embed the molecular structures, which is better at capturing contextual information for different fragments in the molecule and provides a more intelligible representation of the ligand structures for the downstream classification task.


\section{Additional deep learning model results}
\label{SI:additional_dl_results}

Table S\ref{table:model_results} contains the performances of AI-Bind's novel deep learning architectures, a DeepPurpose baseline (Transformer-CNN), and the duplex network configuration model on the network-derived dataset.  We also report the performances for models trained with randomized node features. This removes structural information about the proteins and ligands, helping us understand whether the deep learning models leverage structure to learn binding or take topological shortcuts. We observe that DeepPurpose's performance does not change if the inputs are randomly shuffled, which suggests that DeepPurpose learns the topology of the protein-ligand interaction network instead of the node features (see Table 3 in the main). 

In AI-Bind, network-derived negatives and unsupervised pre-training allow the deep learning models to learn binding patterns using the chemical structures instead of the topology of the protein-ligand interaction network. Thus, we observe diminished performance while using random features to make predictions for unseen nodes (inductive test). In this case network-derived negatives remove the annotation imbalance from the training data and prohibit the ML models from taking topological shortcuts. 

Figure S\ref{fig:training_curve} shows the training curves averaged over 5 data splits ($85:15$ split to create train and validation-test datasets) for AI-Bind's three novel models. We set the stopping criterion for training to maximize the inductive test performance (AUPRC) on the validation set. Figure S\ref{fig:f1_score_threshold} shows the F1-scores for the trained models relative to the classification threshold. We obtain the optimal threshold from this curve, which corresponds to the highest F1-score. This optimal threshold is used to obtain the binary labels from the predicted continuous outputs of the AI-Bind architectures. For AI-Bind's VecNet, we obtain an optimal threshold of $0.09$ ($\pm \ 0.015$) in the inductive test scenario. We observe a low optimal threshold as AI-Bind's VecNet predicts high binding probability ($p_{ij}^{VecNet}$) for a few protein-ligand pairs, but we have roughly the same number of  positive and negative samples in the test data. We recommend to use $p_{ij}^{VecNet}$ values to select the top-N predictions, rather than using this optimal threshold to derive the binary labels for novel protein-ligand pairs absent in AI-Bind test data.


\section{Comparison with MolTrans}
\label{moltrans}

We compare the performance of AI-Bind with the Molecular Interaction Transformer (MolTrans) \cite{Huang2020_Moltrans}, a state-of-the-art protein-ligand binding prediction model which uses a combination of sub-structural pattern mining algorithm, interaction modeling module, and an augmented transformer encoder to better learn the molecular structures. Innovative representation of the molecules improves the transductive test performance upon DeepPurpose. MolTrans achieves transductive AUROC of $0.952 \pm 0.051$ and AUPRC of $0.872 \pm 0.131$ on the BindingDB data, while DeepPurpose achieves transductive AUROC of $0.775 \pm 0.0.25$ and AUPRC of $0.800 \pm 0.025$. However, MolTrans performs poorly in inductive tests, i.e., while predicting over novel proteins and ligands. We observe that AI-Bind's VecNet performs better than both DeepPurpose and MolTrans in transductive, semi-inductive, and inductive tests. VecNet's improved inductive performance validates that unsupervised pre-training improves the generalizability of the protein-ligand binding models. The results are summarized in Table S\ref{table:moltrans_results}.


\section{Interpretability of AI-Bind: Identifying active binding sites} 
\label{SI:trigram_study}


AI-Bind may be used to find active binding sites on the amino acid sequence. We plan to leverage this information to define an optimal search grid for docking simulations. Specifically, we use AI-Bind to identify which  trigrams in the amino acid sequence are most significant in predicting binding, thus indicating potential binding locations. This is achieved via an ablation study \cite{meyes2019ablation}, where each trigram in the input amino acid sequence is mutated, that is, replaced with \textit{xxx}, which maps to the  \textit{unknown} vector $\langle unk \rangle$ in the ProtVec model. The \textit{unknown} vector is a learned 100-dimensional vector to which all out-of-vocabulary entries, i.e., amino acid trigrams not present in the ProtVec training corpus, are mapped. This \textit{unknown} vector is set to the mean of all the other learned vectors. We predict the probable binding locations by mutating each trigram in the amino acid sequence one at a time and observing the fluctuations in the AI-Bind predictions (Figure S\ref{fig:trigram_study_1}A). We then smooth the fluctuations using a moving average with a window size of 10 (to eliminate the auxiliary valleys) and obtain a \textit{binding probability profile}.

The suggested binding sites correspond to the amino acid trigrams in the valleys of the binding probability profile. We validate that ligands bind at these valleys by visualizing the docking outputs (see Results) using PyMOL \cite{pymol} and identifying the region around the ligand with a radius of 5\AA , corresponding to the active binding sites (Figure 6C). These regions enclose the amino acid residues which form different bonds with the ligand molecule. Bond distances are measured between the centers of two atoms. Length of hydrogen bonds are typically between 2.3 and 3.9\AA \cite{Harris1999,Laskowski1993}. London dispersion forces or Van der Waals interaction between non-polar chains have bond length between 3.8 to 4.2\AA \ \cite{Roth1996}. Thus, selecting a sphere with a radius of 5\AA \ around the ligand encloses all possible bonds between the ligand and the protein. We identify the amino acid residues inside this sphere, map them to the regions on the amino acid sequence and compare the results with the valleys in the binding probability profile.


To test this method in a specific case, we identify the active binding sites on the human protein TRIM59. For Pipecuronium, Buprenorphine and Voclosporin, three ligands binding to TRIM59 at three different pockets, we study the valleys in the binding probability profile which predict binding locations on the amino acid sequence (Figure 6C). More generally, considering a broad range of ligands, we predicted a total of four active binding sites on the protein TRIM59, three of which have been validated in the docking simulations (Figure S\ref{fig:trigram_study_1}B). We group the ligands binding to TRIM59 according to the different binding sites (Figure S\ref{fig:trigram_study_2}).
From our analysis, a possible fourth active binding site emerges, based on a valley in the binding probability profile, but not associated with any ligand tested in the docking simulations. The shape of the binding probability profiles remains the  same across different ligands, but the drop from the original VecNet prediction (depth of a valley or $\Delta p_{ij}^{VecNet}$) fluctuates for different ligands (Figure S\ref{fig:trigram_study_2}). We observe a moderate positive correlation between the depth of the valleys and binding affinities ($r_{Spearman}(\Delta p_{ij}^{VecNet},\Delta G) = 0.13$). This indicates the depth of the valleys could be indicative of the binding strength and help in identifying the exact binding site on the protein.

Furthermore, we performed unsupervised hierarchical clustering on the binding probability profiles for different ligands of TRIM59. By clustering first the ligand structures using Tanimoto similarity \cite{Bajusz2015}, we find that the ligands binding to TRIM59 are diverse in structure, irrespective of the pockets they bind to (Figure S\ref{fig:trigram_study_3}A). Thus, we cannot identify the specific binding pocket using only ligand structure. When we compare this result to the clustering emerging from the binding probability profiles, we observe a grouping more correlated with the pocket labeling (Figure S\ref{fig:trigram_study_3}B).

\section{Validation using gene phylogeny and bias in false predictions}
\label{SI:genephylogeny}

As additional validation, we investigate if AI-Bind's VecNet predictions are biased towards certain protein structures. The inductive test sets contain a total of $4{,}583$ proteins which are unseen during training in different splits of the 5-fold cross-validation set-up. 
On $3{,}162$ of these proteins, AI-Bind's VecNet makes at least one false prediction, meaning that our model incorrectly labels at least one ligand as a binder which is not (false positive), or labels a ligand as non-binder which does, in fact, bind (false negative).
Among these targets, we find that only $228$ ($5\%$ of all the proteins) are indeed over-represented (proportions test \cite{proportion_test_1998}; $p_{BH-fdr}$\footnote[2]{$p$-value (Benjamini Hochberg - False Discovery Rate corrected)}-value $\leq 0.05$), meaning that over half the predictions involving these proteins are false.
To assess the nature of these false predictions, we test their bias for false positives or false negatives. We find that $168$ proteins are biased towards false positive predictions, whereas $68$ are biased towards false negatives (proportions test; $p_{BH-fdr}$-value $\leq 0.05$). 

To understand whether these biases are intrinsic to the evolutionary origins of certain proteins and if AI-Bind's biases are associated with certain protein domains, we perform a phylogenetic analysis.
We use MUSCLE \cite{Papadopoulos2007}, a tool for multiple protein sequence alignment, to understand the similarity between these $228$ over-represented protein sequences. We observe only weak similarities between these over-represented proteins. 
We also reconstruct their phylogenetic tree using the neighbor joining tree method \cite{Edgar2004} on their amino acid sequences and visualize the results using \texttt{treeio}  and \texttt{ggtree} R packages \cite{Yu2020,Wang2019}.
The results suggest that the false predictions for AI-Bind's VecNet have no bias towards any particular protein structure (Figure S\ref{fig:SpecieTree}). 


\section{Optimal representation of protein and ligand molecules}
\label{SI:optimal_features}

AI-Bind's VecNet achieves the highest inductive performance, i.e., the performance on never-before-seen proteins and ligands. VecNet uses pre-trained Mol2vec (300-dimensional) and ProtVec (100-dimensional) embeddings. These embeddings encode the structural information from the whole protein and ligand molecules \cite{jaeger2018mol2vec,asgari2015continuous}. However, protein-ligand binding is influenced by specific molecular properties, hence we focused on the structural features that are believed to be important to binding in the literature, the so-called \textit{engineered features} \cite{Rohrer2009}. For ligands, we construct the features using the counts of different atoms in the molecule (B, Br, C, Cl, F, I, P, N, O, S), total count of atoms, count of heavy atoms, rings, hydrogen donors, hydrogen acceptors, chiral centers, molecular weight and solubility. For proteins, we use the count of each amino acid, total number of amino acids, and sum of their overall molecular weight. In this set-up, ligands and proteins are represented using 18- and 22-dimensional features, respectively, instead of the original 300 dimensions for Mol2vec and 100 dimensions in ProtVec. Leveraging these engineered features, we observe inductive performance proximal to VecNet (Table S\ref{table:enginnered_features}). 

We further explore which dimensions of Mol2vec and ProtVec are the best in explaining the engineered features. We do so by learning matrix $E$ through algebraic decomposition, with $VE = F$, $V \in \mathbb{R}^{N_{ligands},300}$ for ligands, and $V \in \mathbb{R}^{N_{proteins},100}$ for proteins. Matrix $F$ encodes the engineered features: for ligands we have $F \in \mathbb{R}^{N_{ligands},18}$, while for proteins $F \in \mathbb{R}^{N_{proteins},22}$ \cite{henderson2012rolx}. We re-scale Mol2vec and ProtVec embeddings, as well as the engineered features, between $[0,1]$ and perform non-negative matrix factorization to obtain $E$. The rows of $E$ explain the relevance of each Mol2vec or ProtVec dimension to the engineered features. While investigating the relation between engineered features and embeddings, we observed that $15$ dimensions of the $300$ for Mol2Vec showed the highest variance, suggesting that relevant information is embedded in a smaller dimensional space compared to the standard dimension used in the literature. Similarly, for ProtVec we found a subset of $16$ dimensions (Figure S\ref{fig:engineered_embed}). On the same note, concatenating the engineered features with Mol2vec and ProtVec does not change the inductive performance of VecNet (Table S\ref{table:enginnered_features}). This experiment suggests that the engineered features do not add any extra information to the binding prediction task, i.e the two representations are highly correlated.

We further investigated the engineered features to understand which of them contribute most to protein-ligand binding as they have an intuitive and straightforward interpretation. SHAP \cite{NIPS2017_7062} values show that count of carbon atoms, hydrogen acceptor count, number of chirals, count of fluorine atoms and count of oxygen atoms are the top 5 properties of a ligand that determine its binding to a protein. Presence of amino acids like Glutamic acid, Tryptophan, Asparagine, Methionine, and Threonine in a protein target, presence of aromatic rings (helps in $\pi$-stacking), presence of R groups, and N or C terminus of the protein molecules, drive protein-ligand binding (see Tables S\ref{table:ligand_shap} and S\ref{table:protein_shap}). 

VecNet with engineered features achieves similar inductive test score as the original version. Yet, the predictions from VecNet using engineered features $\{p_{ij}^{VecNet-Engineered}\}$ show poor negative correlation with $\Delta G$  binding affinities obtained from docking simulations in the Results Section ($r_{Spearman}=-0.10$) when obtain the binary labels by thresholding using the median predicted probabilities, compared to the original VecNet with Mol2vec and ProtVec embeddings ($r_{Spearman}= -0.51$). We also observe a significant reduction in F1-score, from 0.82 to 0.64 (see Results).

Overall, when representing protein and ligand molecules in 2D, we find that only a small subset of the features drive protein-ligand binding and are able to explain intuitive properties of the molecules. Simple molecular descriptors like the presence of R groups in the amino acids, different atom counts, charge distribution in the ligand molecule represented by hydrogen acceptor, and donor counts have significant predictive power for protein-ligand binding. However, these features do not provide insight into the surface structure of the molecules or their rigidity. Indeed, presence of binding pockets on proteins and rotatability of bonds in ligand molecules significantly influence protein-ligand binding. Including these relevant aspects into the prediction task would reduce the number of false positives, often determined by the lack of 3D structures in the model. Adding 3D features of ligands and proteins (e.g., shape of the molecules, rotation of bonds in ligand, location of binding pockets etc.) will help AI-Bind to learn the detailed mechanism behind protein-ligand binding and make more accurate predictions. 


\section{Random Negative Sampling}

Existing binding prediction models do not consider any balancing between the binding and the non-binding pairs. In DeepPurpose, the non-binding pair generation is done via selecting random pairs of proteins and ligands which do not appear as binding pairs in the training data. As a result, an imbalance is created between the positive and negative samples for certain nodes based on their degrees in the network, and the deep learning models learn from the network topology of the protein-ligand network instead of learning the binding patterns from the molecular structures. Researchers are aware of this imbalanced training caused by binding data-sets like Tox21 and have proposed an oversampling-based approach to resolve the issue \cite{Idakwo2020}. This method, however, did not improve prediction accuracy and generalizability since the root cause of degree bias is not resolved via oversampling.

In this section, we propose different methods for generating the negative pairs in a balanced fashion from the protein-ligand bipartite network. As we use a batch size of 16 in AI-Bind training \cite{Bengio_2012}, our ML models observe 16 data instances corresponding to a protein-ligand pair, 15 of which are negative samples, and the remaining being the positive edge.

We generate random negatives for each positive edge $(t,d)$ representing the binding pair of target ($t$) and drug ($d$). This is achieved by randomly selecting drugs with no known binding information to $t$
and randomly selecting proteins with no known binding information to $d$. A list of 15 random negative edges is generated where 7(8) random negatives relate to the target of the positive pair and 8(7) random negatives relate to the drug of the positive pair. Since this method produces negative samples surmounting the number of positives, we use a smaller class weight for the negative samples during training.

In Figure \ref{fig:SI3}, we explore the plausibility of using the network-derived negatives for training ML models instead of the random negative samples. We show that the non-binding (or negative) degrees of the nodes in random negative sampling are correlated with the binding (or positive) degrees. Thus, the random negative samples accommodate the same topological information on the protein-ligand network as the positives, providing no additional information on the negative annotations to training. This issue is resolved by creating the network-derived negative samples, which are based on the shortest path distances in the protein-ligand bipartite network.

Finally, we studied the inductive performance of VecNet on both random negatives and network-derived negatives in a 5-fold cross validation set-up. We observe lower inductive test performance on the random negatives (AUROC of $0.709 \pm 0.011$ and AUPRC of $0.566 \pm 0.013$) compared to the network-derived negatives (AUROC of $0.745 \pm 0,032$ and AUPRC of $0.729 \pm 0.038$).


\section{Gold standard validation of binding probability profile}

In this section we use gold standard protein-ligand binding data to validate the binding probability profiles predicted by AI-Bind. In other words, we validate our hypothesis that ligands bind to proteins at the valleys on the binding probability profile with high confidence gold standard experimental protein-ligand binding data \cite{Cheng2009_Gold}. This validation also shows higher propensity of the $\beta$-sheets and the coils regions to bind with the ligands.

First, we obtain the binding probability profiles generated by AI-Bind for two different ligand-protein pairs. We chose \textit{E. Coli} protein Thymidylate Synthase. The ligands are SP-722 and SP-876. We obtain the experimentally validated secondary structure from the RCSB website, and overlay it over the binding probability profile. We then extract from the PDB file the primary binding sites of the ligand molecules. These binding locations (amino acid residues) are represented by the AC1 keyword in the PDB file. The site lists the amino acids that the ligands bind to, which are represented by red dots on the binding probability profiles (see Figure S\ref{fig:gold_standard_validation}). In both cases, the binding sites lie in the valleys of the probability profile, and overlay on the the $\beta$-sheets and the coils regions. Figure S\ref{fig:gold_standard_validation_human} depicts similar observations on human proteins. We have also compared the binding sites predicted by AI-Bind with p2rank, another state-of-the-art site detection method \cite{Krivk2018_p2rank}.

We compare the binding sites predicted by both AI-Bind and p2rank to the gold standard experimental data. For determining the binding sites from the valleys of AI-Bind's binding probability profile, we fit a sine curve to the valleys and consider the region between the points of inflection of the sine curve as the AI-Bind predicted binding site. On the other hand, p2rank predicts the amino acid residues and the associated pockets as the binding locations. 
Thereafter, we compare the binding pockets predicted by AI-Bind and p2rank to the gold standard experimental data. We observe that the AI-Bind predicted binding sites cover 64.05\% of all pockets on the 195 different proteins mentioned in the gold standard dataset, whereas p2rank is able to identify 53.57\% of all of these pockets.



\bibliographystyle{naturemag}
\bibliography{main_combined}

\newpage

\setcounter{table}{0}


\begin{table}[h!]
    \centering
    \renewcommand{\arraystretch}{1.0}
    \small
    \begin{tabular}{|c|c|c|c|c|c|} 
        \hline
        Variance of $\langle K_d \rangle$ & Annotation distribution & $r_{Spearman}(k, \langle K_d \rangle)$ & $p_{bind}$ & AUROC & AUPRC \\
        \hline
        \hline
        Negligible variance  & Power law & $\approx -0.47$ & $0.16$ & $0.95$ & $0.89$ \\
        \hline
        $k$-independent variance  & Power law & $\approx -0.47$ & $0.16$ & $0.64$ & $0.26$ \\
        \hline
        $k$-dependent variance & Power law & $\approx -0.47$ & $0.16$ & $0.84$ & $0.54$ \\
        \hline
        \hline
        Negligible variance & Poisson & $\approx -0.47$ & $0.16$ & $0.94$ & $0.87$ \\
        \hline
        $k$-independent variance & Poisson & $\approx -0.47$ & $0.16$ & $0.68$ & $0.29$ \\
        \hline
        $k$-dependent variance & Poisson & $\approx -0.47$ & $0.16$ & $0.69$ & $0.27$ \\
        \hline
        \hline
        Negligible variance & Power law & $\approx 0$ & $0.16$ & $0.54$ & $0.17$ \\
        \hline
        $k$-independent variance & Power law & $\approx 0$ & $0.16$ & $0.51$ & $0.17$ \\
        \hline
        $k$-dependent variance & Power law & $\approx 0$ & $0.16$ & $0.49$ & $0.14$ \\
        \hline
        \hline
        Negligible variance & Poisson & $\approx 0$ & $0.16$ & $0.50$ & $0.16$ \\
        \hline
        $k$-independent variance & Poisson & $\approx 0$ & $0.16$ & $0.49$ & $0.16$ \\
        \hline
        $k$-dependent variance & Poisson & $\approx 0$ & $0.16$ & $0.49$ & $0.16$ \\
        \hline
        \hline
    \end{tabular}
    \caption{Transductive test performance of a duplex configuration model on unipartite layers with varying annotation distribution $P(k)$ and correlation $r_{Spearman}(k,\langle K_d \rangle)$. The network has $N=1{,}507$ nodes, same as the number of unique proteins in the BindingDB training data. The constrained features are consistent with the protein sample in BindingDB, e.g., for the power law network we use the degree sequence derived from the protein network in BindingDB data, while the Poisson network has the same average degree $\langle k \rangle = 47$ of the power law network. 
    To achieve $r_{Spearman}(k,\langle K_d \rangle) \approx 0$, we shuffle the edges of the original network, which removes the negative correlation between $k$ and $\langle K_d \rangle$.}
    \label{table:annotation_distribution_correlation}
\end{table}


\begin{table}[h!]
    \centering
    \begin{tabular}{l|ccccccc}
        \hline 
        Path Length $i$ to $j$  & 1 & 3 & 5 & 7 & 9 & 11 & 13  \\
        \hline 
        Mean $\| \bar{u}_j \|$ & 0.045 & 0.035 & 0.014 & 0.004 & 0.001 & 0.0001 & $5 \cdot 10^{-7}$  \\
        Std. dev. $\| \bar{u}_j \|$ & 0.075 & 0.035 & 0.025 & 0.014 & 0.008 & 0.001 & $4 \cdot 10^{-6} $\\ 
        \hline
    \end{tabular}
    \caption{As path length to a fixed protein $i$ increases, the mean and variance of the length of the low-dimensional embedding of the ligand $\|\bar{u}_j\|$ decrease. }
    \label{tab:eigenspokes_mean_var}
\end{table}


\begin{table}[ht]
\renewcommand{\arraystretch}{1.0}
\small
\begin{tabular}{l c c c c c c} 
    \hline
     Model & \multicolumn{6}{c}{\underline{\qquad\qquad\qquad\qquad\qquad Test Data Division \qquad\qquad\qquad\qquad\qquad }} \\
     & \multicolumn{2}{c}{\underline{\quad\quad\quad\quad Transd.\quad\quad\quad\quad }} &  \multicolumn{2}{c}{\underline{\quad\quad\quad\quad Semi-induc. \quad\quad\quad\quad } }  &  \multicolumn{2}{c}{\underline{\quad\quad\quad\quad Induc. \quad\quad\quad\quad }} \\
     & AUROC & AUPRC & AUROC & AUPRC & AUROC & AUPRC \\
    \hline
    Configuration & $.738 \pm .014$ & $.739 \pm .017$ & $.754 \pm .021$ & $.691 \pm .038$ & $.500 \pm .000$ & $.469 \pm .014$ \\
    \hline
     VecNet & $\color{red}.794 \pm .008$ & $\color{red}.817 \pm .013$ & $\color{red}.779 \pm .025$ & $\color{red}.752 \pm .039$ & $\color{red}.745 \pm .032$ & $\color{red}.729 \pm .038$\\
     Siamese & $.664 \pm .027$ & $.637 \pm .003$ & $.666 \pm .031$ & $.621 \pm .032$ & $.639 \pm .026$ & $.583 \pm .025$ \\
     VAENet & $.777 \pm .010$ & $.701 \pm .048$ & $.756 \pm .022$ & $.710 \pm .030$ & $.740 \pm .024$ & $.701 \pm .048$\\
     DeepPurpose & $.775 \pm .025$ & $.800 \pm .025$ & $.642 \pm .022$ & $.591 \pm .042$ & $.642 \pm .025$ & $.583 \pm .016$\\
     \hline 
     \textit{Random Inputs} & & & & \\
     VecNet Unif. & $.668 \pm .012
     $ & $.702 \pm .015$ & $.539 \pm .019$ & $.541 \pm .013$ & $.466 \pm .054$ & $.456 \pm .041$ \\
     VecNet - Targ. & $.704 \pm .008
     $ & $.725 \pm .009$ & $.575 \pm .032$ & $.556 \pm .025$ & $.558 \pm .009$ & $.501 \pm .021$ \\
     \hline
    \hline
\end{tabular}

\caption{\textbf{Results across different models.} We summarize all performances on the network-derived negative samples. We perform 5-fold cross-validation, reporting AUROC and AUPRC averaged over the 5 runs with random initialization and data split.  Results are reported separately on 3 different train-validation-test splits with different data held out in the validation and testing sets: (1) \textbf{Unseen edges (Transducitve test)} - test sets contain unseen edges in the train network, (2) \textbf{Unseen targets (Semi-inductive test)} - test sets contains pairs with proteins that do not appear in train or validation set, (3) \textbf{Unseen nodes (Inductive test)} - nodes in test set pairs are completely disjoint from train set.  \emph{Random Input Tests:} We train and test AI-Bind's VecNet replacing node features with random entries drawn from a uniform distribution in the range $U([-1,1]^{d})$. We run two tests (1) Unif. - All node features are replaced by vectors drawn from a uniform distribution $U([-1,1]^{d})$, (2) Unif.Targ. - Only the target node features are replaced by vectors from $U([-1,1]^{d})$; drug features remaining the same. Note that the transductive (unseen edges) performance is reported based on the models optimized for unseen nodes, except for the case of the Random Inputs, where we report performance based on models optimized for unseen targets.}
\label{table:model_results}
\end{table}


\begin{table}[ht]
\renewcommand{\arraystretch}{1.0}
\small
\begin{tabular}{l c c c c c c} 
    \hline
     Model & \multicolumn{6}{c}{\underline{\qquad\qquad\qquad\qquad\qquad Test Data Division \qquad\qquad\qquad\qquad\qquad }} \\
     & \multicolumn{2}{c}{\underline{\quad\quad\quad\quad Transd.\quad\quad\quad\quad }} &  \multicolumn{2}{c}{\underline{\quad\quad\quad\quad Semi-induc. \quad\quad\quad\quad } }  &  \multicolumn{2}{c}{\underline{\quad\quad\quad\quad Induc. \quad\quad\quad\quad }} \\
     & AUROC & AUPRC & AUROC & AUPRC & AUROC & AUPRC \\
    \hline
     MolTrans\footnote{BindingDB data} & $.952 \pm .051$ & $.872 \pm .131$ & $.653 \pm .041$ & $.415 \pm .095$ & $.575 \pm .059$ & $.430 \pm .098$\\
     
     MolTrans\footnote{Network-derived Negatives} & $.860 \pm .074$ & $.805 \pm .092$ & $.641 \pm .015$ & $.489 \pm .051$ & $.619 \pm .021$ & $.480 \pm .028$\\
     
     DeepPurpose & $.775 \pm .025$ & $.800 \pm .025$ & $.642 \pm .022$ & $.591 \pm .042$ & $.642 \pm .025$ & $.583 \pm .016$\\

     VecNet & $\color{red}.794 \pm .008$ & $\color{red}.817 \pm .013$ & $\color{red}.779 \pm .025$ & $\color{red}.752 \pm .039$ & $\color{red}.745 \pm .032$ & $\color{red}.729 \pm .038$\\

     \hline
    \hline
\end{tabular}

\caption{\textbf{Comparing AI-Bind with MolTrans.} We compare transductive, semi-inductive, and inductive performances of MolTrans with AI-Bind's VecNet. MolTrans uses a combination of sub-structural pattern mining algorithm, interaction modeling module and an augmented transformer encoder to better learn the molecular structures. VecNet performs better compared to MolTrans in semi-inductive and inductive tests. This analysis validates that unsupervised pre-training improves the generalizability of the protein-ligand binding models. We have trained and tested MolTrans on both BindingDB (used in the original paper), and network-derived negatives (AI-Bind data).}
\label{table:moltrans_results}
\end{table}


\newcolumntype{F}[1]{%
    >{\raggedright\arraybackslash\hspace{0pt}}p{#1}}%
\newcolumntype{T}[1]{%
    >{\centering\arraybackslash\hspace{0pt}}p{#1}}%

\begin{table}[h!]
    \centering
    \begin{tabular}{|F{0.2\textwidth}|T{0.14\textwidth}|T{0.14\textwidth}|T{0.18\textwidth}|T{0.18\textwidth}|T{0.15\textwidth}|}
        \hline
        Performance & Original & Engineered features & Mol2vec and ProtVec dimensions explaining Engineered Features & Concatenated: Mol2vec/ProtVec + Engineered features \\ 
        \hline
        AUROC & $0.75 \pm 0.032$ & $0.73 \pm 0.032$ & $0.72 \pm 0.066$ & $0.75 \pm 0.033$ \\
        \hline 
        AUPRC & $0.73 \pm 0.038$ & $0.74 \pm 0.033$ & $0.72 \pm 0.057$ & $0.73 \pm 0.042$ \\ 
        \hline
    \end{tabular}
    \caption{\textbf{Optimal feature selection:} We observe that AI-Bind's VecNet shows similar performances in inductive tests, when Mol2vec and ProtVec are replaced by simple engineered features encoding certain properties of protein and ligand molecules. Furthermore, we observe that only 15 dimensions of Mol2vec and 16 dimensions of ProtVec embeddings encode these molecular properties driving the binding task. Using these feature subsets of Mol2vec and ProtVec helps VecNet achieving similar inductive performance. Concatenating the engineered features with Mol2vec and ProtVec does not improve inductive performance. This suggests that the information encoded in the engineered features strongly correlates with Mol2vec and ProtVec embeddings.} 
    \label{table:enginnered_features}
\end{table}


\begin{table}[h!]
    \centering
    \begin{tabular}{|F{0.6\textwidth}|T{0.4\textwidth}|}
        \hline
        Feature & Average SHAP Importance \\ 
        \hline
        Count of Carbon Atom & 0.012546 \\
        \hline
        Hydrogen Acceptor Count	& 0.012362 \\
        \hline
        Number of Chirals & 0.008750 \\
        \hline
        Count of Flourine Atoms	& 0.006527 \\
        \hline
        Count of Oxygen Atoms & 0.006184 \\
        \hline
        Hydrogen Donor Count & 0.005647 \\
        \hline
        Number of Atoms & 0.004165 \\
        \hline
        Count of Heavy Atoms & 0.003468 \\
        \hline
        Solubility in Water ($\log p$) & 0.003202 \\
        \hline
        Molecular Weight & 0.003161 \\
        \hline
        Count of Nitrogen Atoms & 0.002007 \\
        \hline
        Count of Chlorine Atoms	& 0.001720 \\
        \hline
        Count of Sulphur Atoms & 0.001483 \\
        \hline
        Number of Rings	& 0.000525 \\
        \hline
        Count of Phosphorus Atoms & 0.000191 \\
        \hline
        Count of Iodine Atoms & 0.000130 \\
        \hline
        Count of Bromine Atoms & 0.000083 \\
        \hline
        Count of Boron Atoms & 0.000070 \\
        \hline
    \end{tabular}
    \caption{\textbf{Engineered feature importance for ligands:} We tabulate the engineered features in descending order of average absolute SHAP importance over AI-Bind data. A higher SHAP value represents more relevance of the molecular property in predicting protein-ligand binding.} 
    \label{table:ligand_shap}
\end{table}


\begin{table}[h!]
    \centering
    \begin{tabular}{|F{0.6\textwidth}|T{0.4\textwidth}|}
        \hline
        Feature & Average SHAP Importance \\ 
        \hline
        Count of Glutamic acid (E) & 0.036747 \\
        \hline
        Count of Tryptophan (W) & 0.033210 \\
        \hline
        Count of Asparagine (N)	& 0.024770 \\
        \hline
        Count of Methionine (M) & 0.022734 \\
        \hline
        Count of Threonine	(T) & 0.021194 \\
        \hline
        Count of Glycine (G) & 0.020832 \\
        \hline
        Count of Arginine (R) & 0.019599 \\
        \hline
        Count of Phenylalanine (F) & 0.017040 \\
        \hline
        Count of Cysteine (C) & 0.016428 \\
        \hline
        Count of Isoleucine (I) & 0.016215 \\
        \hline
        Count of Alanine (A) &	0.015732 \\
        \hline
        Count of Histidine (H) & 0.014813 \\
        \hline
        Count of Leucine (L) & 0.014026 \\
        \hline
        Count of Tyrosine (Y) & 0.013844 \\
        \hline
        Count of Proline (P) & 0.013014 \\
        \hline
        Count of Valine (V) & 0.011152 \\
        \hline
        Count of Serine (S) & 0.010930 \\
        \hline
        Count of Lysine (K) & 0.008689 \\
        \hline
        Count of Aspartic acid (D) & 0.008303 \\
        \hline
        Total amino acid count & 0.003381 \\
        \hline
        Count of Glutamine (Q) & 0.002957 \\
        \hline
        Molecular Weight & 0.002088 \\
        \hline
    \end{tabular}
    \caption{\textbf{Engineered feature importance for proteins:} We tabulate the engineered features in descending order of average absolute SHAP importance over AI-Bind data. A higher SHAP value represents more relevance of the molecular property in predicting protein-ligand binding.
    } 
    \label{table:protein_shap}
\end{table}


\newpage

\setcounter{figure}{0}   

\begin{figure}[ht!]
    \centering
    \includegraphics[clip,angle=0,width=0.7\textwidth,height=0.5\textheight]{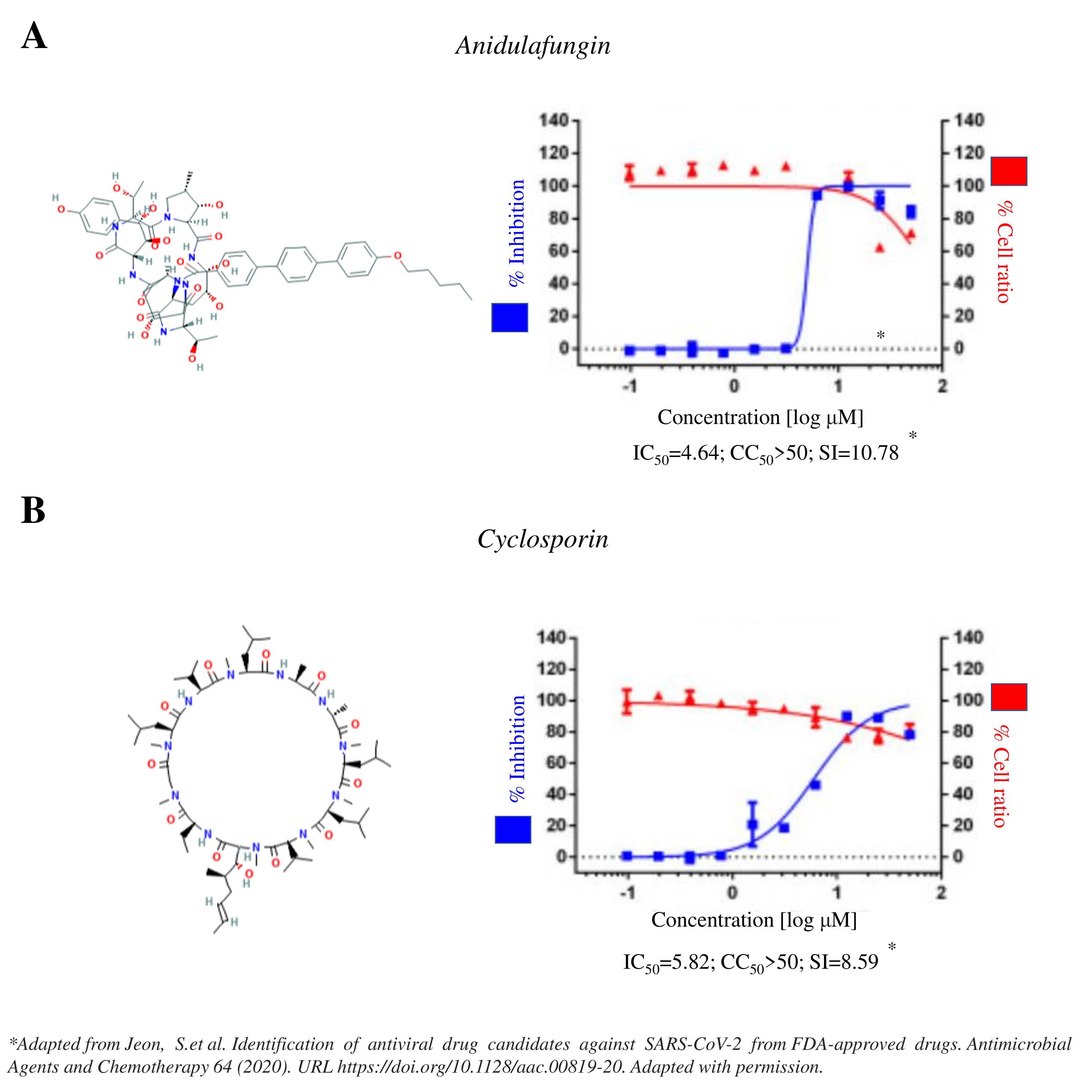}
    \caption{\textbf{Experimental evidence to validate AI-Bind predictions.} \textbf{(A)-(B)} Anidulafungin and Cyclosporin, two FDA approved anti-fungal agents predicted by AI-Bind, show potential antiviral activities against SARS-CoV-2, with $IC_{50}$ values $4.64\mu M$ and $5.82\mu M$, respectively.}
    \label{fig:SI27}
\end{figure}

\begin{figure}[ht!]
    \centering
    \includegraphics[clip,angle=0,width=1.0\textwidth,height=0.5\textheight]{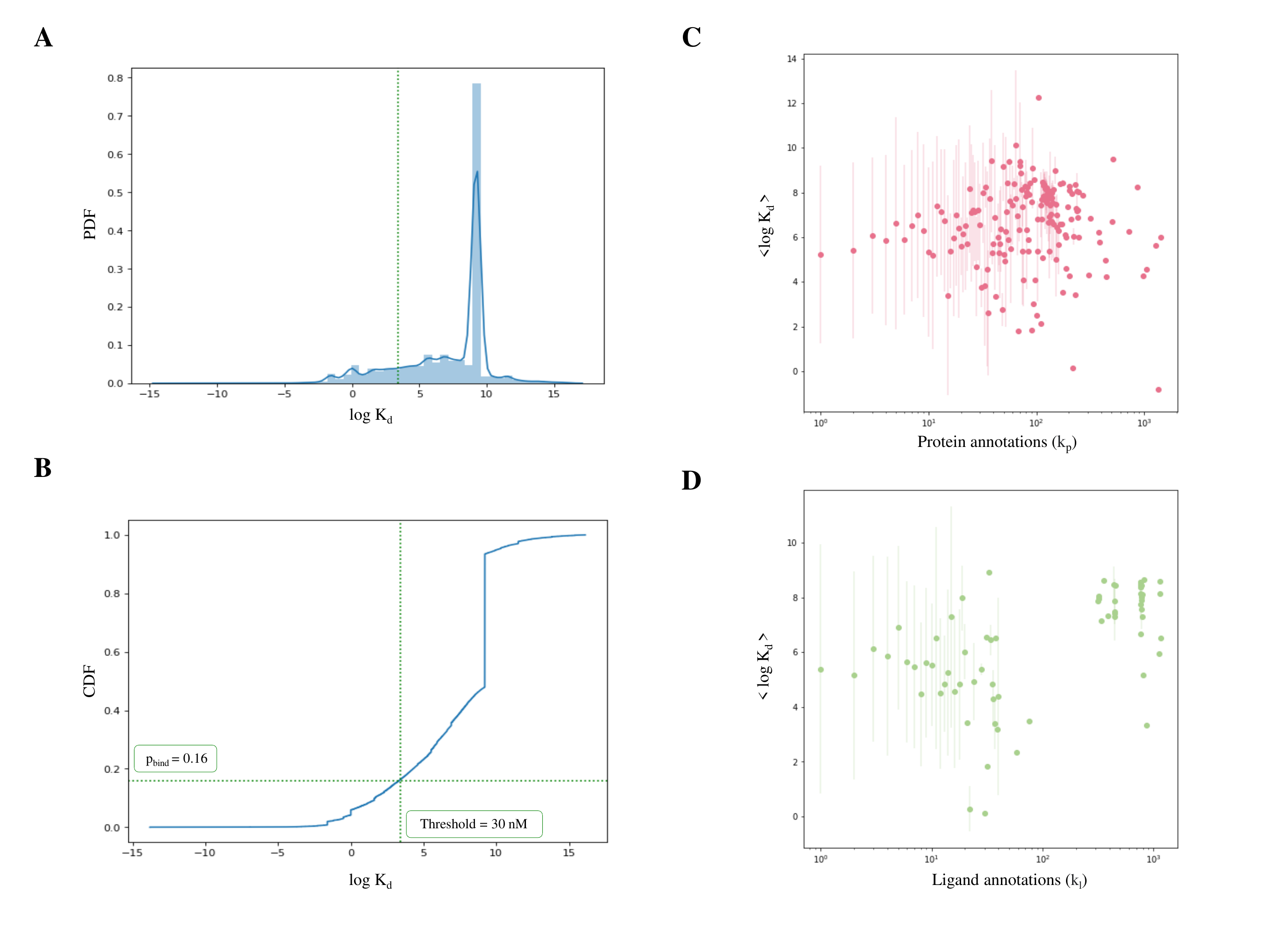}
    \caption{\textbf{Disassociation constant $K_d$ and its relation with the number of annotations/records $k$ in BindingDB.} \textbf{(A)-(B)} Density distribution and cumulative distribution of $\log K_d$ in BindingDB training data. With threshold $30nM$, we obtain an average binding probability of $p_{bind} = 0.16$. \textbf{(C)-(D)} Each node is characterized by the number of annotations $k$, and the average $\langle K_d \rangle$ over its records. We select the $K_d$ values in the log-space for creating the toy model. We do not observe significant correlation between $k$ and $\langle \log K_d \rangle$ opposed to the anti-correlation observed in the linear space, but $k$ and the variance of $\langle \log K_d \rangle$ values are highly anti-correlated with are  $r_{Spearman}(k,  \sigma_{\langle \log K_d \rangle})=-0.71$. This observation implies that the lower degree nodes have higher fluctuations in the associated $\langle \log K_d \rangle$ values compared to the higher degree nodes and hubs.}
    \label{fig:SI8}
\end{figure}

\begin{figure}[ht!]
    \centering
    \includegraphics[clip,angle=0,width=\textwidth,height=\textheight]{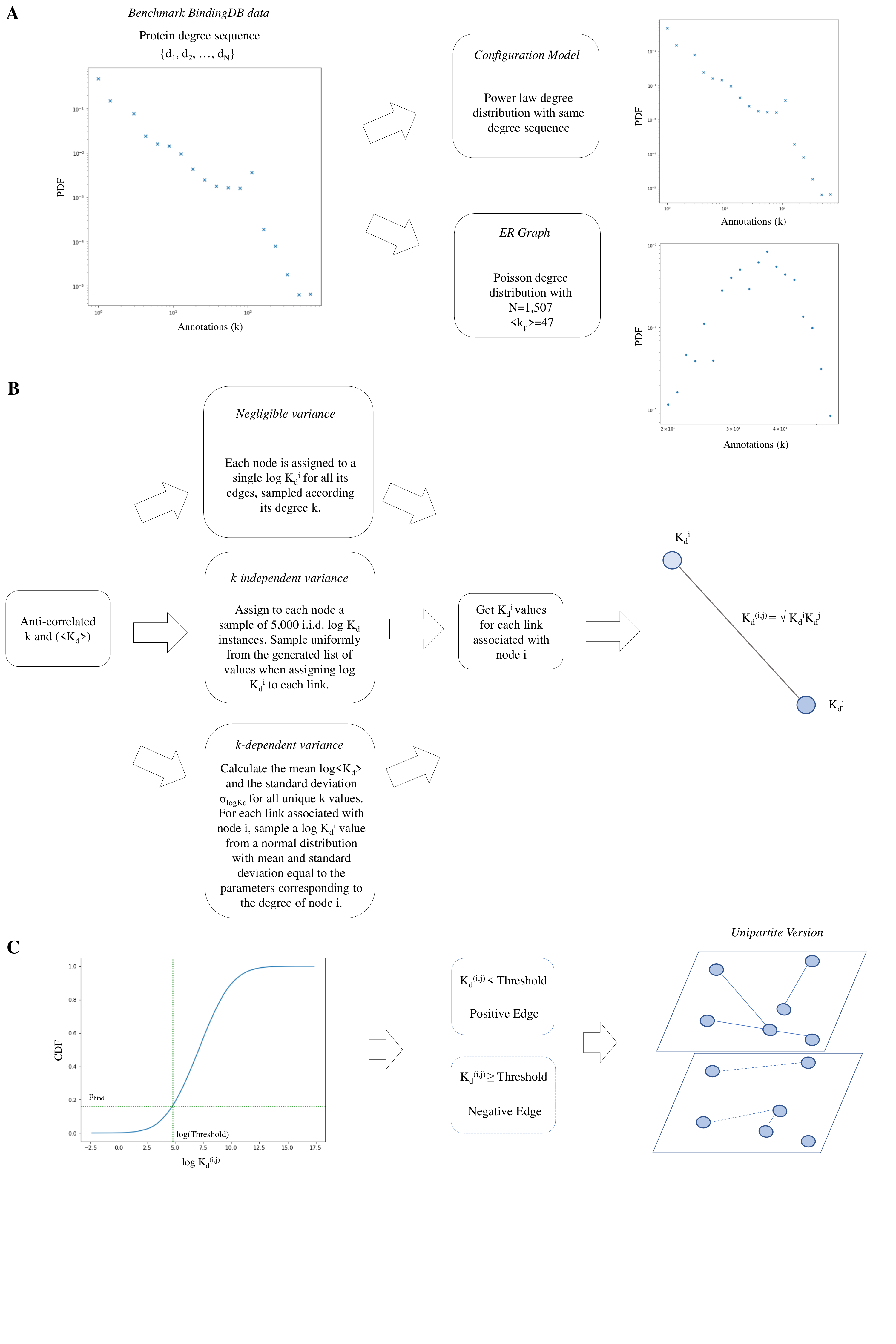}
\end{figure}

\begin{figure*}
    \caption{\textbf{Experimental set-up for studying the emergence of topological shortcuts.} \textbf{(A)} We generate random unipartite networks inspired by the topological and kinetic features of the protein sample in BindingDB. In particular, we fix the size of the network to $N=1,507$ and use the same degree distribution as in BindingDB, while for the Poisson case the link density is constrained by the average number of annotations in the power law network.    
    \textbf{(B)} We explore three different strategies of sampling the kinetic constants: 
    (a) sampling without any variance in the $\log K_d^i$ values contributed by node $i$ to its links, (b) sampling with variance in the $\log K_d^i$ values, independent from the degree $k$ of node $i$ and equal to logarithmic variance of $\log K_d$ in the BindingDB protein sample, and (c) sampling with variance in the $\log K_d^i$ values, decreasing as a function of $k$,  as observed in the BindingDB data. According to the sampling strategy, each node contributes to all its edges with a different extent of variability. In particular, in (a) each node is assigned to a single $\log K_d^i$ for all its edges, sampled according its degree $k$. In (b) we follow a similar approach to (a), but instead of sampling a single value, we assign to each node a sample of $5{,}000$ i.i.d. $\log K_d$ instances. Thereafter, when assigning $\log K_d^i$ to each link associated with node $i$, we sample uniformly from the generated list of values. In (c), the scenario observed in BindingDB data, we first calculate the mean $\langle \log K_d \rangle$ and the standard deviation $\sigma_{\langle \log K_d \rangle}$ for all unique $k$ values. Then, for each link associated with node $i$ we sample a $\log K_d^i$ value from a normal distribution with mean and standard deviation equal to the parameters corresponding to the degree of node $i$. The final disassociation constant $K_d^{(i,j)}$ assigned to edge $(i,j)$ is the geometric average of the contribution $K_d^i$ from node $i$ and the contribution $K_d^j$ from node $j$. In the uncorrelated scenario, we randomly shuffle the $K_d$ values associated with the links, which removes the anti-correlation between $k$ and $\langle K_d \rangle$. \textbf{(C)} We select as threshold for $K_d^{(i,j)}$ the value for which a fixed percentage of the annotations become positive or binding, enforcing the constraint on the observed $p_{bind}=0.16$. Based on this threshold, we generate the duplex layers with positive and negative edges and calculate the multilink degree sequences, input to the configuration model (see Methods).}
\end{figure*}

\begin{figure}[ht!]
    \centering
    \includegraphics[clip,angle=0,width=\textwidth,height=0.4\textheight]{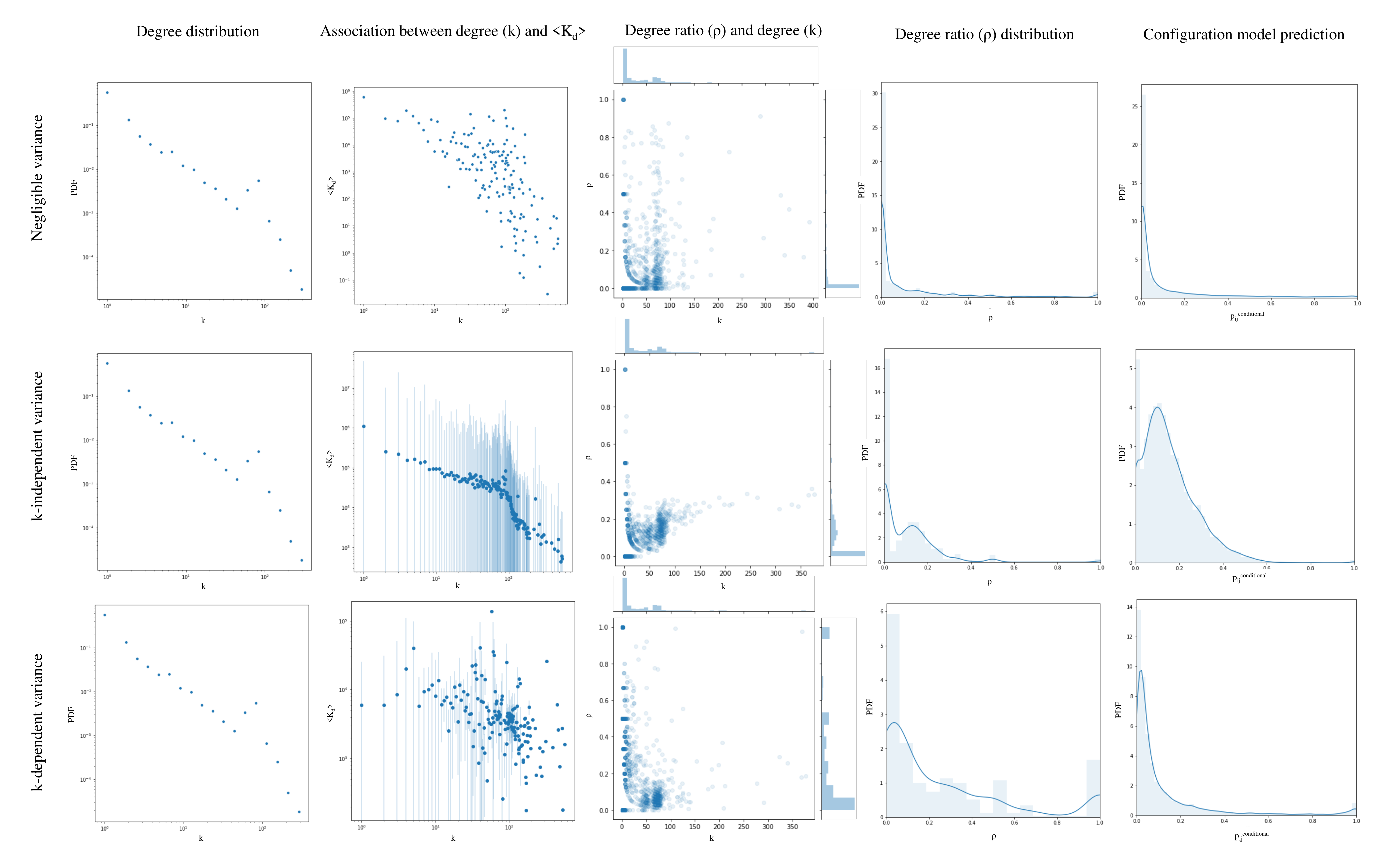}
    \caption{\textbf{Emergence of topological shortcuts in scale-free networks.} In absence of variance in the $\langle K_d \rangle$ values, the relation between $k$ and $\langle K_d \rangle$ is monotonic and the configuration model is able to predict the link types using only the degree information of the nodes. When variance is introduced, the monotonicity of the relation between $k$ and $\langle K_d \rangle$ is disrupted. Thus, the configuration model is unable to learn the link types only using the degree information. The scenario with varying variance resembles the data in BindingDB. Less fluctuations for the hubs makes the link classification task easier for the hubs. Since majority of the links in the protein-ligand interaction network are associated with the hubs, we observe the configuration model achieving excellent transductive test performance.}
    \label{fig:SI10_1}
\end{figure}

\begin{figure}[ht!]
    \centering
    \includegraphics[clip,angle=0,width=1.0\textwidth]{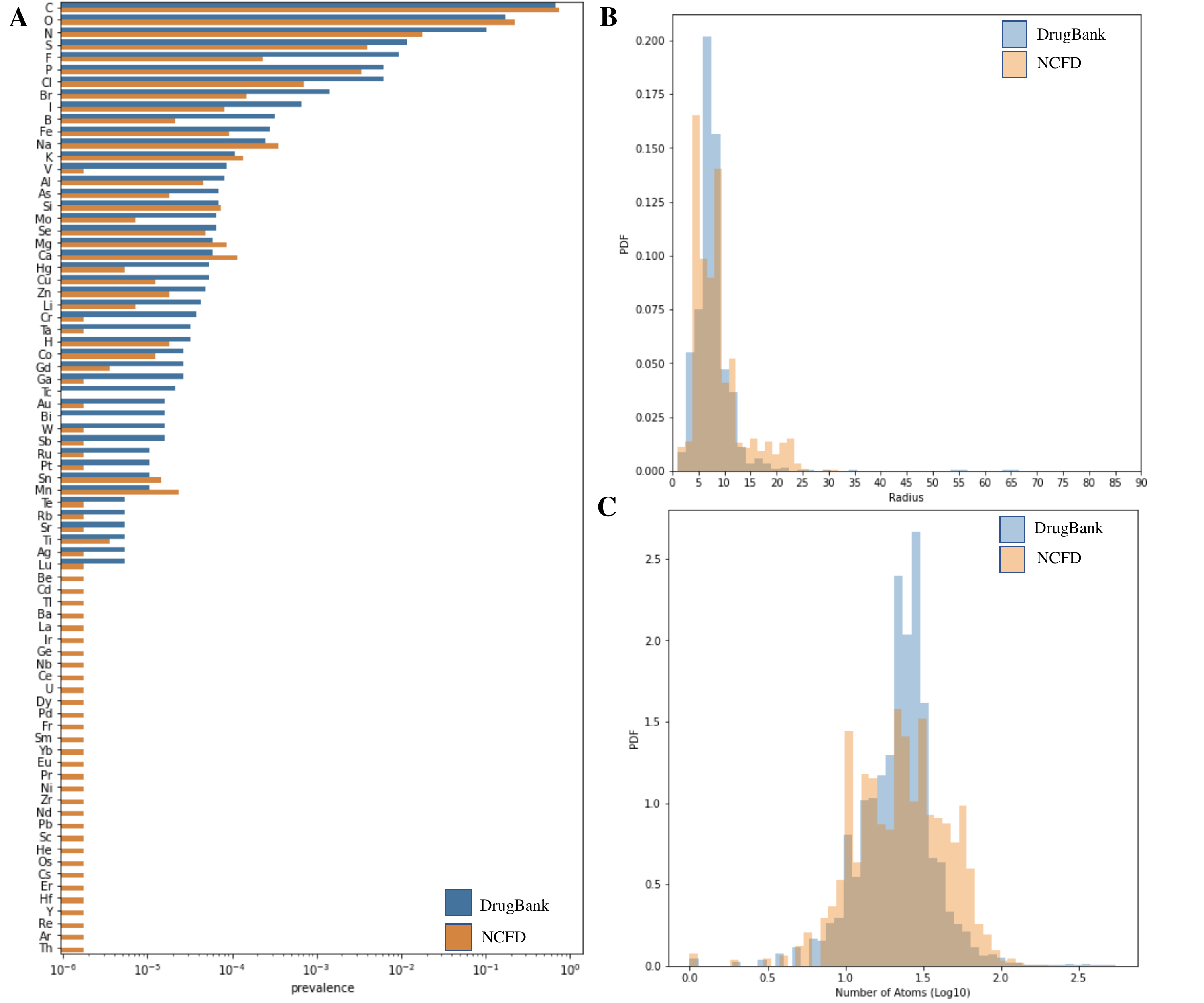}
    \caption{\textbf{Naturally occurring compounds are structurally more complex than drugs.} \textbf{(A)} Prevalence of different atoms in DrugBank and natural ligands present in NCFD. Natural ligands show more diversity in terms of the constituent atoms. \textbf{(B)} The distribution of the radius across the ligand molecules in DrugBank and NCFD, and \textbf{(C)} The distribution of the number of atoms across the ligand molecules in DrugBank and NCFD. }
    \label{fig:SI1}
\end{figure}

\begin{figure}[ht!]
    \centering
    \includegraphics[clip,angle=0,width=1.0\textwidth]{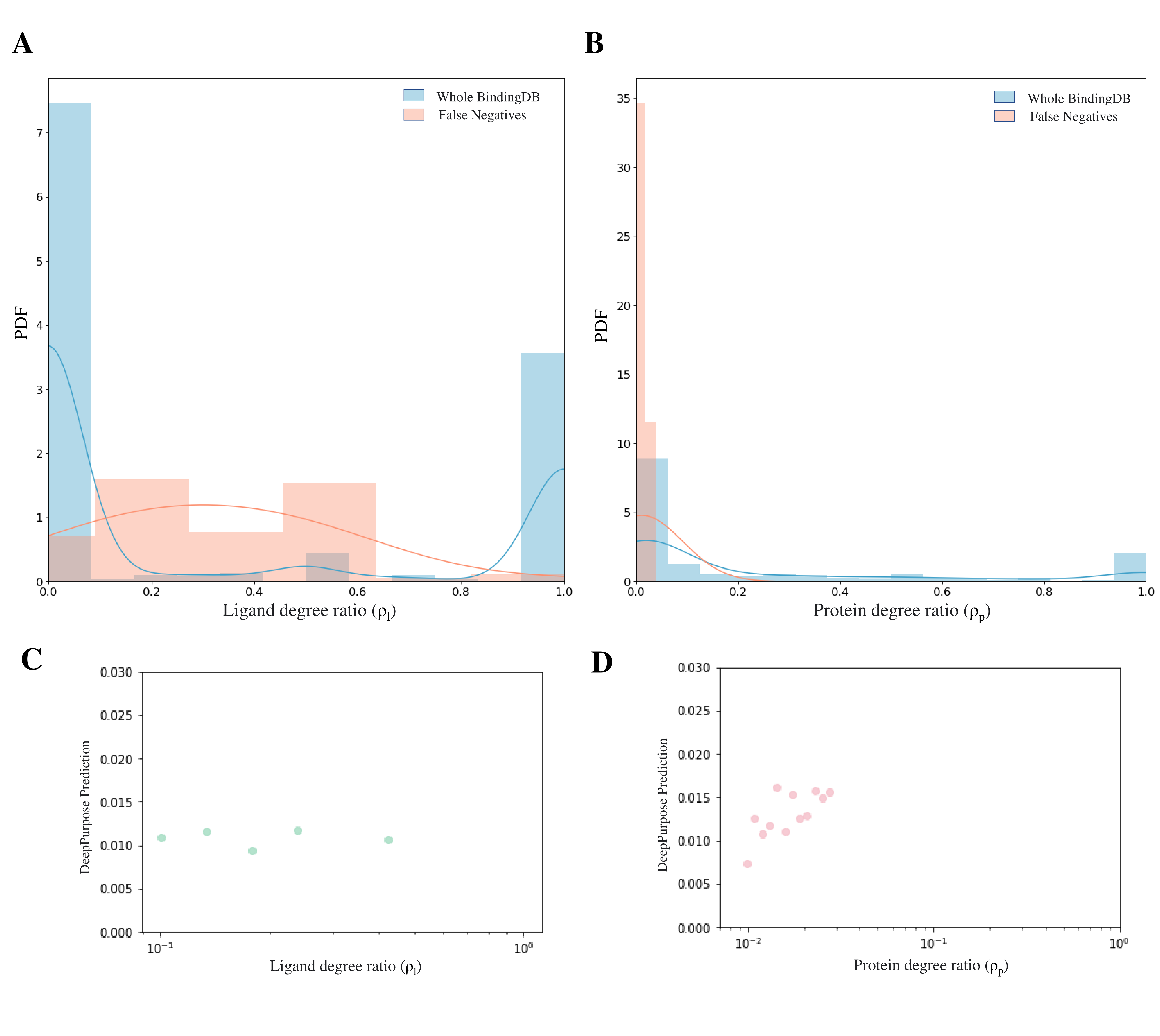}
    \caption{\textbf{Annotation bias in top 100 false negative predictions made by DeepPurpose.} \textbf{(A)-(B)} Degree ratio distribution of the nodes involved in the false negative predictions is shown compared to all the nodes in the BindingDB data. The false negative predictions correspond to proteins and ligands with low degree ratios. \textbf{(C)-(D)} DeepPurpose predicts lower binding probabilities for the nodes with lower degree ratios.}
    \label{fig:SI5}
\end{figure}

\begin{figure}[ht!]
    \centering
    \includegraphics[clip,angle=0,width=1.0\textwidth]{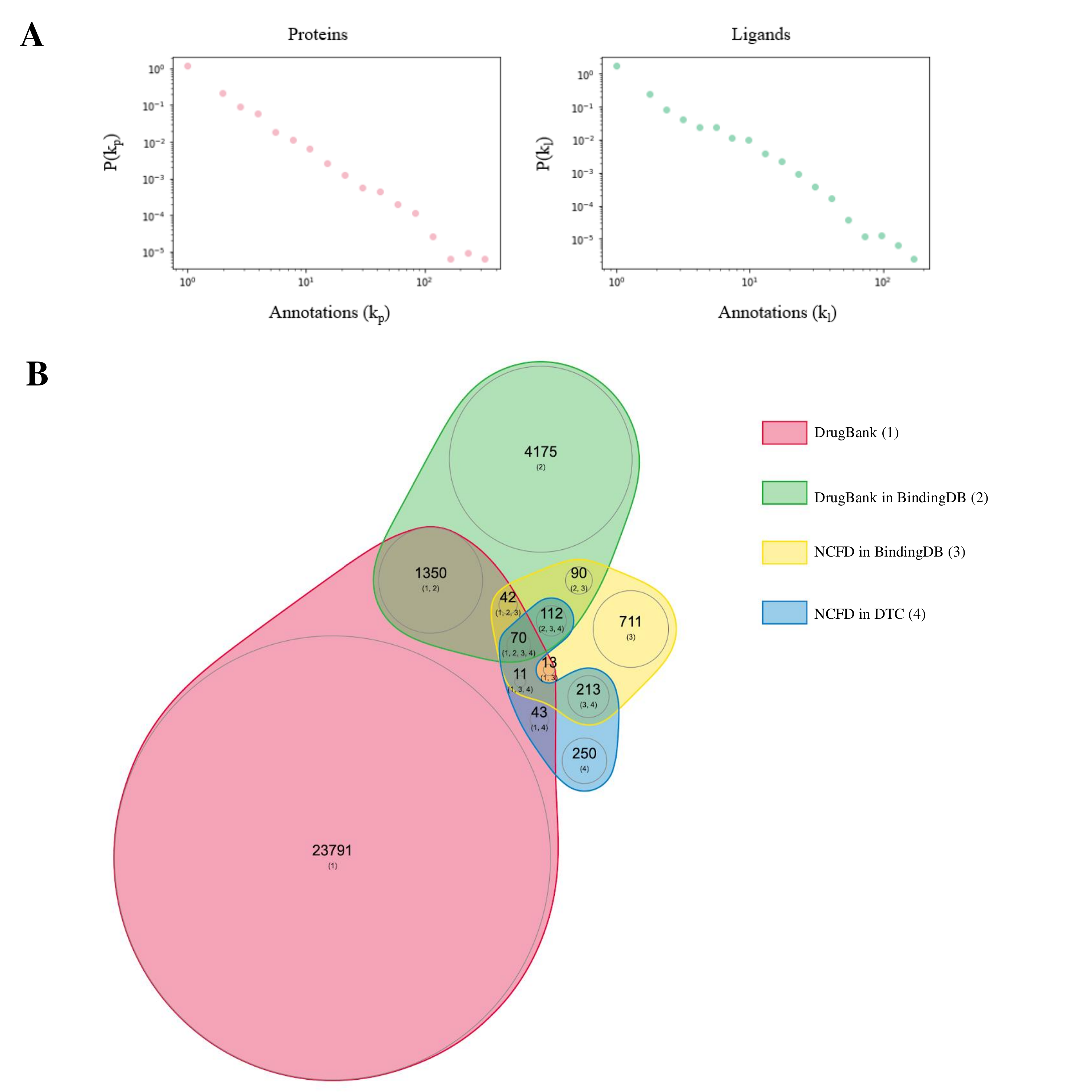}
    \caption{\textbf{Network property of the DrugBank DTI and Venn diagram of positive binding samples across different databases.} \textbf{(A)} Annotation distribution of the proteins and the drugs in DrugBank are fat-tailed. The nature of the annotation distribution is similar to our observations in BindingDB. \textbf{(B)} AI-Bind training data combines protein-ligand binding data from three databases: DrugBank, BindingDB, and Drug Target Commons (DTC). Majority of the binding examples are taken from DrugBank. BindingDB and DTC are used to obtain additional protein-ligand pairs, especially to maximize the binding information involving naturally occurring ligands.}
    \label{fig:SI2}
\end{figure}

\begin{figure}
    \centering
    \includegraphics[width=1\textwidth]{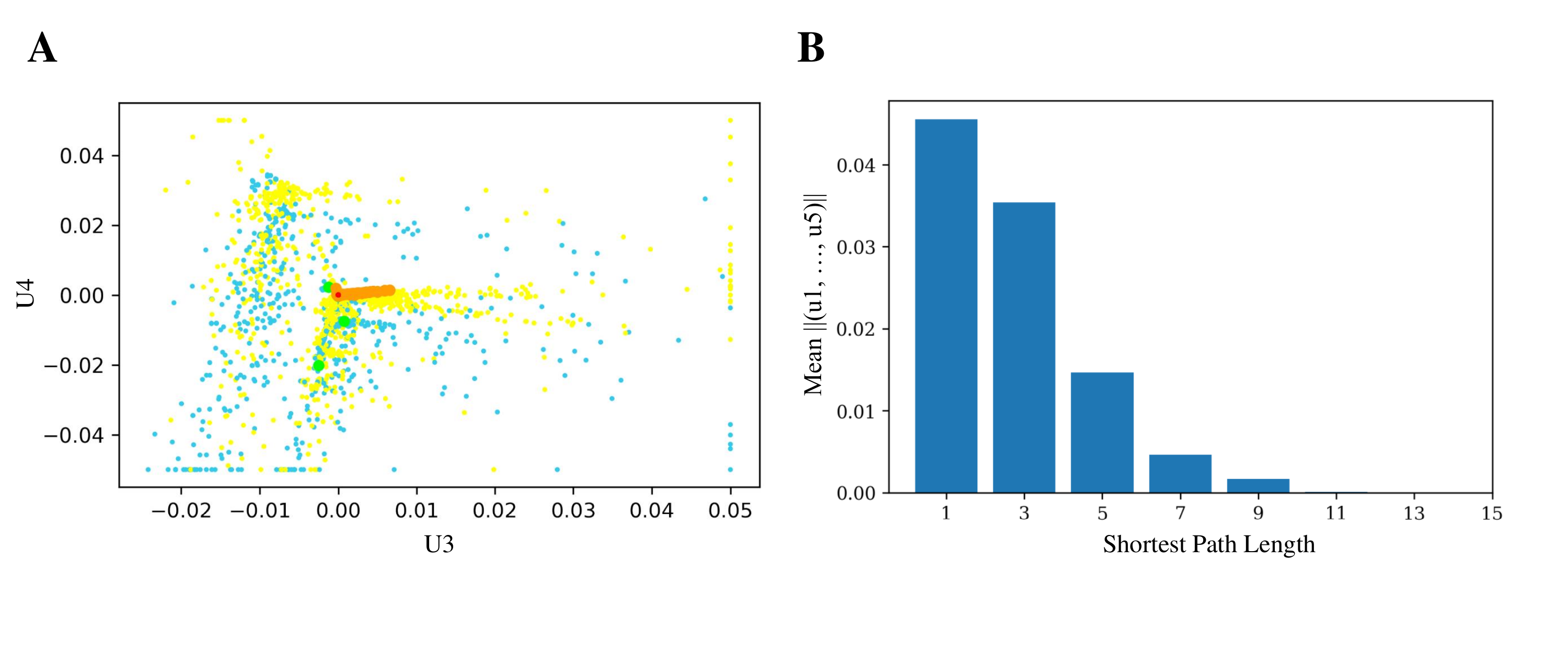}
    \caption{\textbf{EigenSpokes Analysis.} \textbf{(A)} Network-based dimension reduction of nodes in the full protein-ligand network.  Node $i$ is represented by the vector $\bar{u}_i = (u_1,u_2,u_3,u_4,u_5) \in \mathbb{R}^5$.  Here we visualize $(u_3,u_4)$ for only the ligands.  Coloring is based on the hop-distances from an example target BPT4: Green = 1 hop, Blue = 3 hops, Yellow = 5 hops, Orange = 7 hops, Red $\geq$ 9 hops.  We see that at $> 7$ hop, most nodes are very close to the origin. \textbf{(B)} Mean of all reduced vector magnitudes $\|\bar{u}_j\|$ averaged over all pairs $(i,j)$ of a given path length. We see a significant decrease in magnitude as the shortest path length increases.}
   \label{fig:eigenspokes}
\end{figure}

\begin{figure}[ht!]
    \centering
    \includegraphics[clip,angle=0,width=1.0\textwidth]{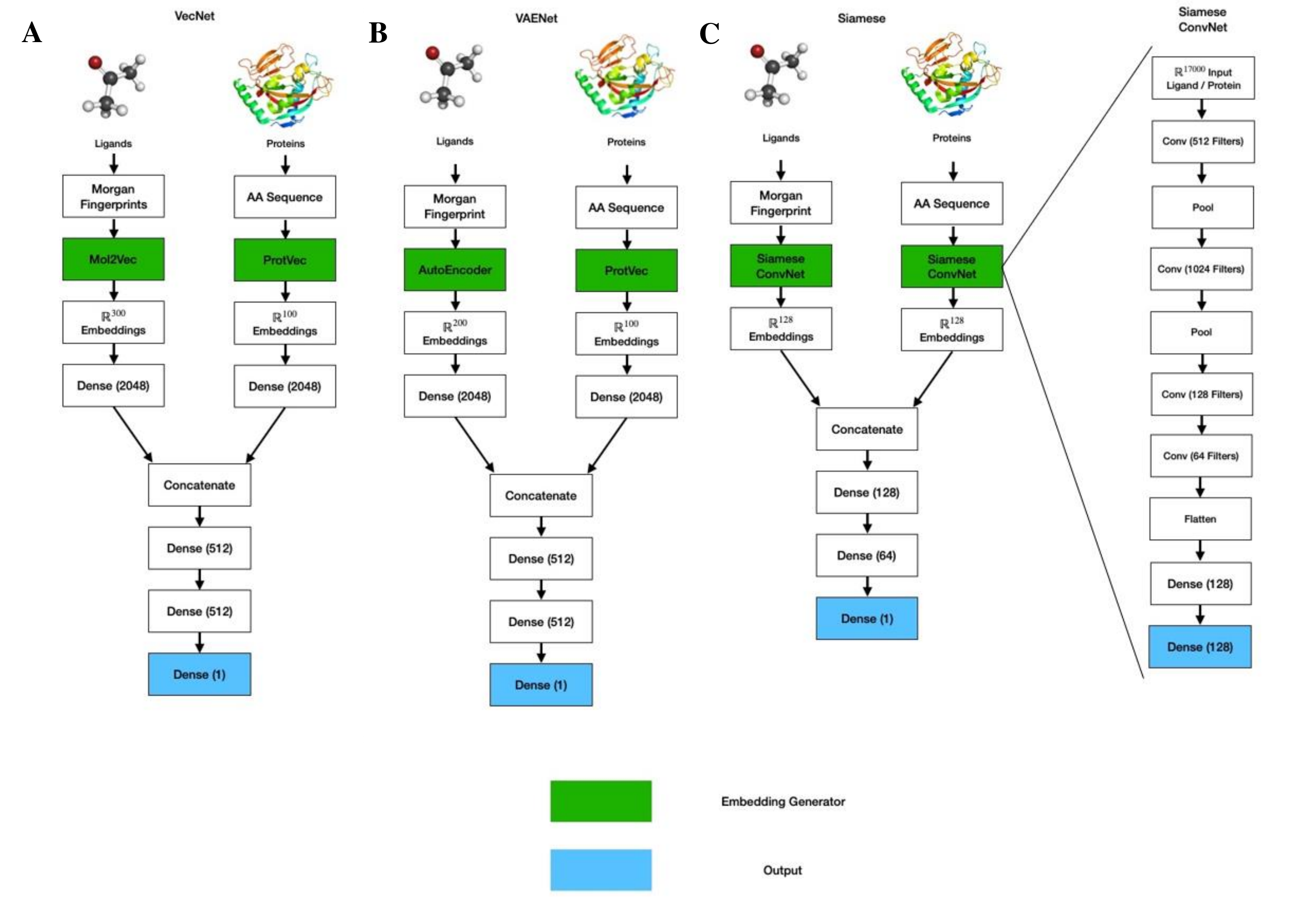}
    \caption{\textbf{Deep architectures of VecNet, VAENet, and Siamese model.} \textbf{(A)} VecNet uses Mol2vec and ProtVec as the unsupervised pre-trained models for ligand and protein embeddings respectively. The dense layers act as decoders, and are trained using the network-derived dataset. \textbf{(B)} VAENet architecture is similar to VecNet, where Mol2vec embeddings are replaced with embeddings obtained from a variational auto-encoder. This auto-encoder is trained on $\approx 9.5$ million compounds from the ZINC database. \textbf{(C)} Siamese model embeds both proteins and ligands onto the same latent space. Siamese ConvNet blocks minimize the triplet loss between the proteins binding to the same ligand. We follow a similar approach for generating the ligand embeddings. }
    \label{fig:Deep_Architectures}
\end{figure}

\begin{figure}[ht!]
    \centering
    \includegraphics[clip,angle=0,width=1.0\textwidth]{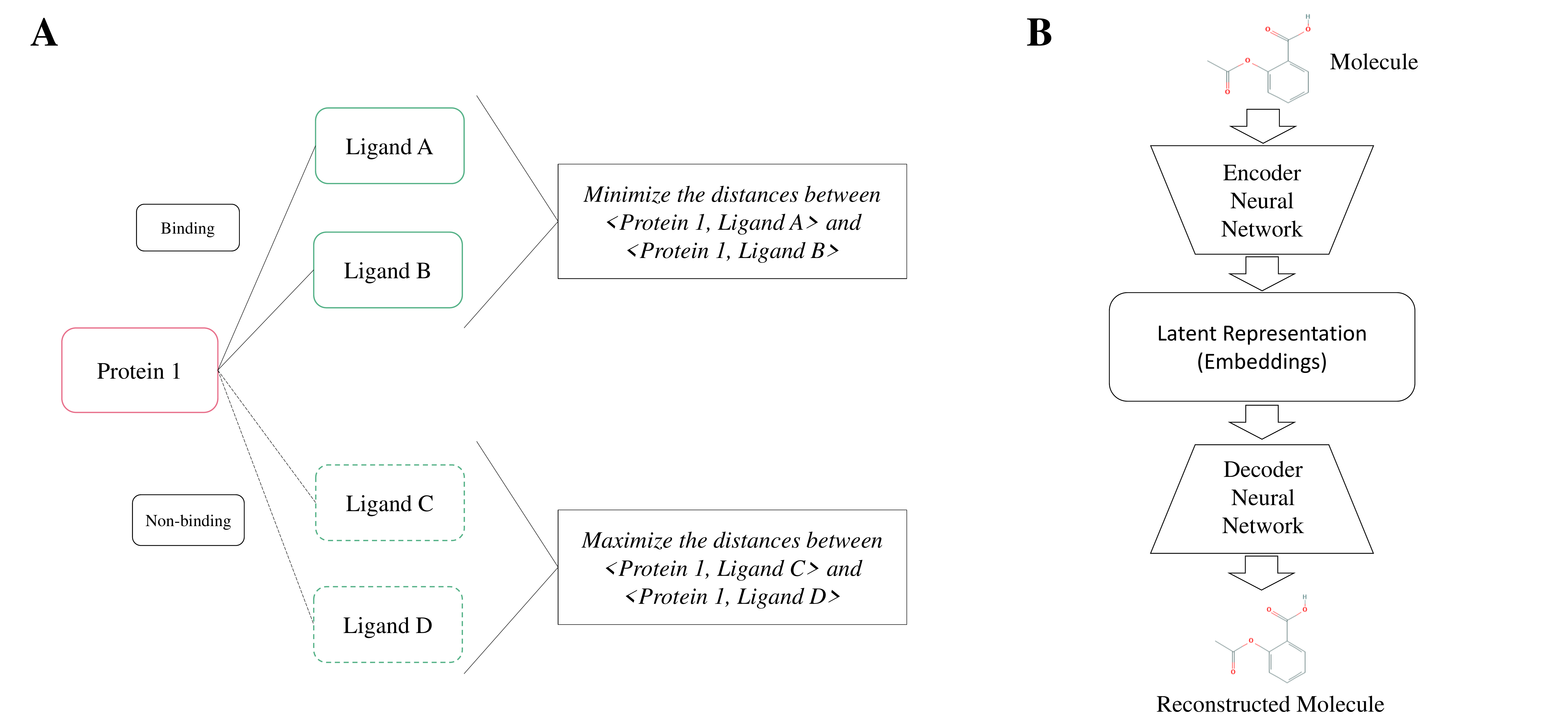}
    \caption{\textbf{Logical flow of the Siamese Model and Variational Auto-Encoder}. \textbf{(A)} We minimize the embedded Euclidean distances between the proteins which bind to the same ligand, and maximize the distance between the non-binding ones. Similar logic is applied for creating the ligand embeddings. \textbf{(B)} Variational auto-encoder minimizes the reconstruction loss for the ligands to create a latent space embeddings. We generate Morgan fingerprints from the isomeric SMILES and feed that to the auto-encoder. The auto-encoder generates latent space representations by minimizing reconstruction loss on the fingerprints.}
    \label{fig:model_flow}
\end{figure}

\begin{figure}[htb]
    \centering
    \includegraphics[clip,angle=0,width=0.8\textwidth,height=0.9\textheight]{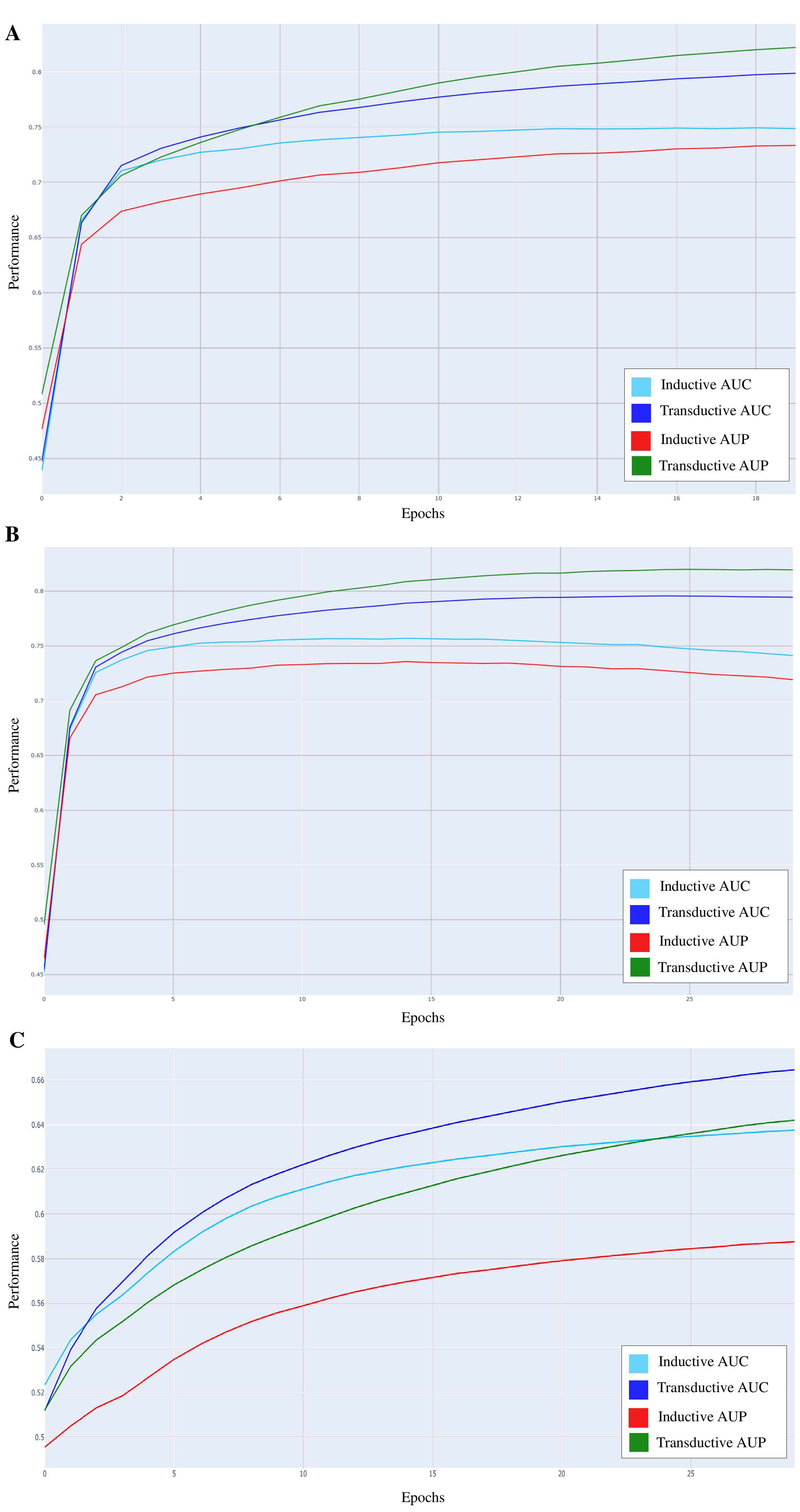}
    \caption{\textbf{Training curves for three AI-Bind architectures.}  We plot the training curves for \textbf{(A)} VecNet, \textbf{(B)} VAENet, and \textbf{(C)} Siamese model over 30 epochs. The AUROC and the AUPRC are separately shown for the transductive (unseen edges) and inductive (unseen nodes) test scenarios.}
    \label{fig:training_curve}
\end{figure}

\begin{figure}[htb]
    \centering
    \includegraphics[clip,angle=0,width=0.8\textwidth,height=0.8\textheight]{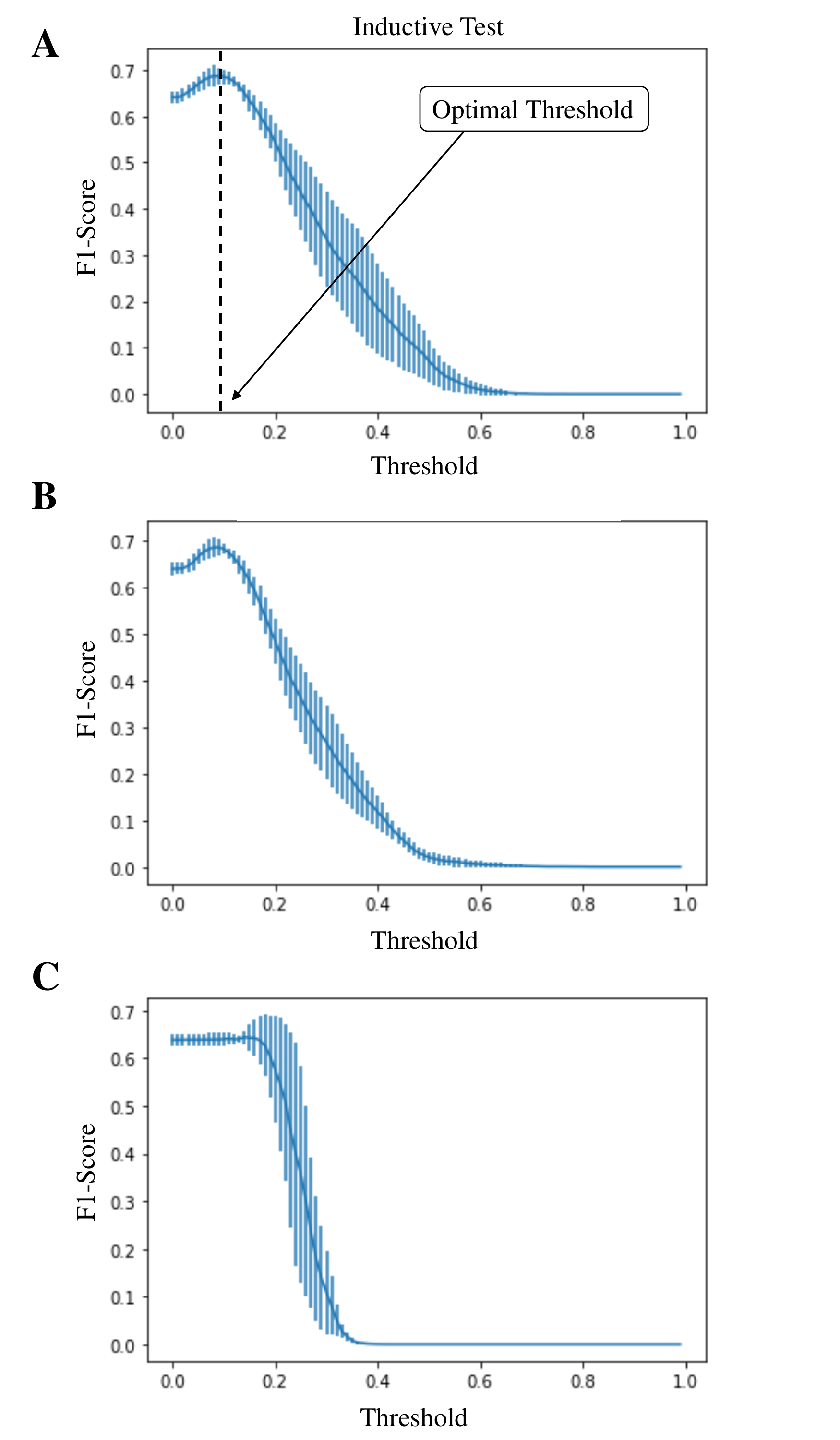}
    \caption{\textbf{F1-Score and Optimal Threshold.} We plot the F1-scores for the trained \textbf{(A)} VecNet, \textbf{(B)} VAENet, and \textbf{(C)} Siamese models relative to the classification threshold in the inductive test scenario. The threshold value corresponding to the highest F1-score is considered as the optimal threshold, and is used to obtain the binary labels from the predicted binding probabilities. For VecNet, we obtain an optimal threshold of $0.09$.}
    \label{fig:f1_score_threshold}
\end{figure}

\begin{figure}[h!]
    \centering
    \includegraphics[clip,angle=0,width=\textwidth,height=0.7\textheight]{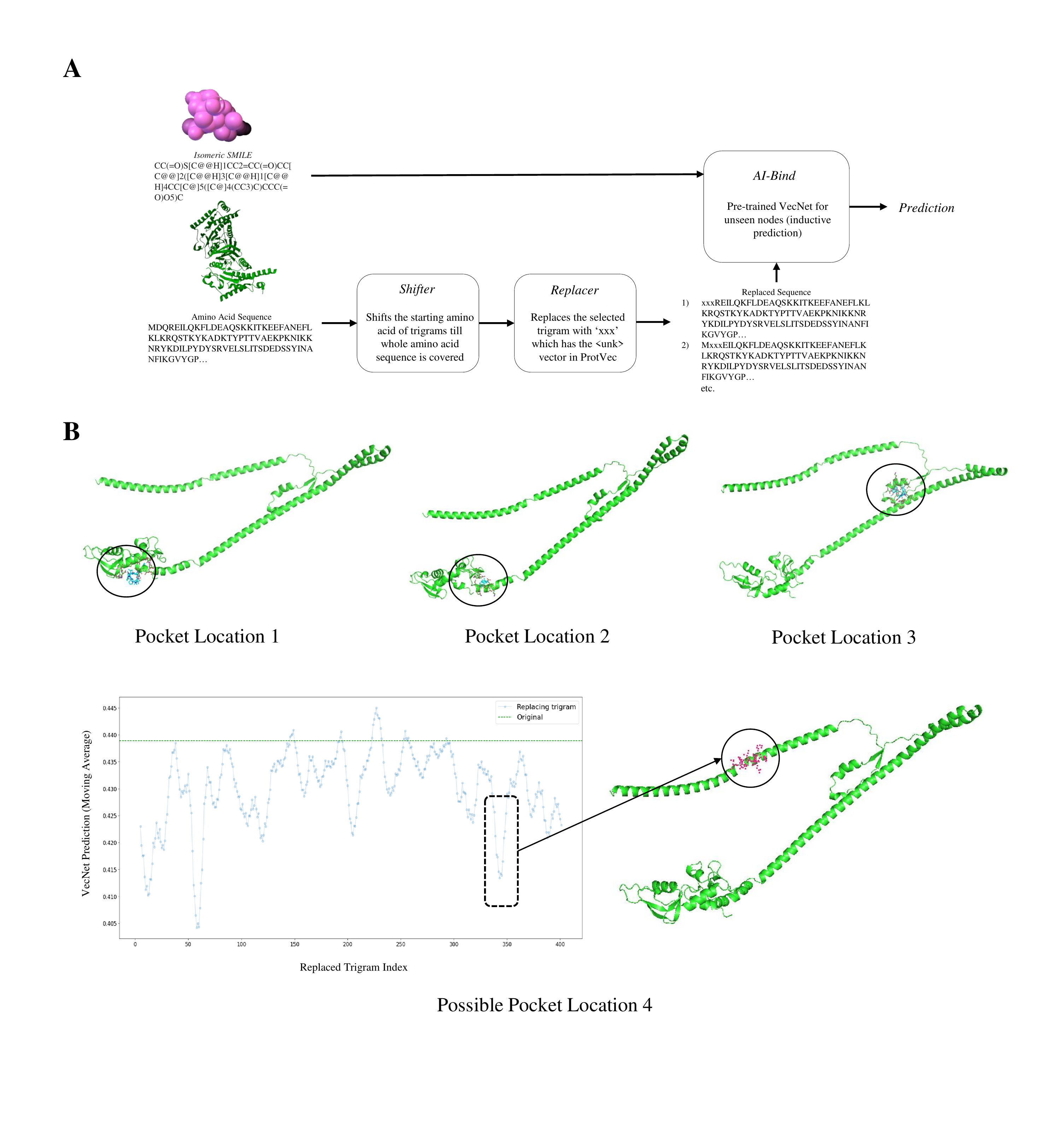}
    \caption{\textbf{Methodology for interpreting AI-Bind predictions.} \textbf{(A)} We replace each amino acid trigram in the amino acid sequence with \textit{xxx}, which maps to the out-of-vocabulary \textit{$\langle$unk$\rangle$} vector in ProtVec, and observe the fluctuations in AI-Bind prediction. Removal of some trigrams affect AI-Bind prediction more than others. These trigrams are indicative of the binding location(s) on the protein. \textbf{(B)} We identify three active binding sites on the protein TRIM59 from the docking simulations and map them to AI-Bind's binding probability profile. We also identify a possible pocket location from the binding probability profile and visualize that on the 3D protein structure.}
    \label{fig:trigram_study_1}
\end{figure}

\begin{figure}[h!]
    \centering
    \includegraphics[clip,angle=0,width=0.65\textwidth,height=0.7\textheight]{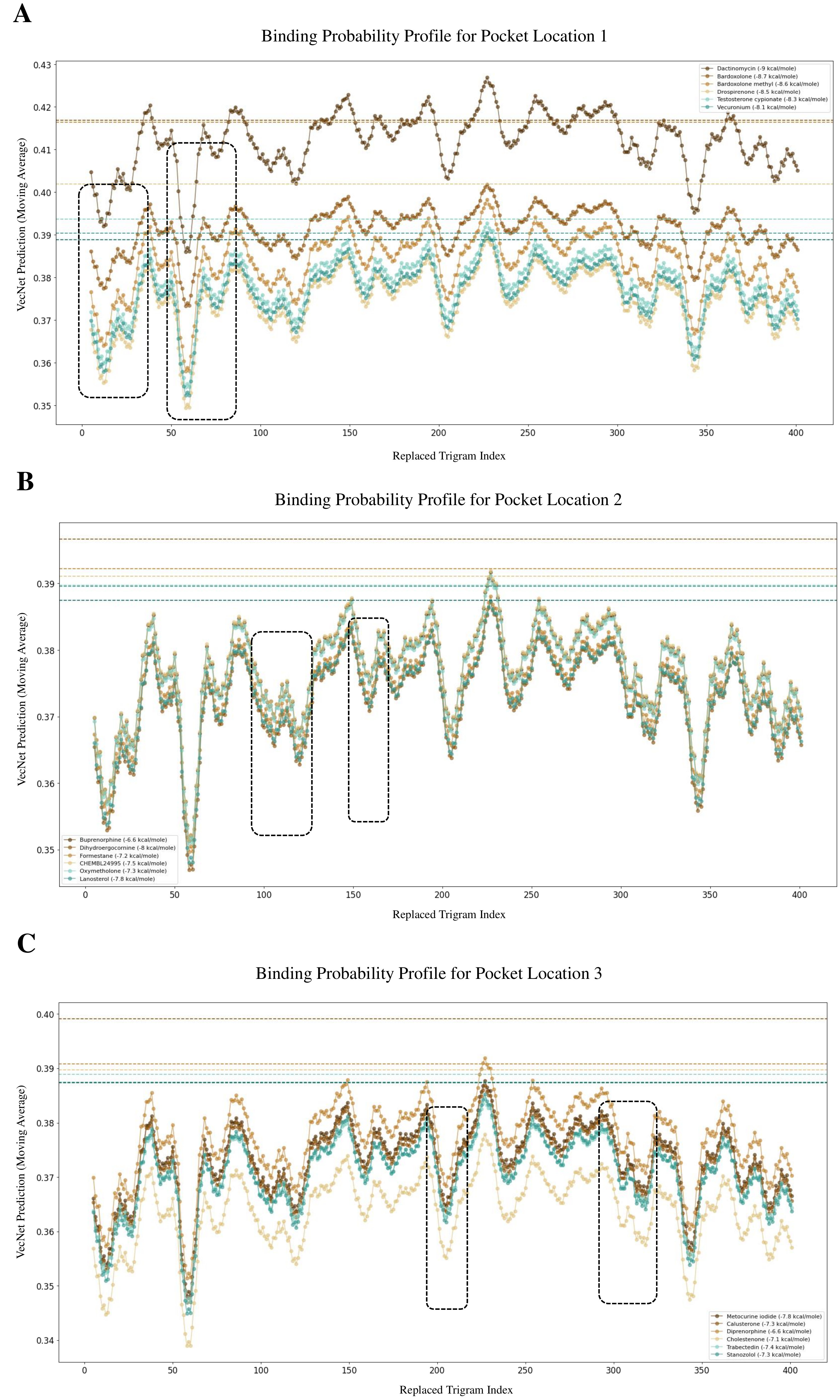}
    \caption{\textbf{Binding probability profiles for different active binding sites on TRIM59.} \textbf{(A)-(C)} We group the ligands based on the binding pockets on TRIM59 and plot the binding probability profiles, highlighting the binding locations on the amino acid sequence. We observe similar shape of the binding probability profiles for different ligands, but the deviation from the original AI-Bind prediction varies across the ligands, which conveys the dependency of the binding probability profile on the ligand structure.}
    \label{fig:trigram_study_2}
\end{figure}


\begin{figure}[h!]
    \centering
    \includegraphics[clip,angle=0,width=0.8\textwidth,height=0.85\textheight]{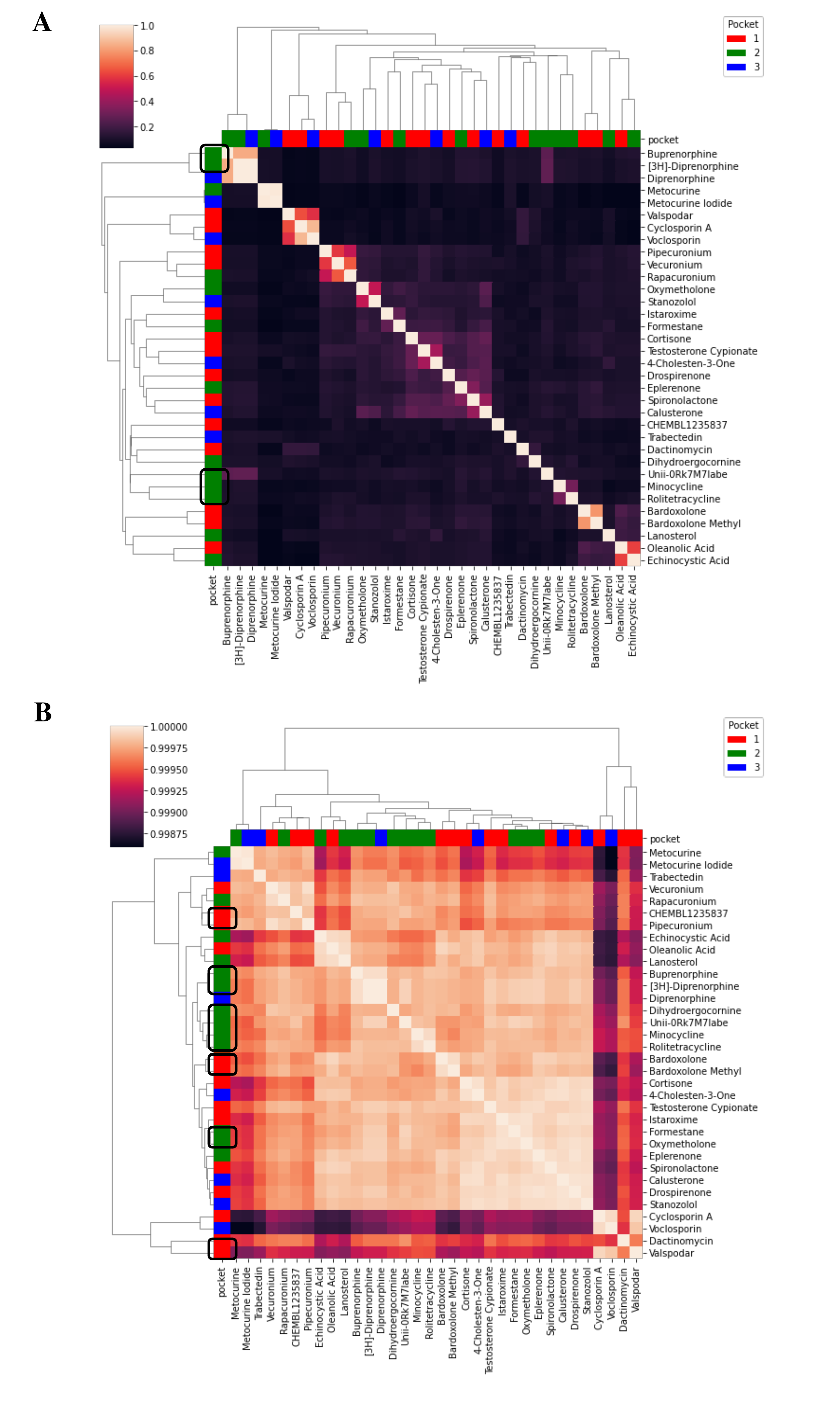}
    \caption{\textbf{Hierarchical clustering on binding probability profiles.} \textbf{(A)} We plot the heatmap of the Tanimoto similarities between the ligands binding to TRIM59. We do not observe significant grouping of the ligands solely based on their molecular structures. \textbf{(B)} We cluster the ligands based on he similarities of their binding probability profiles. We observe that multiple ligands binding to the same pocket are clustered together in the clustermap. Thus, the binding probability profiles generated by AI-Bind are not only specific to a protein, but carry information about the ligand structures.}
    \label{fig:trigram_study_3}
\end{figure}

\begin{figure}[h!]
    \centering
    \includegraphics[clip,angle=0,width=\textwidth,height=0.7\textheight]{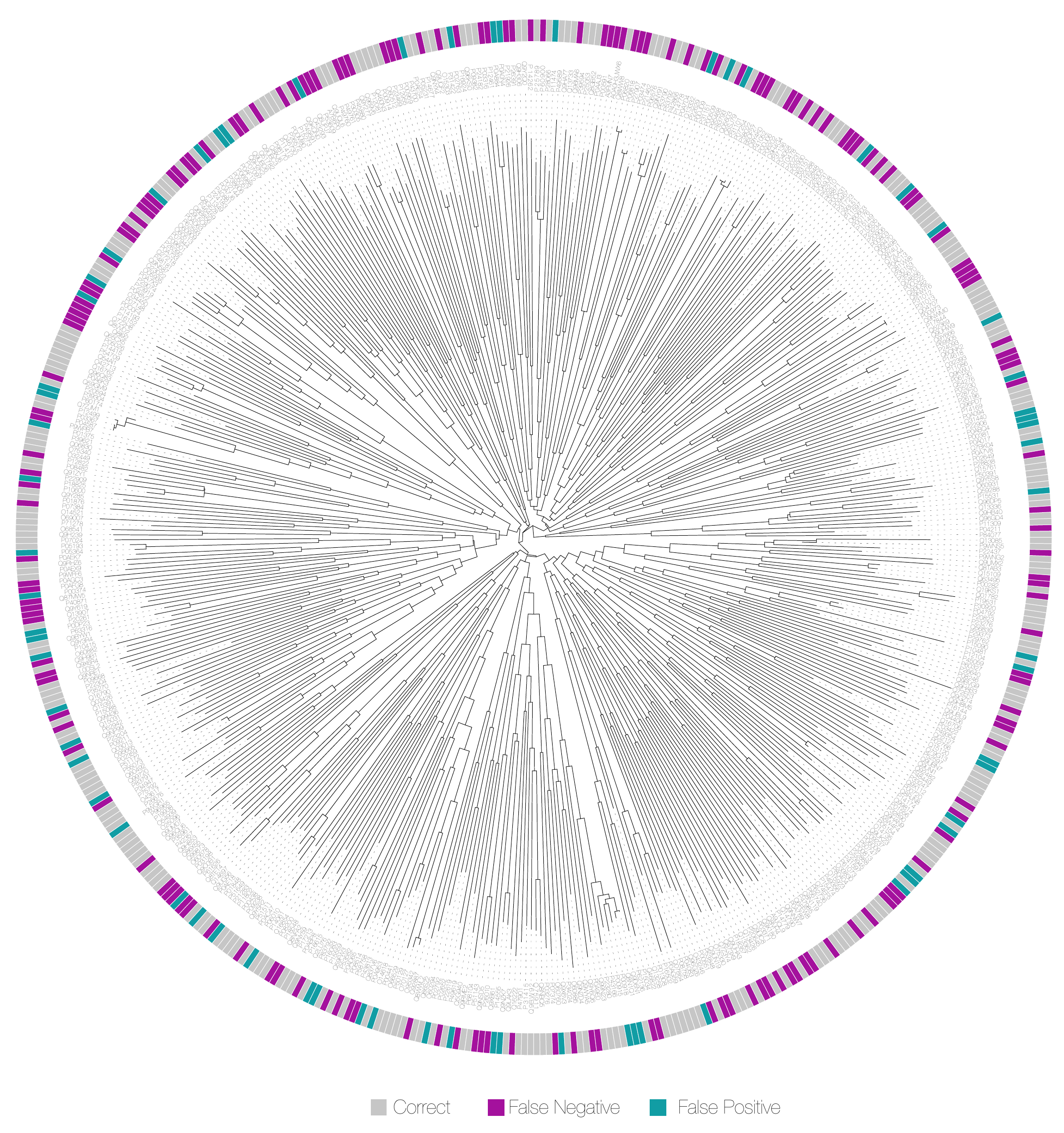}
    \caption{\textbf{Phylogenetic tree of genes enriched towards prediction bias.} We compare proteins associated with the false predictions (both false positives and false negatives) made by AI-Bind's VecNet to uncover structural similarities. AI-Bind does not show any bias towards certain protein structures in the false predictions, and can be used for binding prediction involving protein structures emerging from different organisms.} 
    \label{fig:SpecieTree}
\end{figure}

\begin{figure}[h!]
    \centering
    \includegraphics[clip,angle=0,width=0.6\textwidth,height=0.9\textheight]{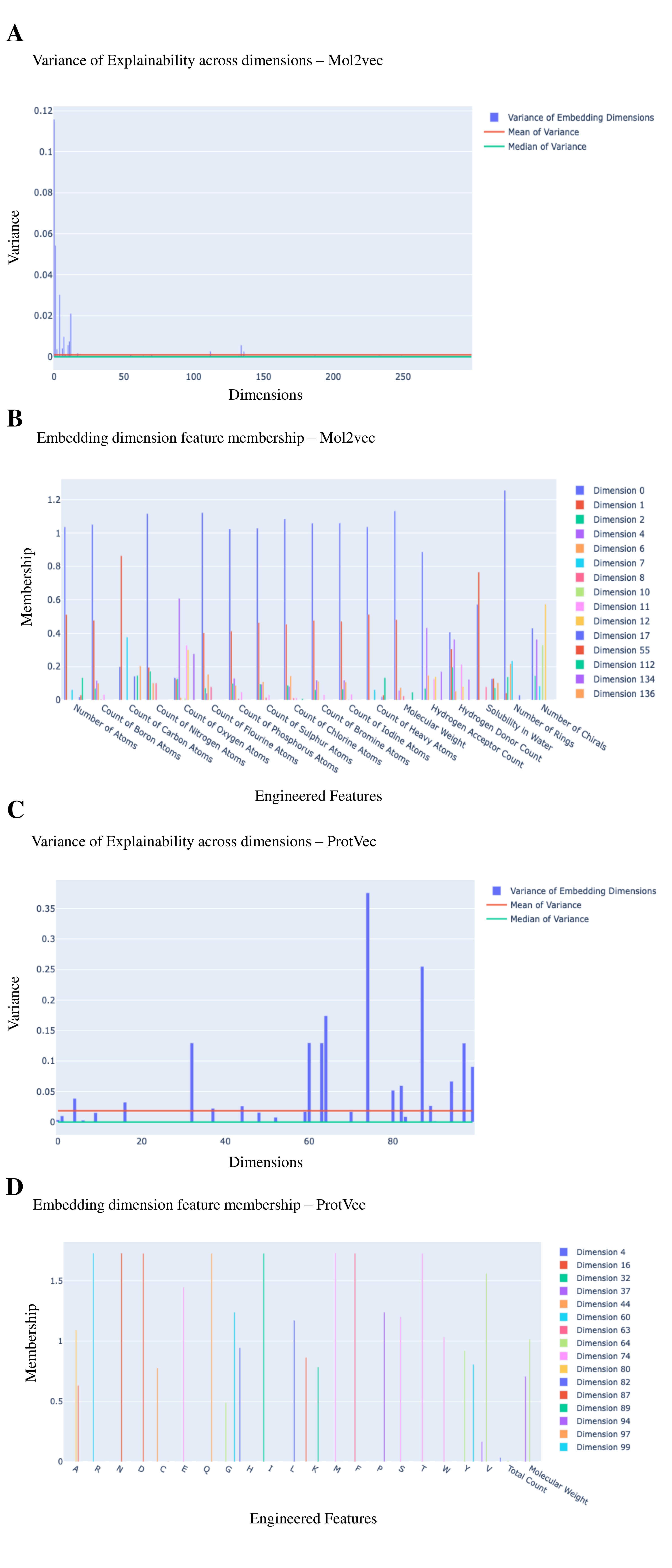}
    \caption{\textbf{Dimensions of Mol2vec and ProtVec contributing to protein-ligand binding.} \textbf{(A)-(B)} Only 15 Mol2vec dimensions show high variability when explaining the engineered features representing ligand molecules. \textbf{(C)-(D)} We find similar results for 16 ProtVec dimensions.}
    \label{fig:engineered_embed}
\end{figure}


\begin{figure}[ht!]
    \centering
    \includegraphics[clip,angle=0,width=1.0\textwidth]{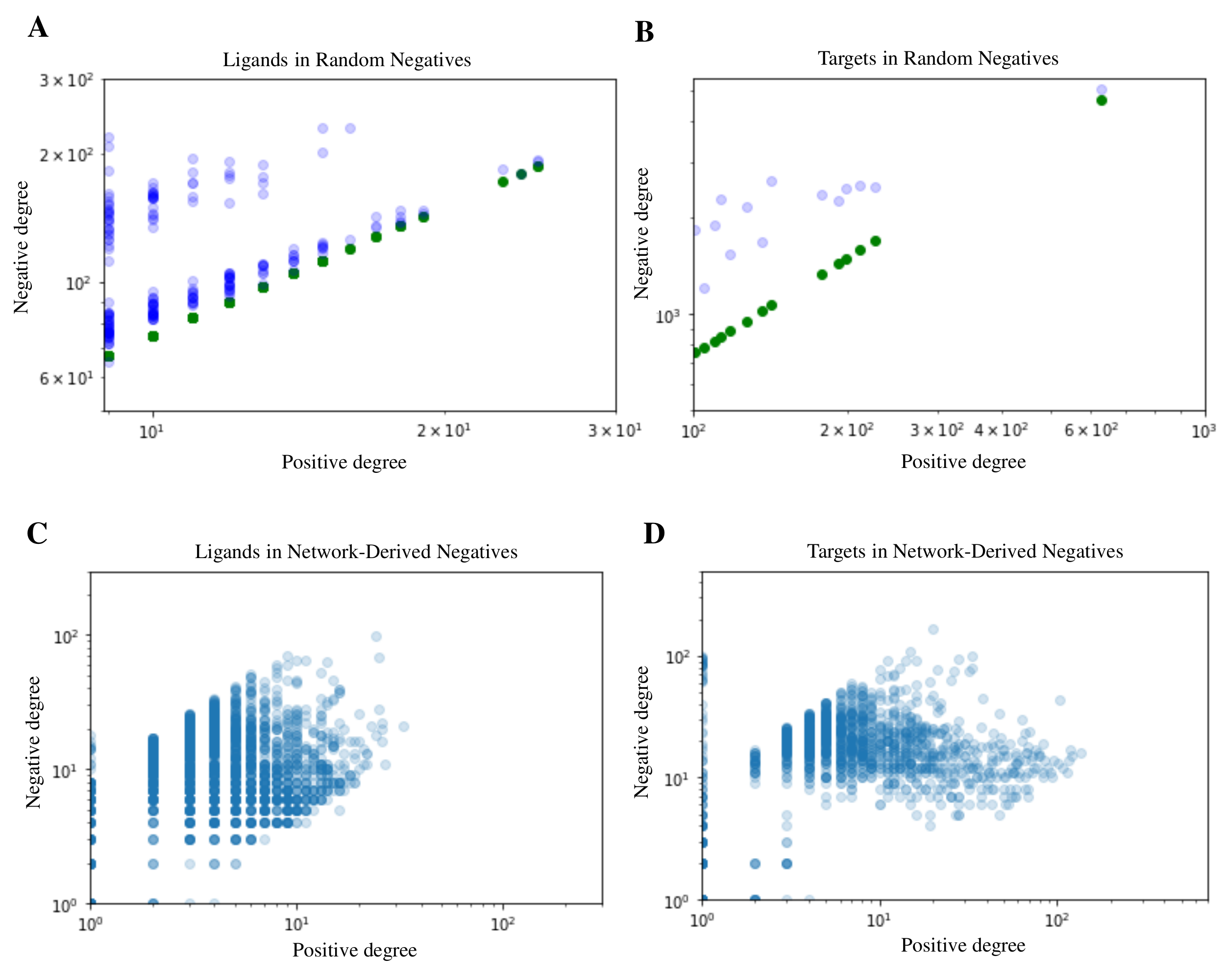}
    \caption{\textbf{Random vs. network-derived negatives}. \textbf{(A)-(B)} In random negative sampling, both the ligand and the protein on a positive edge has the lower bound of negative degree equal to 7.5 times its positive degree. Higher positive degree nodes have lower probabilities of being present in a random negative sample, as they are present in many positive edges and are discarded more often from getting included in a negative sample. Thus, the negative degree diminishes as the positive degree increases. \textbf{(C)-(D)} We observe less correlation between positive and negative degrees for the network-derived negatives. This helps in  removing the annotation imbalance we observe in the existing protein-ligand databases.}
    \label{fig:SI3}
\end{figure}

\begin{figure}[ht!]
    \centering
    \includegraphics[clip,angle=0,width=1.0\textwidth]{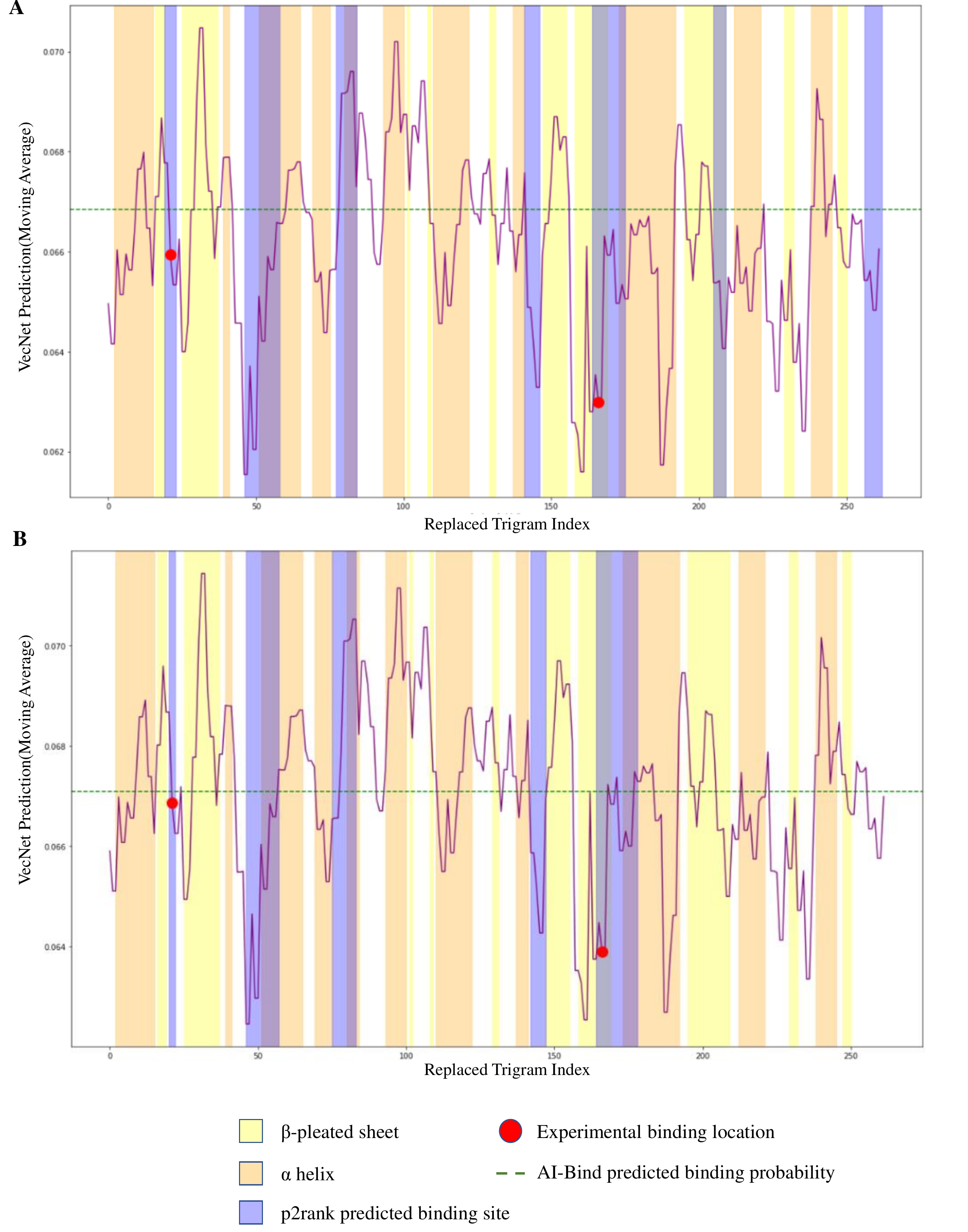}
    \caption{\textbf{Experimental binding sites lie on the valleys of the binding probability profile.} We plot the binding probability profile for the \textit{E. Coli} protein Thymidylate Synthase, and the ligands \textbf{(A)} SP-722 and \textbf{(B)} SP-876. We observe that the experimentally obtained binding sites are in the valleys of the binding probability profile, and overlay on the the $\beta$-sheets and the coils regions. These binding locations also overlap with the binding locations predicted by p2rank, a state-of-the-art binding site prediction algorithm.}
    \label{fig:gold_standard_validation}
\end{figure}

\begin{figure}[ht!]
    \centering
    \includegraphics[clip,angle=0,width=1.0\textwidth]{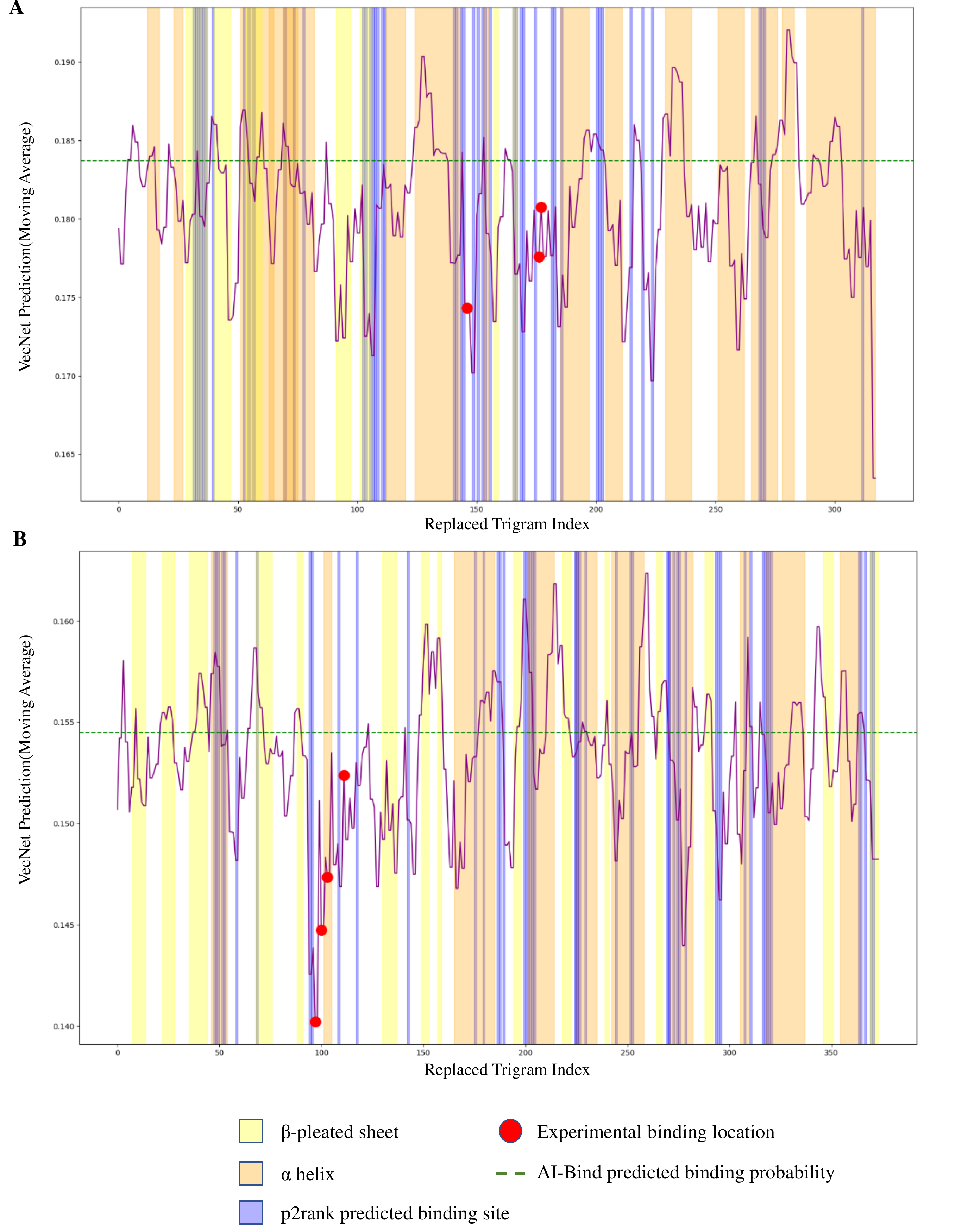}
    \caption{\textbf{Experimental binding sites lie on the valleys of the binding probability profile for human proteins TAO3 Kinase and Human Alcohol Dehydrogenase.} We plot the binding probability profile for the human protein and ligand pairs \textbf{(A)} Human TAO3 Kinase and ADP \textbf{(B)} Human Alcohol Dehydrogenase and Nicotinamide Adenine Dinucleotide. We observe that the experimentally obtained binding sites are in the valleys of the binding probability profile, and often overlay on the the $\beta$-sheets and the coils regions. These binding locations also overlap with the binding locations predicted by p2rank, a state-of-the-art binding site prediction algorithm.}
    \label{fig:gold_standard_validation_human}
\end{figure}

\end{document}